\documentclass{article}

\usepackage{arxiv}

\usepackage{amsmath}
\usepackage[utf8]{inputenc} 
\usepackage[T1]{fontenc}    
\usepackage{hyperref}       
\usepackage{url}            
\usepackage{booktabs}       
\usepackage{amsfonts}       
\usepackage{nicefrac}       
\usepackage{microtype}      
\usepackage{cleveref}       
\usepackage{lipsum}         
\usepackage{graphicx}
\usepackage{natbib}
\usepackage{doi}
\usepackage{bbm}
\usepackage{multirow}
\usepackage{bm}

\usepackage{amsfonts}
\usepackage{mathrsfs}
\usepackage{graphicx}
\usepackage{caption}
\usepackage{subcaption}
\usepackage[skip=1pt]{subcaption}
\usepackage{changepage}
\usepackage{booktabs}
\usepackage[shortlabels]{enumitem}
\usepackage{caption}
\usepackage{algorithm}
\usepackage{algpseudocode}

\DeclareMathOperator*{\argmin}{arg\,min}

\newtheorem{assumption}{Assumption}
\newtheorem{lemma}{Lemma}
\newtheorem{proposition}{Proposition}

\usepackage{color, colortbl}	
\definecolor{Gray}{gray}{0.9}

\title{A New Framework for Bayesian Function Registration}

\author{ Yijia Ma  \qquad\qquad Wei Wu\\ 
	Department of Statistics, 
	Florida State University\\
	Tallahassee, FL 32306 \\
	\texttt{ym19f@my.fsu.edu }\quad\texttt{wwu@stat.fsu.edu}
}

\date{}

\hypersetup{
pdftitle={A New Framework for Bayesian Function Registration},
}

\begin{document}
\maketitle

\begin{abstract}
	Function registration, also referred to as alignment, has been one of the fundamental problems in the field of functional data analysis.  Classical registration methods such as the Fisher-Rao alignment focus on estimating optimal time warping function between functions.  In recent studies, a model on time warping has attracted more attention, and it can be used as a prior term to combine with the classical method (as a likelihood term) in a Bayesian framework. The Bayesian approaches have been shown improvement over the classical methods. However, its prior model on time warping is often based a nonlinear approximation, which may introduce inaccuracy and inefficiency.  To overcome these problems, we propose a new Bayesian approach by adopting a prior which provides a linear representation and various stochastic processes (Gaussian or non-Gaussian) can be effectively utilized on time warping.  No linearization approximation is needed in the time warping computation, and the posterior can be obtained via a conventional Markov Chain Monte Carlo approach.  We thoroughly investigate the impact of the prior on the performance of functional registration with multiple simulation examples, which demonstrate the superiority of the new framework over the previous methods.  We finally utilize the new method in a real dataset and obtain desirable alignment result. 
\end{abstract}

\keywords{Bayesian registration \and isometric isomorphism \and isometric isomorphism \and time warping functions}


\section{Introduction}

In functional data analysis (FDA), one of the fundamental and extensively studied issues is function registration. The goal of function registration is to separate phase and amplitude variabilities, where amplitude variation signifies changes in the magnitude of function, while phase variation represents shifts or timing differences in the data. Time warping is an important notion to characterize the phase variability. A common space of time warping functions is defined as 
\begin{equation}
\Gamma=\{\gamma:[0,1]\rightarrow [0,1]|\gamma (0)=0, \gamma(1)=1,0<\dot\gamma(t)<\infty\},
\label{eq:gamma}
\end{equation}
which is an infinite-dimensional nonlinear manifold. Function registration is a process to identify the optimal time warpings to properly align functional observations. Over the past two-to-three decades, a variety of methods have been developed for estimating optimal time warping functions. Early approaches formulated a least-square problem by representing warping function through a linear combination of B-spline basis functions, and the warping can be obtained by estimating the corresponding coefficients \citep{ramsay1998curve, gervini2004self, james2007curve, eilers2004parametric}. More recent approaches conducted registration by minimizing the Fisher-Rao metric \citep{srivastava2011registration,wu2014analysis}. However, these methods are primarily loss-function-based rather than model-based. 

Bayesian registration is a relatively new concept that involves integrating prior knowledge about time warping, and incorporates the classical methods as a likelihood component to conduct function alignment \citep{cheng2014analysis, cheng2016bayesian, lu2017bayesian, kurtek2017geometric,tucker2021multimodal,matuk2021bayesian}. However, it is worth noting that existing studies in Bayesian registration have either focused on employing simple models, such as the Dirichlet distribution \citep{cheng2016bayesian}, or implemented within the SRVF (Squared Root Velocity Function) framework \citep{kurtek2017geometric, lu2017bayesian}. However, there are several basic problems with the SRVF framework: 1. It relies on a nonlinear exponential mapping between the positive orthant of a hyper-sphere and its tangent linear space. The functions within the tangent space are approximations and could not accurately represent the original time warpings.  2. This exponential mapping is not bijective, and these two spaces are not isometrically isomorphic, meaning that functions in the tangent space might not correspond to appropriate time warping functions \citep{happ2019general}. 3. In Monte Carlo computations for registration, an extra step is needed to discard inappropriate warping proposals, which can result in both inaccuracy and inefficiency in the estimation process.

To address these problems, we propose to adopt a linear framework on the warping function \citep{ma2024stochastic}, which offers several advantages: Firstly, it enables an exact representation of the time warping function in the conventional $\mathbb{L}^2$ space, thereby effectively eliminating approximation errors. Secondly, the isometric isomorphism between the warping space and a conventional $\mathbb{L}^2$ subspace ensures the reliability of linear operations within this framework. These characteristics offer multiple benefits for Bayesian registration. For instance, all procedures are streamlined within the conventional $\mathbb{L}^2$ space, simplifying both the theoretical framework and its implementation. The isometric isomorphism ensures the warpings proposed in Monte Carlo simulations are appropriate for use, thereby improving the efficiency of the algorithm. Additionally, the linear space facilitates the exploration of various Gaussian priors with different means and covariances, aiding the registration process and yielding desirable results. The use of non-Gaussian priors would also be feasible, which opens the door to discovering new possibilities. Moreover, the method can be easily extended from pairwise to multiple function registration, enhancing its applicability.

Previous studies in the Bayesian registration have predominantly focused on comparing the alignment performance with the dynamic programming estimation in the Fisher-Rao method, aiming to demonstrate that Bayesian registration outperforms dynamic programming \citep{cheng2016bayesian, lu2017bayesian, kurtek2017geometric,tucker2021multimodal,matuk2021bayesian}. However, from our perspective, dynamic programming is adept at finding the globally optimal warping without any prior model, a challenging feat to surpass. In this paper, our attention is directed towards the inherent advantages of Bayesian methods, particularly their capacity to provide variability in results. Additionally, Bayesian methods enable incorporating prior information to constrain outcomes, allowing us to manipulate the prior to exert control over the alignment results. Therefore, our focus in this paper diverges from previous approaches as we concentrate on thoroughly exploring these aspects.

The rest of this manuscript is organized as follows: In Section \ref{Sec: BayesReg}, we will present a comprehensive description of the linear-model-based Bayesian registration for pairwise function registration, by introducing a new Gaussian Process prior. In particular, our framework will be expanded to include the Non-Gaussian prior, demonstrating its utility in scenarios involving non-continuity. In Section \ref{Sec:prior}, we conduct an exploration to delineate the effects of the prior and conduct a comparative analysis between our proposed method and existing approaches, highlighting the distinct advantages of our methodology. Section \ref{Sec: mul} will further extend our framework to multiple function registration. A real-world application is given in Section \ref{Sec:realdata}.  Finally, we will summarize our study and outline future work in section \ref{Sec: disc}.

\section{Bayesian registration }
\label{sec: pairwise}
Our new Bayesian framework is based on a recently proposed linear, inner product representation on time warping functions.  We will at first provide a brief review on this representation.

\label{Sec: BayesReg}
\subsection{Time warping and its $\mathbb{L}^2$ representation }
In the Bayesian framework, we focus on constructing a proper prior model on the time warping functions.  However, this is a challenging task for functions in the domain $\Gamma$ in Eqn. \eqref{eq:gamma}.  As the derivative $\dot \gamma$ is unbounded, it may have infinite first and second order moments and conventional stochastic processes such as Gaussian process can not be directly applied.  To address this issue, a subset of $\Gamma$ with bounded derivative was proposed in \cite{ma2024stochastic}.  That is,  
\begin{equation}
\Gamma_1=\{\gamma:[0,1]\rightarrow [0,1]|\gamma (0)=0, \gamma(1)=1, 0<m_\gamma< \dot\gamma(t)<M_\gamma<\infty\}.
\label{eq:gamma1}
\end{equation}
In this new domain the two bounds, $m_\gamma$ and $M_\gamma$, vary with respect to the function $\gamma$.   It is easy to verify that $\Gamma_1$ is a {\em group} with the action of function composition.   In this paper, we will focus on time warping functions from $\Gamma_1$.  This boundedness assumption can always be satisfied in practical functional observations.  

It was shown in \cite{ma2024stochastic} that $\Gamma_1$ is a linear, inner-product space with the defined addition, scale multiplication, and inner-product operations.  
In addition, $\Gamma_1$ can be transformed via an {\em isometric isomorphism} to a subspace of the conventional $\mathbb L^2$ space, i.e., 
$$H(0,1)=\Big\{h\in \mathbb L^2([0,1])|\int_{0}^{1} h(t)\,dt =0, -\infty<m_h< h(t)<M_h<\infty\Big\},$$  
where the two bounds, $m_h$ and $M_h$, also vary with respect to the function $h$.
This transformation $\psi_B: \Gamma_1 \rightarrow H(0,1)$, referred to as Centered Log-Ratio (CLR)  \citep{egozcue2006hilbert},  is given in the following form: 
$$h(t)=\psi_B(\gamma)(t)=\log(\dot{\gamma}(t))-\int_{0}^{1} \log(\dot{\gamma}(s))\,ds. $$ 
The inverse transformation $\psi_B^{-1}: H(0,1) \rightarrow \Gamma_1$ can be written as:
$$\gamma(t)=\psi_B^{-1}(h)(t)=\frac{\int_{0}^{t}\exp(h(s))\,ds}{\int_{0}^{1}\exp(h(\tau))\,d\tau}.$$
With this CLR transformation, both linear and inner-product operations in $\Gamma_1$ are converted to the classical linear and inner-product operations in the conventional $\mathbb L^2$ space. Once a time warping function is transformed to the $\mathbb L^2$ space, we can model it using a stochastic process. Because the mapping is an isometric isomorphism, the stochastic framework indeed directly models the time warping functions in $\Gamma_1$.  This is in contrast to the SRVF-based methods where time warping space is a nonlinear manifold and its model is approximated with the classical tangent projection \citep{srivastava2011registration, lu2017bayesian}.  

\begin{figure}[ht!]
\centering
\begin{subfigure}[h]{0.25\textwidth}
	\includegraphics[width=\textwidth]{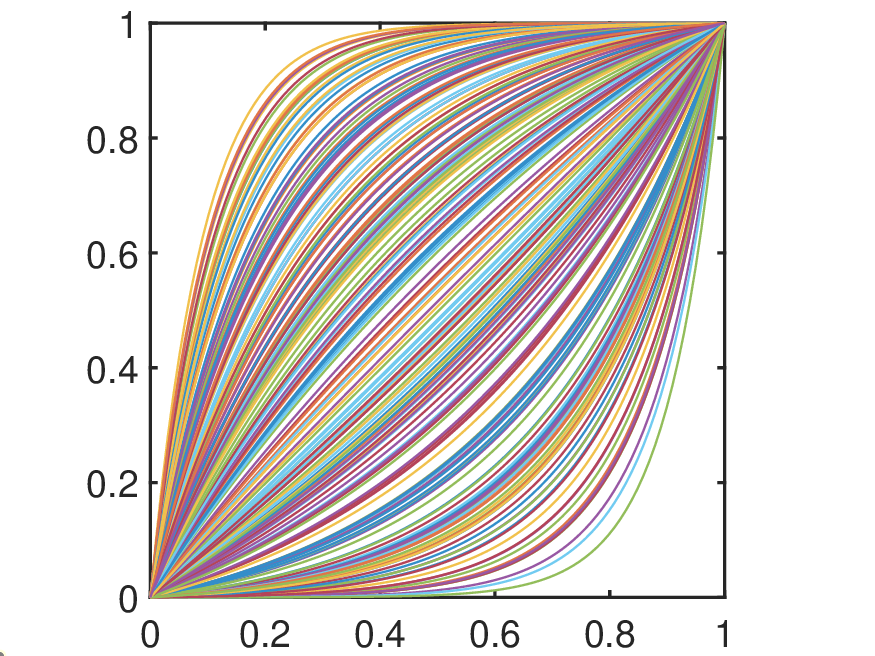}
	\caption{I. obs}
\end{subfigure}\hspace{-0.7cm}
\begin{subfigure}[h]{0.25\textwidth}
	\includegraphics[width=\textwidth]{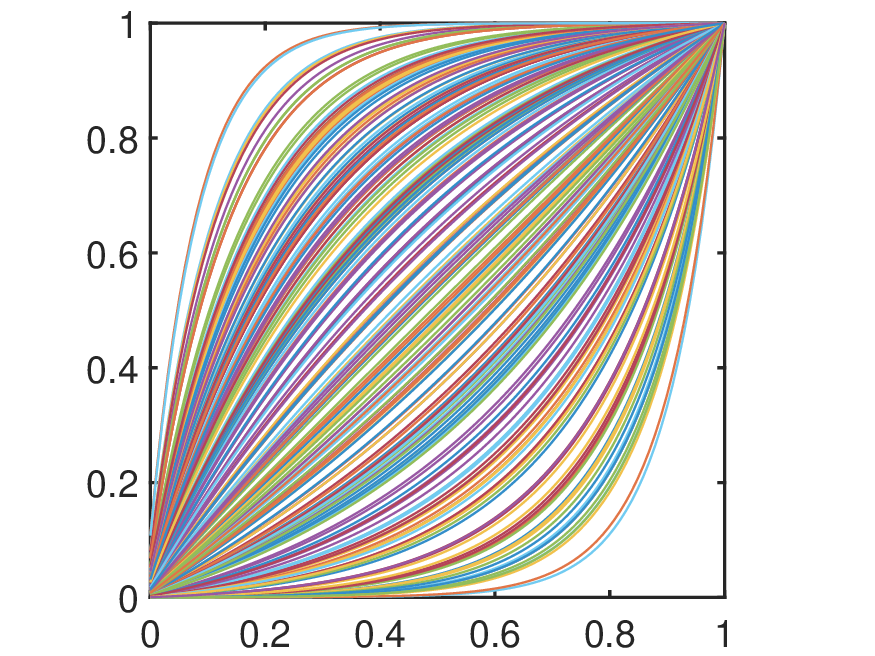}
	\caption{I. CLR}
\end{subfigure}\hspace{-0.7cm}
\begin{subfigure}[h]{0.25\textwidth}
	\includegraphics[width=\textwidth]{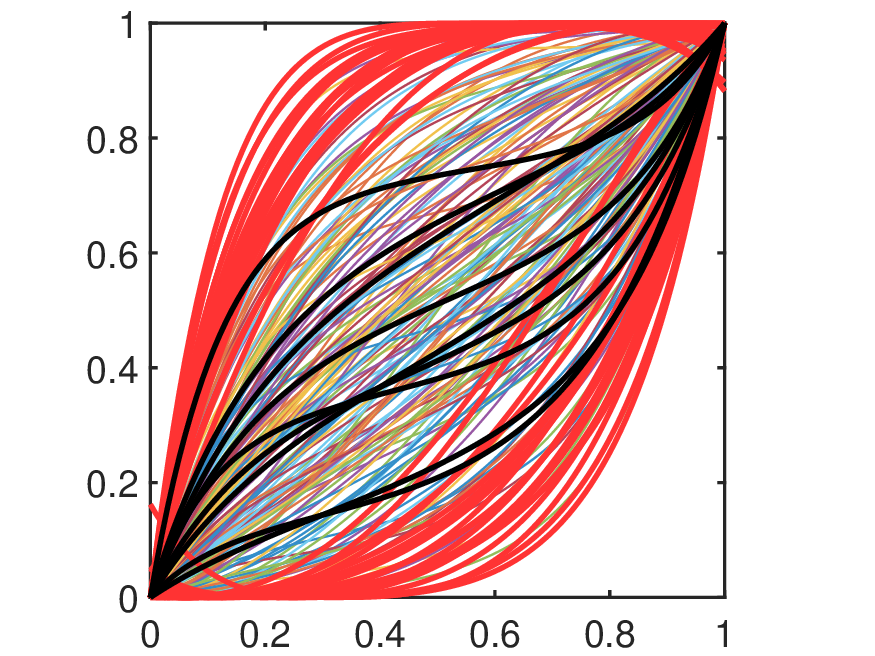}
	\caption{I. SRVF}
\end{subfigure}

\begin{subfigure}[h]{0.25\textwidth}
	\includegraphics[width=\textwidth]{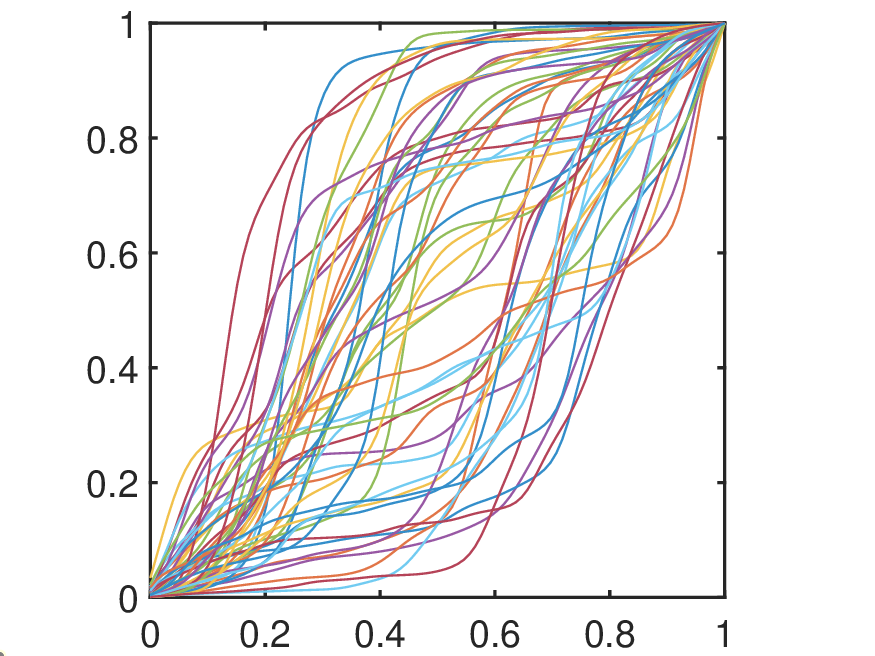}
	\caption{II. obs}
\end{subfigure}\hspace{-0.7cm}
\begin{subfigure}[h]{0.25\textwidth}
	\includegraphics[width=\textwidth]{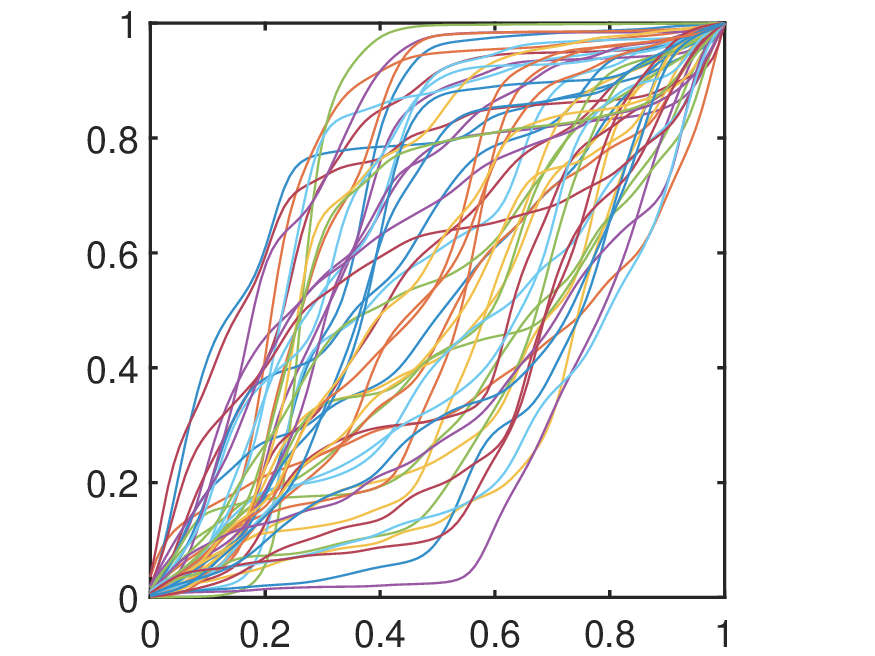}
	\caption{II. CLR}
\end{subfigure}\hspace{-0.7cm}
\begin{subfigure}[h]{0.25\textwidth}
	\includegraphics[width=\textwidth]{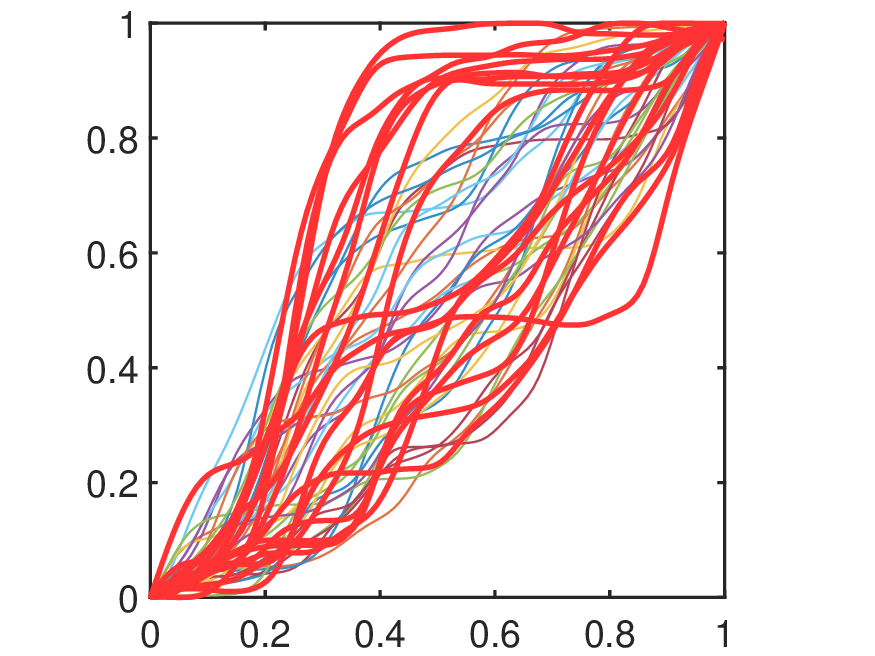}
	\caption{II. SRVF}
\end{subfigure}

\caption{Two examples to visually compare time warping models. (a) First example: observed time warpings. (b) First example:  resampled warpings via the CLR-based model. (c) First example: resampled warpings via the SRVF-based model. (d)-(f) Same as (a)-(c) except for the second example.  Bold black and bold red curves indicate improper resampled warpings. }
\label{fig:srvf}
\end{figure}

We now use two examples to illustrate the performance of the CLR-based model, and compare it with the SRVF-based model.  In the first example, we simulate 200 warping functions $\gamma_i(t) = \frac{e^{a_it}-1}{{e^{a_i}-1}}, i= 1,\cdots, 200$, where $a_i\sim \mathcal{N}(0, 16)$. These functions are shown in Figure \ref{fig:srvf}(a).  We then fit the CLR-based model on the observations, and conduct resampling (see the detailed procedure in \citep{ma2024stochastic}).  The resampled 200 warping functions are shown Figure \ref{fig:srvf}(b).  Similarly, the SRVF-based model can be adopted to fit the data and conduct resampling.  The result is shown in Figure \ref{fig:srvf}(c).   We can see that the CLR-based resampled warpings closely resemble the original observations in terms of function patterns and variation ranges.  On the other hand, the resampling results obtained through the SRVF-based model exhibit significant differences.  There is less variability and some resampled warpings display unusual patterns that are not present in the original observations (shown in bold black in Figure \ref{fig:srvf}(c)).  Moreover, some resampled warpings are even not proper warping functions (i.e., not bijective between $[0, 1]$ and $[0,1]$, shown in bold red in Figure \ref{fig:srvf}(c)).  

In the second example, we aim to generate warping functions with more complicated structure. We
utilize basis functions $\{\psi_{2j-1}=\sqrt{2}\sin(2j\pi t), \psi_{2j}=\sqrt{2}\cos(2j\pi t)$,  $j=1,\cdots,10, t \in [0, 1]\}$, and simulate 50 warping functions $\gamma_i(t) = \psi_B^{-1}(\sum_{k=1}^{20}c_{ik}\phi_k)(t), i= 1,\cdots, 50$, where $c_{ik}\sim \mathcal{N}(0, \frac{1}{k^2})$.  Consistent to the first example, the simulated observations, the resampled warpings with the CLR-based model, and the resamples warpings with the SRVF-based model are shown in Figure \ref{fig:srvf}(d)-(f), respectively.   Again, the CLR-based resampled warpings closely resemble the original observations, whereas the resampling results obtained through the SRVF-based model exhibit significant differences.  In particular, the SRVF warpings exhibit less variability and some resampled functions even may not
be strictly increasing on $[0,1]$ (shown in bold red in Figure \ref{fig:srvf}(f)).


\subsection{Mathematical theory on registration with the $\mathbb L^2$ representation}
\label{mp}
We have used a linear, inner-product system to describe the time warping functions in $\Gamma_1$.  That is, we can have addition, scaling multiplication, and inner-product operations on the time warping functions.  To be useful for function registration, we also need to examine the function composition on time warpings (note that $\Gamma_1$ is a {\em group} with the action of function composition).   

Let $f$ be an absolutely continuous function on the domain $[0,1]$. Its Square-Root-Velocity-Function (SRVF) is defined as $q:[0,1]\rightarrow \mathbb{R}$, $q(t)=sign(\dot{f}(t))\sqrt{|\dot{f}(t)|}$  \citep{srivastava2011registration}. For $\gamma \in \Gamma_1$, the SRVF of $f \circ \gamma$ is given by: $(q,\gamma)(t)=\sqrt{\dot{\gamma}(t)}q(\gamma(t))$.  The famous Fisher-Rao framework provides a superior alignment performance as it has an important property on isometry.  That is, for any $q_1, q_2 \in \mathbb L^2$ and warping function $\gamma$, 
\[ \| (q_1, \gamma) - (q_2, \gamma) \| = \|q_1 - q_2\|, \]
where $\| \cdot \|$ indicates the conventional $\mathbb L^2$ norm.
This property is critically important for the notion of the-center-of-the-orbit, which is a necessary step in the Fisher-Rao alignment method. Basically, if the Karcher mean of $\{\gamma_i\}_{i=1}^n$ is $\gamma^*$, then for any warping $\gamma_0$ the Karcher mean of $\{\gamma_i \circ \gamma_0 \}_{i=1}^n$ is $\gamma^* \circ \gamma_0$ \citep{srivastava2011registration}.

We point out that the CLR-transformation does not have the above desired isometry property.  That is, in general, 
\[ \| \psi_B(\gamma_1 \circ \gamma) - \psi_B(\gamma_2 \circ \gamma) \| 
\neq
\| \psi_B(\gamma_1) - \psi_B(\gamma_2) \| . \]
However, we can still have the mean property aforementioned on the Fisher-Rao method.  This will be clearly given in a lemma below, where the detailed proof is given in \ref{app:lemma}.    

\begin{lemma}
Assume $\{\gamma_1, \cdots, \gamma_n\}$ is a set of $i.i.d.$ time warping functions in $\Gamma_1$ and $\mathbb E(\psi_B(\gamma_i)) = \psi_B(\gamma^*)$, with $\gamma^* \in \Gamma_1$.  Let $\gamma_0$ be any given warping in $\Gamma_1$.  Then
$$ \mathbb E(\psi_B(\gamma_i \circ \gamma_0)) = \psi_B(\gamma^* \circ \gamma_0).$$   
\label{lemma1} 
\end{lemma}

In the Fisher-Rao framework, the notion of the-center-of-the-orbit is used to adjust the alignment result, ensuring the center of the warping functions is the identity.  This property can also be achieved in the CLR-based linear system.  Let us consider the observation model $f_i = c_i (g \circ \gamma_i) + e_i$, $i=1, \dots, n,$ where $g$ is an unknown, absolutely continuous function on $[0, 1]$, and $c_i \in
\mathbb{R}^+$, $\gamma_i \in \Gamma_1$ and $e_i \in \mathbb{R}$ are all i.i.d. random scalings, warpings, and translations, respectively. Given the observations $\{f_i\}$, the goal is to estimate the underlying signal $g$. To make the system identifiable, we have the following constraints: 1) $\mathbb E(\psi_B(\gamma_i^{-1})) = 0$, 2) $\mathbb E(c_i) = 1$, and 3)
$\mathbb E(e_i) = 0$.  The estimation process can be given in the following Estimation algorithm: \\ 

\noindent {\bf Estimation Algorithm}: Given a set of functions $\{f_i\}_{i=1}^n$ on $[0,1]$, let $\{q_i\}_{i=1}^n$ denote the SRVFs of $\{f_i\}_{i=1}^n$, respectively.  
\begin{enumerate}
\item Take any $q_k (k \in \{1, \cdots, n\})$  as the initial estimate of the SRVF of $g$.  Note that $q_k = \sqrt{c_k}(q_g, \gamma_k)$, where $q_g$ is the true SRVF of $g$.  Denote $q_0 = q_k.$
\item For $i=1,2,\dots,n$, find $\gamma_i^*$  by solving: $\gamma_i^* = \argmin_{\gamma \in \Gamma} \| q_0 - (q_i, \gamma)\|$.
\item Compute the sample mean $\hat \gamma_0 = \psi_B^{-1}\big(\frac{1}{n}\sum_{i=1}^n \psi_B(\gamma_i^*)\big)$
\item Return the warping functions $\gamma^{**} = \gamma_i^* \circ \hat \gamma_0^{-1}$
and aligned functions $\tilde{f}_i = f_i \circ \gamma_i^{**}, i = 1, \cdots, n.$
\item Return the estimated signal $\hat g_n =  \frac{1}{n}\sum_{i=1}^n \tilde{f}_i$.
\end{enumerate}


The above algorithm indeed provides a consistent estimator to the underlying signal $g$.  This is under a regularity condition on $g$, referred to as the {\em warping continuity}:

\begin{assumption}
For any warping functions $\gamma, \gamma_1 \in \Gamma_1$, 
$$\|g\circ \gamma - g\| \le C_g \| \psi_B(\gamma \circ \gamma_1) - \psi_B(\gamma_1)\|, $$ 
where $C_g$ is a positive constant dependent on $g$, and $\| \cdot\|$ is the conventional $\mathbb L^2$ norm.
\label{assp:cont}
\end{assumption}

The consistency is formally given in the following proposition, with the detailed proof given in \ref{app:prop}.     
\begin{proposition}
Let $f_i = c_i g \circ \gamma_i(t) + e_i, i = 1, \cdots, n$ be a set of functions generated from and absolutely continuous function $g$ on $[0, 1]$ with random warping, scaling, and translation, where $\gamma_i \in \Gamma_1$, $c_i \in \mathbb R^+$ and $e_i \in \mathbb R$.  Let $\hat g_n$ denote the estimate from the above Estimation Algorithm.  If 1) the underlying signal $g$ satisfies Assumption \ref{assp:cont}, and 2) $\mathbb E(\psi_B(\gamma_i^{-1})) = 0, \mathbb E(c_i) = 1$, and $\mathbb E(e_i) = 0$, then
$\hat g_n \rightarrow g \ \ (a.s.) \ \ \ (n \rightarrow \infty).  $ 
\label{prop}
\end{proposition}

%
%
%

\begin{figure}[ht!]
\centering
\begin{subfigure}[h]{0.22\textwidth}
	\includegraphics[width=\textwidth]{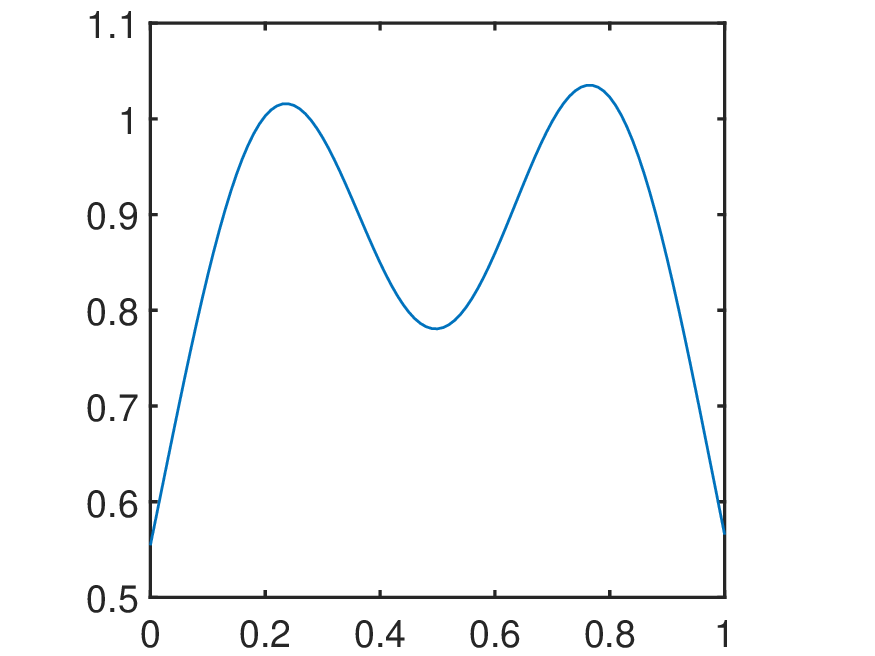}
	\caption{$g$}
\end{subfigure}\hspace{-0.6cm}
\begin{subfigure}[h]{0.22\textwidth}
	\includegraphics[width=\textwidth]{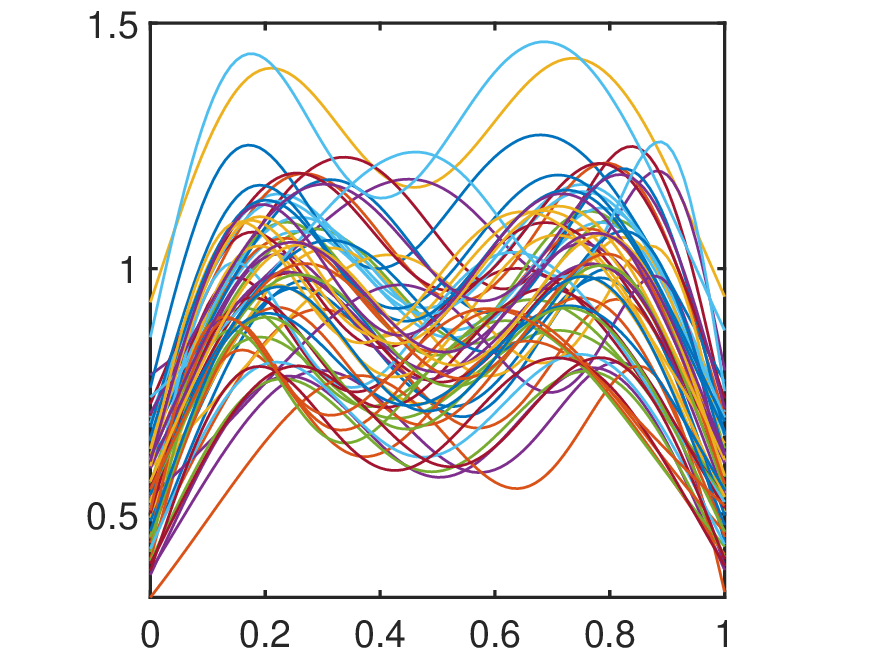}
	\caption{$\{f_i\}$}
\end{subfigure}\hspace{-0.6cm}
\begin{subfigure}[h]{0.22\textwidth}
	\includegraphics[width=\textwidth]{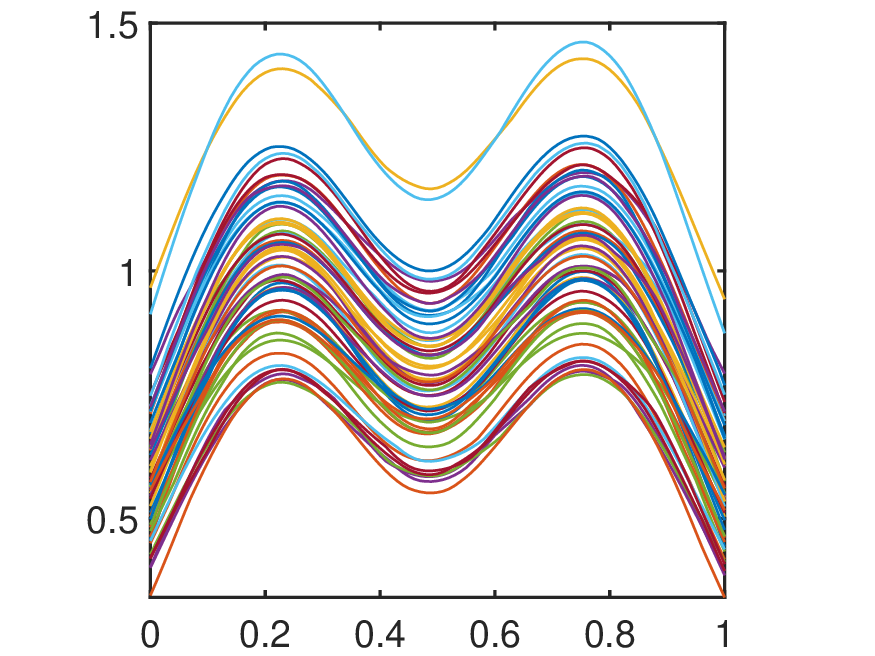}
	\caption{$\{\tilde {f}_i\}$}
\end{subfigure}\hspace{-0.6cm}
\begin{subfigure}[h]{0.22\textwidth}
	\includegraphics[width=\textwidth]{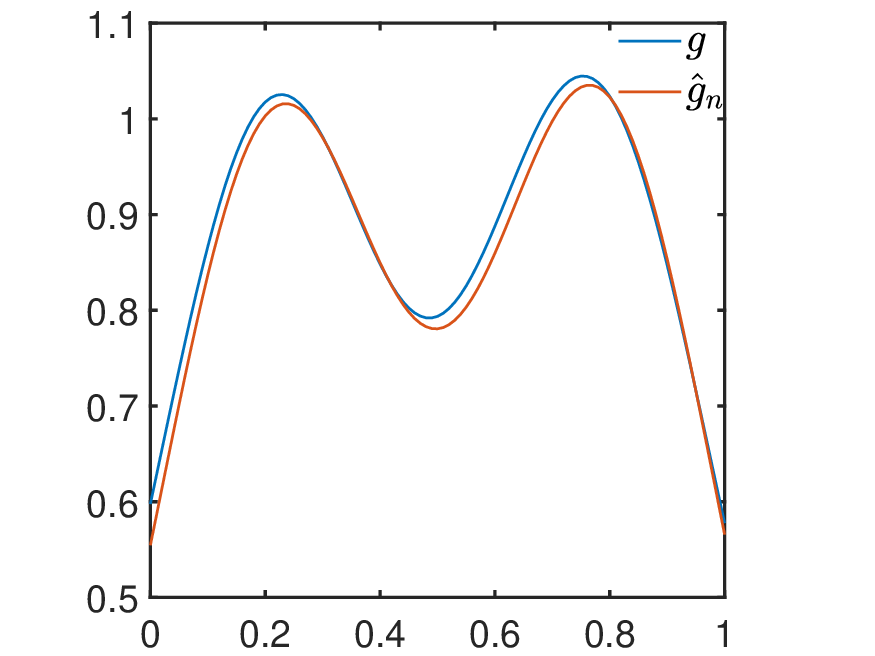}
	\caption{$\hat{g}_n$}
\end{subfigure}\hspace{-0.6cm}
\begin{subfigure}[h]{0.22\textwidth}
	\includegraphics[width=\textwidth]{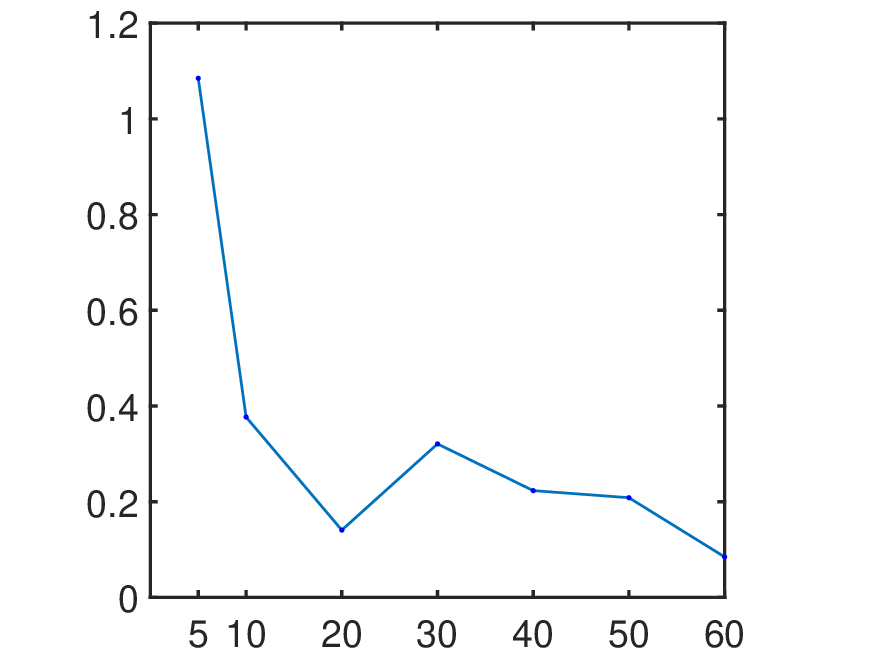}
	\caption{error w.r.t. $n$}
\end{subfigure}

\caption{Result on Simulation 1. (a) $g$. (b) Functions $\{f_i\}$. (c) Aligned functions $\{\tilde f_i\}$. (d) The estimated function $\hat g_n$. (e) The difference between $\hat g_n$ and $g$ w.r.t. $n$.} 
\label{fig:lemma}
\end{figure}

\textbf{Simulation 1:} We here use one simulation example to illustrate the above  Estimation Algorithm.  At first, a bimodal function is given as $g(t)=0.99 \exp(-12(t-0.22)^2) + 1.01 \exp(-12(t-0.78)^2), t\in [0, 1]$. We then randomly generate $n=60$ warping functions $\gamma_i = \psi_B^{-1}(a_i\cdot b_1), i=1,\cdots, n$, where ${b_1 = \frac{\sqrt{3}}{2}}(t-0.5)$ and $a_i\sim\mathcal{N}(0, 0.2)$. We then compute functions $f_i = c_i g \circ \gamma_i+e_i$, with both $c_i\sim\mathcal{N}(1, 0.1)$ and $e_i\sim\mathcal{N}(0, 0.1)$, which forms the functional data. In Figure \ref{fig:lemma}, Panel (a) shows the function $g$, and Panel (b) shows the data $\{f_i\}$. By applying the Estimation Algorithm, we obtain the aligned functions $\tilde f_i$, shown in Panel (c).  We also show both $\hat g_n$, the estimated $g$ function, and the true $g$ in Panel (d). The close resemblance between the estimated and original $g$ demonstrates the effectiveness of this algorithm. Moreover, we examine the performance of the estimator with respect to the sample size, by performing this estimation for $n$ equal to 5, 10, 20, 30, 40, 50, and 60, respectively. The estimation errors, computed using the $\mathbb{L}^2$ norm between $\hat g_n$'s and the true $g$, are shown in Panel (e). As expected from the Proposition 1, this estimate converges to the true $g$ when sample size $n$ grows large.  

Once the model on time warping is known, it can be used as a prior in the Bayesian framework for function registration \citep{cheng2016bayesian,lu2017bayesian}.  
In this paper, our registration is based on the same likelihood term in a Bayesian framework, whereas we propose a new prior on the CLR transformed warping space to moderate the degree of phase variation. Unlike previous covariance representation \citep{cheng2016bayesian, lu2017bayesian}, our covariance is well defined in the CLR transformed $\mathbb L^2$ subspace so that we can adopt any 2nd order stochastic process as the prior.  We will provide the detail of our Bayesian framework in the following subsections. To simplify the description, we will at first use a Gaussian process in the prior. 

\subsection{Bayesian framework with a new Gaussian process prior}
\label{gp}
In the renowned Fisher-Rao registration method, the main criterion for aligning two functions, $f_1$ and $f_2$, is to minimize the loss function, expressed as $\|q_1-(q_2,\gamma)\|$, which quantifies the difference between their corresponding SRVF functions $q_1$ and $q_2$.   Same as in the previous Bayesian methods \citep{lu2017bayesian, cheng2014analysis, cheng2016bayesian}, conditioned on the warping function, the difference $q_1-(q_2,\gamma)$ is assumed as a Gaussian process with mean 0.  That is, $q_1-(q_2,\gamma) | \gamma \sim GP(0, \Sigma)$, where $\Sigma$ denotes the covariance function.  

In practical problems, observed functional data consist of function values only at discrete points in the domain. Let $[t]= \{t_1, \cdots, t_k\}$ denote a set of discrete points on $[0, 1]$. 
We use $q_1([t])$ and $(q_2,\gamma)([t])$ to denote vectors evaluated at the same finite points for functions $q_1(t)$ and $(q_2,\gamma)(t)$, respectively.  
By the Gaussian process assumption, the joint distribution of these finite differences $q_1([t])-(q_2,\gamma)([t])|\gamma$ is a multivariate normal distribution.  That is, $\Big\{q_1([t])-(q_2,\gamma)([t])|\gamma\Big\}\sim N_k(0_k,\Sigma_{k\times k})$, where $k$ is the number of discrete points. 
Consistent to previous Bayesian methods \citep{lu2017bayesian, cheng2014analysis, cheng2016bayesian}, we assume an isotropic covariance $\Sigma_{k\times k}=\sigma I_{k\times k}$. 
Hence, in the Bayesian framework, the likelihood is given as: 

\begin{eqnarray}
\pi(q_1,q_2|\gamma) &\propto& \exp\Big(-\frac{1}{2\sigma^2} \|q_1([t])-(q_2,\gamma)([t])\|^2\Big).
\label{eq:likelihood}
\end{eqnarray}  
The key term in the above likelihood is $\|q_1([t])-(q_2,\gamma)([t])\|$, which measures the alignment performance.   
In the Fisher--Rao method \citep{srivastava2011registration}, the goal is to find an optimal $\gamma$ that can minimize this alignment term. This is equivalent to finding the maximizer in the likelihood Eqn. \eqref{eq:likelihood} with respect to $\gamma$. Computationally, this can be achieved via a dynamic programming algorithm.
In addition, one can easily see that $\sigma$ is a normalizing parameter which describes the importance of the alignment term -- the smaller the value of $\sigma$ is, the more dominant of the alignment term will be in the Bayesian framework. In the next section, we will assess the impact of $\sigma$ on the outcome of the Bayesian alignment.

In previous Bayesian methods, a Dirichlet prior is assigned to model the discrete version of time warping $\gamma$ \citep{cheng2016bayesian}, or a Gaussian process prior with mean zero is used to model SRVF-based time warping in the tangent space \citep{lu2017bayesian}.  In this study, we propose to use a Gaussian process prior, denoted by $\mu_0$, to model the CLR-transformed warping functions in the $\mathbb L^2$ space, i.e., $h(t) = \log(\dot{\gamma}(t))-\int_{0}^{1}\log(\dot{\gamma}(s))ds\sim GP(\mu_h, C_h)$. Since there is a one-to-one deterministic correspondence between $\gamma$ and $h$, the randomness in $\gamma$ can be fully characterized by the randomness in $h$, and the posterior measure, denoted by $\mu$, is absolutely continuous with respect to the prior with density :

\begin{eqnarray}
\frac{d \mu}{d \mu_0}(\gamma_h) &\propto& \pi(q_1,q_2|\gamma_h) \\ \nonumber
&\propto& \exp\Big(-\frac{1}{2\sigma^2} \|q_1([t])-(q_2,\gamma_h)([t])\|^2\Big).
\label{eqn:postG}
\end{eqnarray} 
Here, $\gamma_h= \psi_B^{-1}(h)$ is introduced to simplify the notation. 

The posterior given above is thoroughly investigated in \cite{cotter2013mcmc} via a Markov Chain Monte Carlo (MCMC) procedure.  In principle, any type of Metropolis-Hastings framework with an appropriate proposal density can be used to sample the warping functions.  In this paper, we will implement the Z-mixture pCN algorithm to obtain the posterior of the time warpings, as described in \cite{lu2017bayesian}. This algorithm is particularly advantageous because it is robust to the choice of discretization grid. Additionally, the pCN algorithm facilitates the identification of multiple modes in the posterior.  In contrast, the Hamiltonian Monte Carlo (HMC) has recently preferred to sample functions  for its efficiency \citep{neal2011mcmc, beskos2011hybrid}. However, this method poses challenges in identifying appropriate hyperparameters, making it less practical in scenarios involving multimodal cases.  The specific steps are outlined in Algorithm \ref{alg:pwbayesian}. 

\begin{algorithm}[ht!]
\caption{Pairwise Bayesian Registration with a Gaussian Process Prior}
\begin{algorithmic} 
	\Require SRVFs $q_1, q_2$, pre-given distribution for $\beta \in [0, 1]$, mean $\mu_h$ and covariance $C_h$ for the prior Gaussian process. 
	\State Random pick $h_{(1)}\sim GP(\mu_h, C_h)$
	\For{$k = 1: N$}
	\State Pick $\beta$ from a pre-given distribution 
	\State Propose $h'_{(k)} = \sqrt{1-\beta^2}(h_{(k)}-\mu_h) + \beta\xi_{(k)}$, where $\xi_{(k)}\sim GP(\frac{\mu_h}{\beta}, C_h)$
	\State Set $h_{(k+1)}=\begin{cases}
		h'_{(k)} & \text {with probability $\rho = min\Big(1, \frac{\pi(q_1, q_2|\gamma_{h'_{(k)}})}{\pi(q_1, q_2|\gamma_{h_{(k)}})}\Big)$} \\
		h_{(k)} & \text {with probability $1 -\rho$} 
	\end{cases}$
	\State \ \ \ \ \ where the function $\pi(\cdot, \cdot | \cdot)$ is the likelihood given in Eqn. \eqref{eq:likelihood}.
	\State $\gamma_{k+1}(t) =\psi_B^{-1}(h_{(k)})$
	\EndFor
	\State Output $\{\gamma_{k}(t)\}_{k=2}^{N+1}$  
\end{algorithmic}
\label{alg:pwbayesian}
\end{algorithm}

\noindent \textbf{Remark 1:} Regarding the weight parameter $\beta$ in Algorithm \ref{alg:pwbayesian}, it is used to enhance the efficiency of the MCMC algorithm within given iterations and its specific value does not influence the final posterior estimation. This parameter is required to have a value ranging between 0 and 1. Our recommendation is to select $\beta$ from a distribution that permits values near zero to keep proposed $h'_{(k)}$ close to the previously accepted proposal and values near one to have the capacity to escape from the current local area. This setting allows the pCN to effectively solve multimodal problems. For instance, we can let $\beta$ follow a discrete distribution, presenting two potential outcomes 0.9 and 0.01, with probabilities 0.5 and 0.5, respectively.   \\

Under this new Bayesian framework, the primary distinctions and advantages introduced by the linear representation of warping are threefold. Firstly, it enables us to sample time warpings directly from the prior model without any approximations, in contrast to the SRVF framework. Due to the lack of isometric isomorphism between the SRVF space and the tangent space, sampling in the tangent space could yield non-increasing functions in the warping space, necessitating an additional filtering step to discard these inappropriate samples. This implies our method is both more {\em efficient} and {\em effective}. Secondly, there is no longer a requirement to confine the warping mean to zero. This change allows for the manipulation of the mean function to create various priors. We can use different mean functions to modulate the alignment process, and will elaborate this point in the next section. Thirdly, unlike the SRVF framework that exclusively uses the basis method for constructing covariance functions, our method allows for various types of covariance by directly specifying variability and covariability in the time domain. This flexibility provides a more comprehensive understanding of covariance, allowing for the selection of diverse covariance structures at will. We will use one example to illustrate this new prior as follows. 

\begin{figure}[ht!]
\centering
\begin{subfigure}[h]{0.3\textwidth}
	\includegraphics[width=\textwidth]{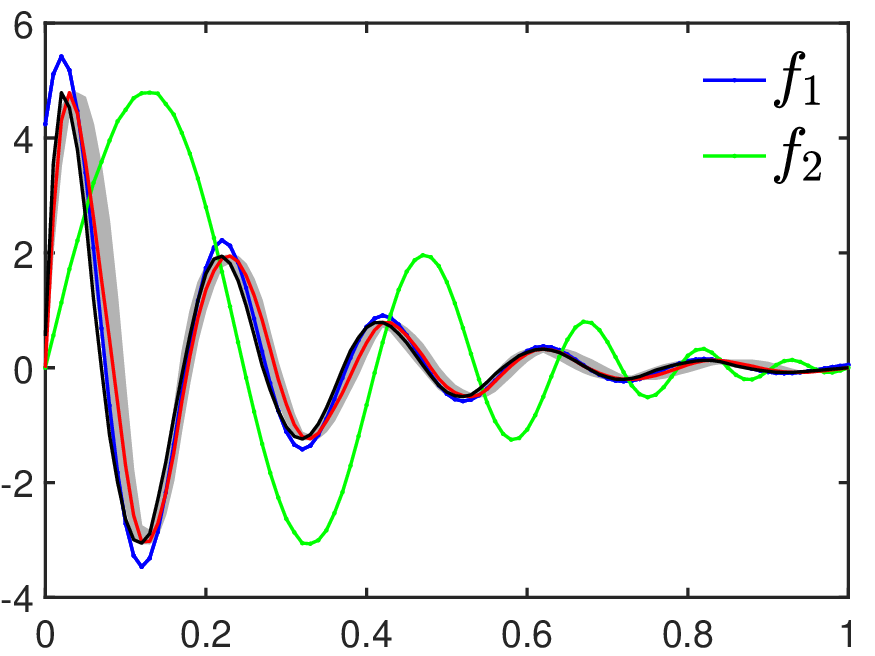}
	\caption{observations}
\end{subfigure}
\begin{subfigure}[h]{0.3\textwidth}
	\includegraphics[width=\textwidth]{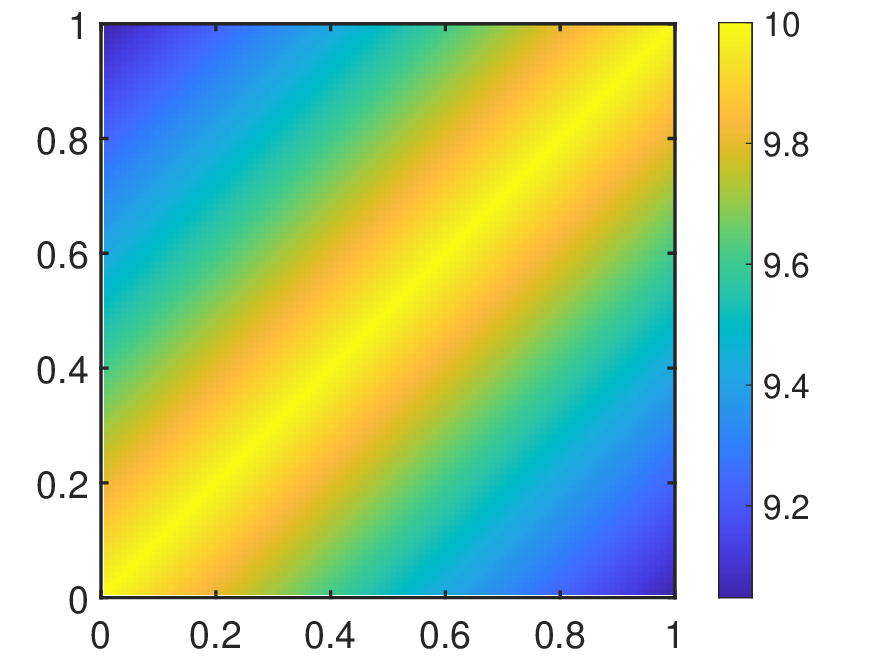}
	\caption{$C_h$}
\end{subfigure}\hspace{-0.8cm}
\vfill
\begin{subfigure}[h]{0.3\textwidth}
	\includegraphics[width=\textwidth]{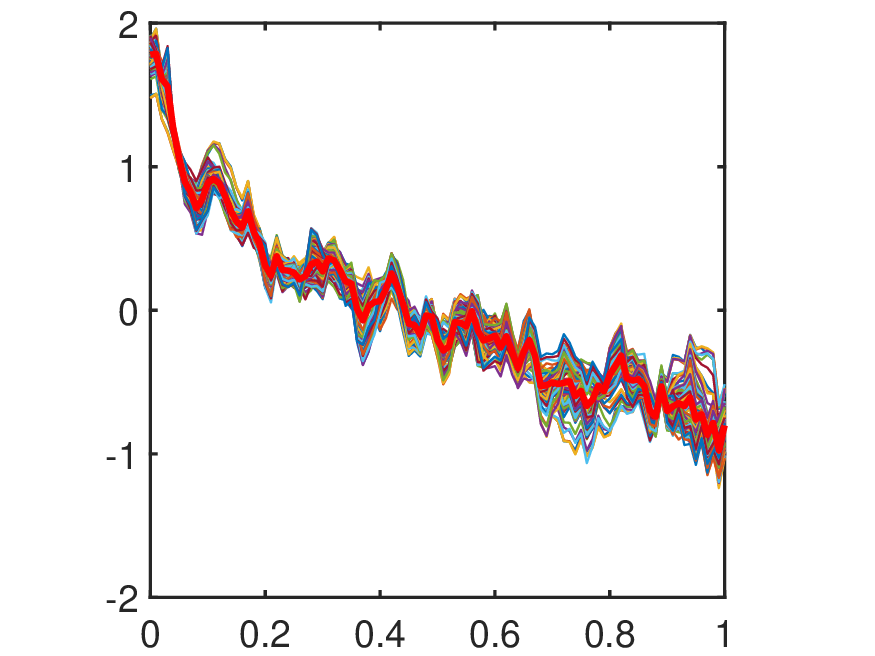}
	\caption{sample in $H(0, 1)$}
\end{subfigure}	
\begin{subfigure}[h]{0.3\textwidth}
	\includegraphics[width=\textwidth]{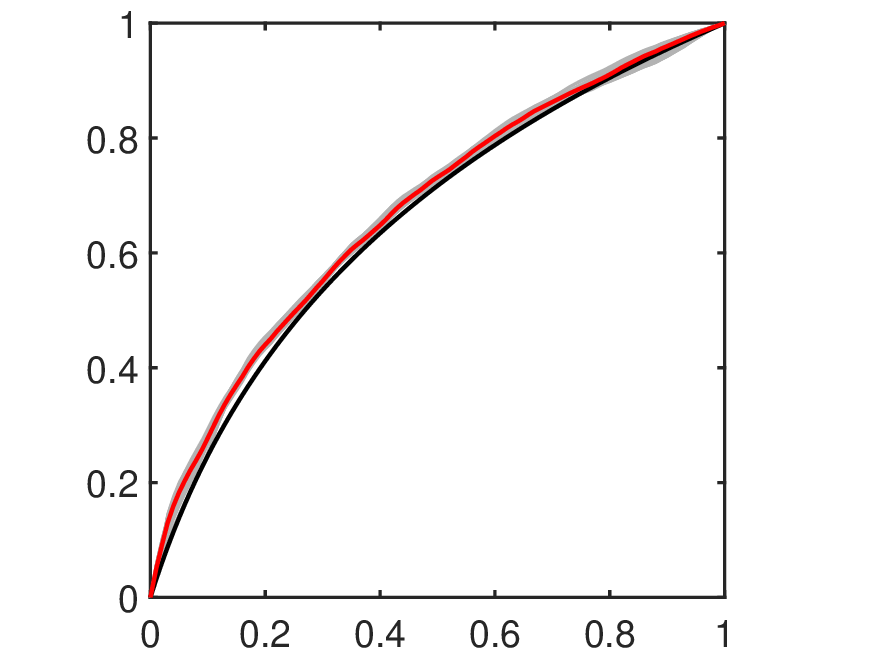}
	\caption{warpings}
\end{subfigure}	
\caption{Result on Simulation 2. (a) Original and aligned functions. The blue and green curves are the two given functions $f_1$ and $f_2$. The black curve is the aligned $f_2$ via the dynamic programming method. The grey area highlights the range of aligned $f_2$ with all sampled warping via proposed Bayesian registration, with the red curve represents aligned $f_2$ with the averaged warping.  (b) The heatmap of the $C_h$. (c) Sampled functions in $H(0,1)$,and their mean is shown in bold red. (d) Sampled warping functions in Bayesian alignment are shown in grey shade, and their mean is shown as the red curve. The bold black curve is the warping estimated using the dynamic programming method. } 
\label{fig:sim2}
\end{figure}

\textbf{Simulation 2: } We at first simulate two multimodal functions, i.e., $f_1(t)=6 \cdot 0.8^{20t} \cdot \cos (10\pi t-\frac{\pi}{4})$ and $g(t)=6 \cdot 0.8^{20t} \cdot \sin (10\pi t), t\in [0, 1]$. Then we generate a warped version of $g(t)$ by defining  $f_2(t) = g(\gamma(t))$ with the warping function $\gamma(t)=\frac{e^{2t}-1}{e^{2}-1}$. The functions $f_1(t)$ and $f_2(t)$ are shown as blue and green solid curves in Figure \ref{fig:sim2}(a), respectively. For the prior Gaussian process, we set $\mu_h=0$  and $C_h(s,t)=10 \cdot 0.999^{100|s-t|}, s, t \in[0,1]$ (see the heat-map graph of the covariance in Figure \ref{fig:sim2}(b)). Through this specific definition, we establish the covariance to ensure a uniform variance of the CLR-transformed warping $h$ across various time points, while also ensuring the correlation between any two points decreases as the time difference between them increases. Moreover, we set the parameter $\sigma$ in Eqn. \eqref{eq:likelihood} a relatively small value of $2$, thereby increasing the emphasis on the likelihood term within the posterior. This assignment ensures that the warpings derived from our Bayesian approach are expected to closely align with the result obtained via the dynamic programming method in the Fisher-Rao framework which maximizes the likelihood term.   

Using Algorithm \ref{alg:pwbayesian}, we obtain the Bayesian alignment, with 100 sampled posterior CLR-transformed time warping functions and its corresponding time warping functions displayed in Figure \ref{fig:sim2}(c) and (d), respectively. The red curves in these figures represent the means. Within this linear framework, to derive the warping mean, we first compute the mean of their CLR-transformed functions, and then project this mean function back into warping space to obtain the warping mean.  With a small $\sigma$, we see there is little variability in the posterior sample  (i.e., a narrow band in grey) as the likelihood term is more dominant.  The aligned $f_2$ functions with the sampled warping functions are also shown in Figure \ref{fig:sim2}(a) in grey, which properly align with $f_1$, especially the red curve which is the aligned $f_2$ with the mean of all the sampled warpings.   
We can also see the aligned $f_2$ with the warping obtained through the dynamic programming, illustrated in black, also stays inside the grey region, and it stays close to the mean of our sampled warpings. This shows our method can get consistent alignment result to the dynamic programming in the Fisher-Rao framework, and clearly demonstrates the effectiveness of our new prior within the Bayesian registration framework.  When the value of $\sigma$ gets larger, the posterior will exhibit more importance in the prior. This will be further explored in the next section. 


\subsection{Extension to non-Gaussian process prior}
\label{Sec: ngp}

In principle, we can assign other stochastic processes for the prior under the linear representation system, and they do not have to be a conventional Gaussian process. According to the Karhunen-Lo\`eve (KL) expansion, any mean-square continuous stochastic process in $H(0, 1)$ can be expressed as an infinite linear combination of a set of orthonormal basis in $H(0, 1)$. In practical use, approximations are typically made by using a finite subset, with cardinality $M \in \mathbb N$, of the basis functions, and by assuming independence on the basis coefficients.  Therefore, given the eigenpairs $(b_i(t), \lambda_i, i\ge 1)$ of the covariance kernel $C_h$, with $\{b_i(t), i\ge 1\}$ being the orthonormal basis functions of $H(0, 1)$, we can simulate any mean-square continuous stochastic process $h(t)$ with covariance kernel $C_h$ by sampling independent variates $\xi_i$ from any distributions with mean 0 and variance $\lambda_i$, and then setting $h(t) = \sum_{i=1}^{M}\xi_ib_i(t)$. 

In this case, we fully characterizes the randomness in $h(t)$ by exploring all possible variabilities of the coefficients $\xi_i$. It facilitates the straightforward application of a non-Gaussian prior by specifying that any of $\xi_i$ follow a non-Gaussian distribution. It is obvious that this procedure can also be used for a Gaussian process prior, where we can simply let all $\xi_i$ follow a Gaussian distribution. Let $\bm\xi = (\xi_1, \cdots, \xi_M)^T \in \mathbb R^M$, and then its posterior can be written as: 

\begin{eqnarray}
L(\bm\xi|q_1, q_2)&\propto& \pi(q_1,q_2|\gamma_{\bm\xi} ) \pi(\bm\xi) \nonumber \\ 
&\propto& \exp\Big(-\frac{1}{2\sigma^2} \|q_1([t])-(q_2,\gamma_{\bm\xi})([t])\|^2\Big)\prod_{i=1}^{M}\pi(\xi_i)
\end{eqnarray}  
Here, $\gamma_{\bm\xi} = \psi_B^{-1}(\sum_{i=1}^{M}\xi_ib_i)$ is introduced for simplicity in expression. Similar to Algorithm \ref{alg:pwbayesian}, we can apply the Metropolis-Hastings algorithm to sample from the posterior. The specific steps are outlined in Algorithm \ref{alg:pwbayesian_nonG} as follows.

\begin{algorithm}[ht!]
\caption{Pairwise Bayesian Registration with a Non-Gaussian Process Prior}
\begin{algorithmic} 
	\Require SRVFs $q_1, q_2$, basis $\{b_i\}_1^M$, pre-given distribution $f_i$ for $\xi_i, i = 1,\cdots, M$, $\sigma_\xi$, and $p$.  
	\State Set $\bm\xi^{(1)}=(\xi_1^{(1)}, \xi_2^{(1)},\cdots,\xi_M^{(1)})^T$ with  $\xi_i^{(1)} \sim f_i,  i = 1,\cdots, M$, and calculate $\gamma_{\bm\xi^{(1)}}$. 
	\For{$k = 1: N$}
	\For{$i = 1: M$}
	\State Simulate $\xi^*_{i} \sim f_i$.
	\State Propose $\xi_i' = \beta \mathcal{N} (\xi_{i}^{(k)}, \sigma_\xi) + (1-\beta)\xi^*_{i}$, where $\beta\sim Bernoulli(p)$.		
	\EndFor
	\State Set $\bm\xi'=(\xi_1', \xi_2',\cdots,\xi_M')^T$, and calculate $\gamma_{\bm\xi'}$. 
	\State Set $\bm\xi^{(k+1)}=\begin{cases}
		\bm\xi' & \text {with probability $\rho = min\Big(1, \frac{\pi(q_1, q_2|\gamma_{\bm\xi'})\pi(\bm\xi')q(\bm\xi^{(k)}|\bm\xi')}{\pi(q_1, q_2|\gamma_{\bm\xi^{(k)}})\pi(\bm\xi^{(k)})q(\bm\xi'|\bm\xi^{(k)})}\Big)$} \\
		\bm\xi^{(k)} & \text {with probability $1 -\rho$} 
	\end{cases}$,
	\State \ \ \ 	where 	$q(a|b) = (1-p)\pi(a) + \frac{p}{\sqrt{2\pi} \sigma_\xi}\exp(-\frac{\|a-b\|^2}{2\sigma_\xi^2})$ \ for \ $a, b \in \mathbb R^M$.
	\State $\gamma_{\bm\xi^{(k+1)}} =\psi_B^{-1}(\sum_{i=1}^{M}\xi_i^{(k+1)}b_i)$.
	\EndFor
	\State Output $\{\gamma_{\bm\xi^{(k)}}(t)\}_{k=2}^{N+1}$  .
\end{algorithmic}
\label{alg:pwbayesian_nonG}
\end{algorithm}

\begin{figure}[ht!]
\centering
\begin{subfigure}[h]{0.23\textwidth}
	\includegraphics[width=\textwidth]{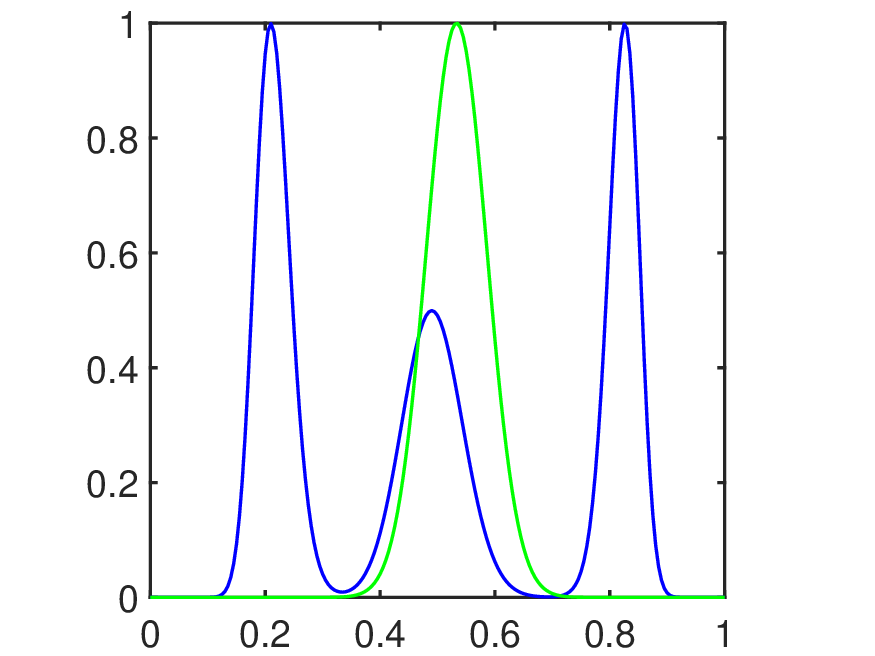}
	\caption{obs}
\end{subfigure}	
\begin{subfigure}[h]{0.23\textwidth}
	\includegraphics[width=\textwidth]{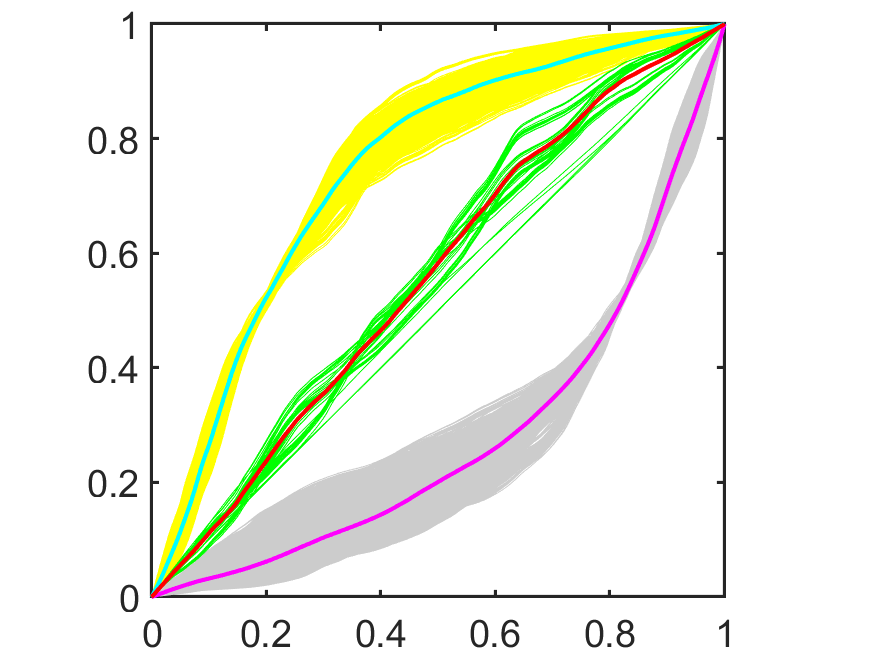}
	\caption{warping (G)}
\end{subfigure}	
\begin{subfigure}[h]{0.23\textwidth}
	\includegraphics[width=\textwidth]{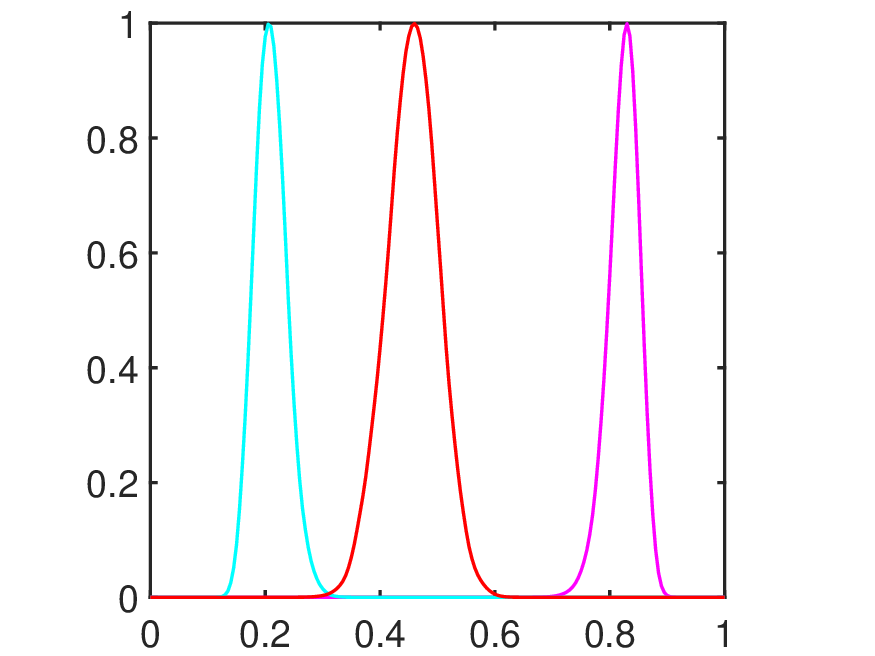}
	\caption{aligned $f_2$ (G)}
\end{subfigure}
\vfill
\begin{subfigure}[h]{0.23\textwidth}
	\includegraphics[width=\textwidth]{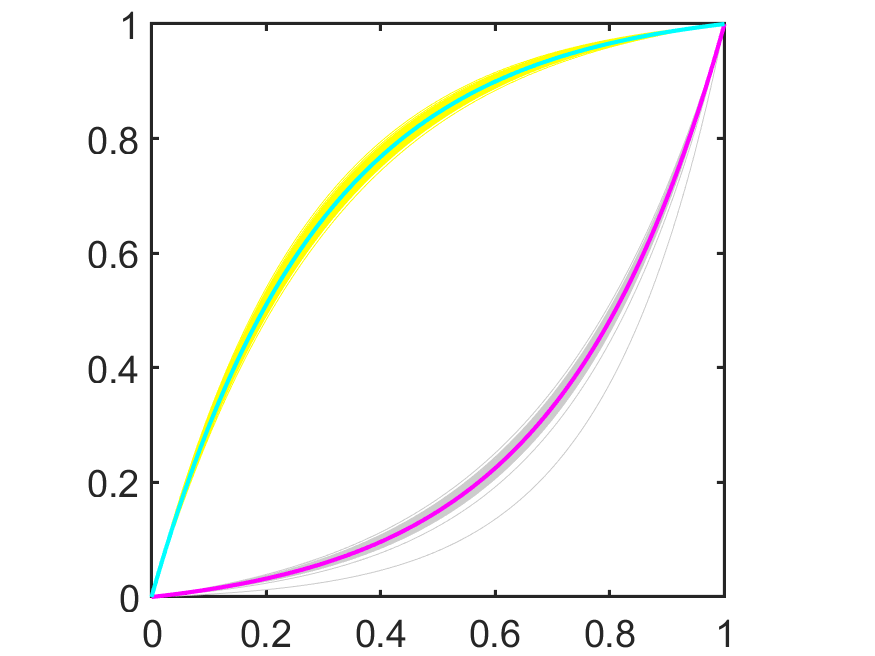}
	\caption{warping (NG)}
\end{subfigure}
\begin{subfigure}[h]{0.23\textwidth}
	\includegraphics[width=\textwidth]{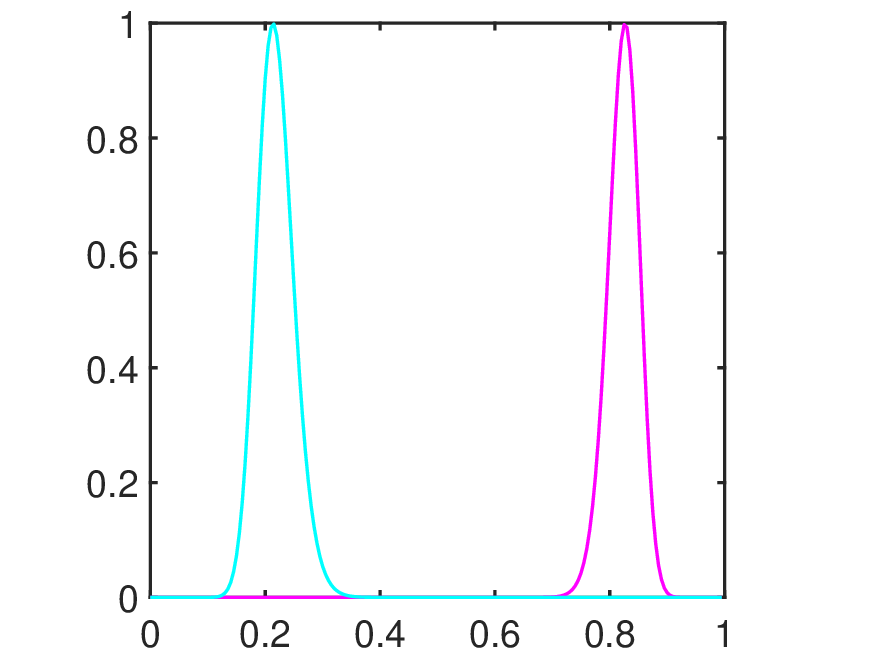}
	\caption{aligned $f_2$ (NG)}
\end{subfigure}

\caption{Result on the non-Gaussian process prior. (a) Original functions $f_1$ (in blue) and $f_2$ (in green). (b) Sampled warping functions with a Gaussian process prior: Warping functions are categorized into three clusters, colored in yellow, green and grey, with the cluster averages in cyan, red and magenta, respectively. (c) The cyan, red and magenta curves are the aligned $f_2$ with the cluster averaged warpings in (b), respectively. (d) Sampled warping functions with a non-Gaussian process prior: Warping functions are categorized into two clusters, colored in yellow and grey, with the cluster averages in cyan and magenta, respectively. (e) The cyan and magenta curves are the aligned $f_2$ with the cluster averaged warpings in (d), respectively. }
\label{fig:nonG}
\end{figure}

\textbf{Simulation 3: }
We here use one simulation to demonstrate Algorithm \ref{alg:pwbayesian_nonG} and showcase the effectiveness of the non-Gaussian process prior in the Bayesian registration, in contrast to a Gaussian process prior. We at first simulate a function $f_2(t)= \exp(-180(t-0.53)^2), t\in [0, 1]$. Using the basis function ${b_1 = \frac{\sqrt{3}}{2}}(t-0.5)$, we then generate 3 warping functions $\gamma_1, \gamma_2, \gamma_3$ with coefficients $\xi_1 = -1$, $\xi_2 = -0.2$ and $\xi_3 =1$. Using these warpings, we can obtain the other function $f_1 = f_2(\gamma_1(t)) + 0.5f_2(\gamma_2(t))+f_2(\gamma_3(t))$. The goal of this simulation is to align $f_2$, characterized as a single-peaked curve, to $f_1$, another curve featuring three peaks. The second peak in $f_1$ is much smaller and can be treated as a noise feature, as depicted in Figure \ref{fig:nonG} (a).  

As a comparison, we at first use a Gaussian process prior on the time warping, with the mean $\mu_h=0$  and covariance $C_h(s,t) = 8 \cdot 0.999^{200|s-t|}, s, t \in[0,1]$. The Gaussian prior, when applied, enables identification of warpings that align $f_2$ with each of these three peaks of $f_1$. The outcomes of this approach are presented in Panels (b), which notably form three distinct clusters.  To understand the alignment in each cluster, we conduct the conventional K-means clustering method (more details on $K$-means will be provided in the next section) and categorize the warping functions into 3 groups. 
Panel (c) illustrates three different alignments of $f_2$ using the mean warping of each cluster shown in Panel (b). 

However, if the second (relatively insignificant) peak is considered as a noise feature, then the associated middle warping cluster should not be desired in the Bayesian registration.  In this case, the Gaussian process prior becomes inappropriate due to its continuous nature, and a Non-Gaussian process prior can be naturally adopted to address the issue. We here choose the basis function ${b_1 = \frac{\sqrt{3}}{2}}(t-0.5)$, and set the prior distribution of $\xi_1\sim U((-1.5, -0.8)\cup(0.8, 1.5))$. The alignment results with the non-Gaussian process prior are presented in Panels (d) and (e), where Panel (d) exhibits the sampled warpings with two cluster and Panel (e) displays the aligned $f_2$ by the averaged warping in each cluster.  It can be seen that in contrast to the outcomes achieved with a Gaussian process prior, the sampled warpings effectively exclude those close to the identity warping that aligns with the second peak. \\

\noindent \textbf{Remark 2:}  This approach is more effective when we have prior knowledge or certain constraint on the desirable warping functions. In the above simulation, we assume the orthonormal basis functions are pre-known.  When such information is not available, we can employ the functional Principal Component Analysis (fPCA) on the covariance kernel to achieve a more accurate representation of the basis.

\section{Effects of the prior}
\label{Sec:prior}
In this section, we will explore the effects of the prior selection on the final outcomes of the Bayesian registration. We will focus on Gaussian process prior for its standard representation form.  Examining these effects enables a deeper understanding of how to strategically choose priors to achieve desired results. Rather than directly comparing the dynamic programming as an alternative method for optimal alignment, our objective is to gain a comprehensive understanding of the Bayesian registration. We aim to leverage its capabilities fully, focusing on its capacity to offer diverse results for user selection and the incorporation of priors to enhance practical outcomes, transcending mere mathematical considerations.

%

\subsection{Effect of $\sigma$}

In previous studies, $\sigma$ is often considered as a random variable, typically with a prior of the inverse Gamma distribution for its sampling and estimation. This can be done within our framework as well. However, the comprehensive impact of $\sigma$ has yet to be fully explored. Therefore, in this section, we treat it as a deterministic parameter and aim to investigate its effect in regulating function alignment. It can be seen from Eqn. \eqref{eqn:postG}, by manipulating $\sigma$ within this posterior measure, we can influence the role of the prior in shaping the posterior. Specifically, a smaller $\sigma$ amplifies the impact of the likelihood term, so that the sampled $\gamma$ remains close and densely distributed around the warping, obtained directly using the dynamic programming method within the SRVF framework. Conversely, a larger $\sigma$ intensifies the influence of the prior term, yielding a posterior of $\gamma$ that diverges from the warping obtained via the dynamic programming, and more closely aligns with the prior.   In the following part, we will employ a simulation to demonstrate its influence on function alignment. 

\begin{figure}[h]
\centering
\begin{subfigure}[h]{0.25\textwidth}
	\includegraphics[width=\textwidth]{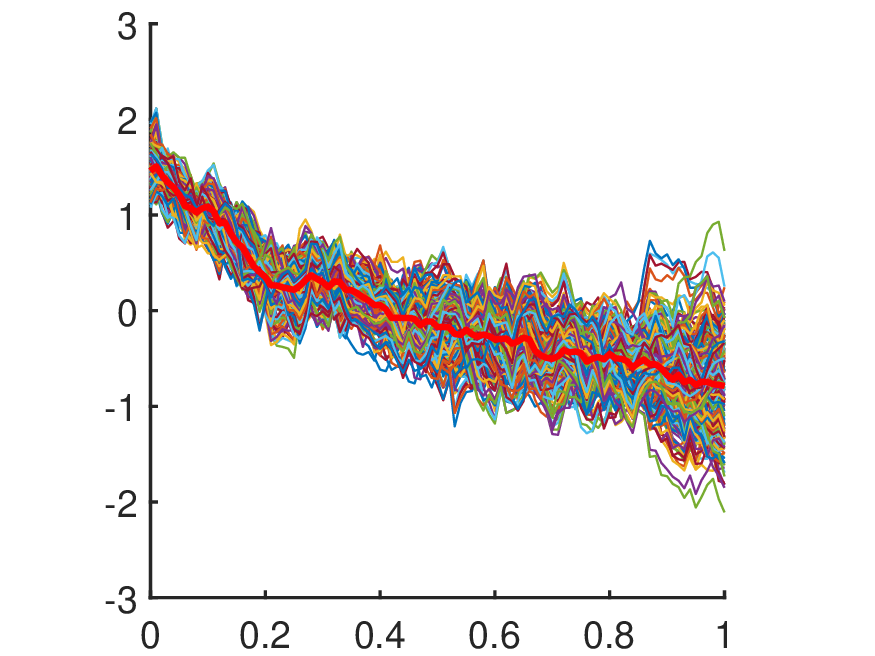}
	\includegraphics[width=\textwidth]{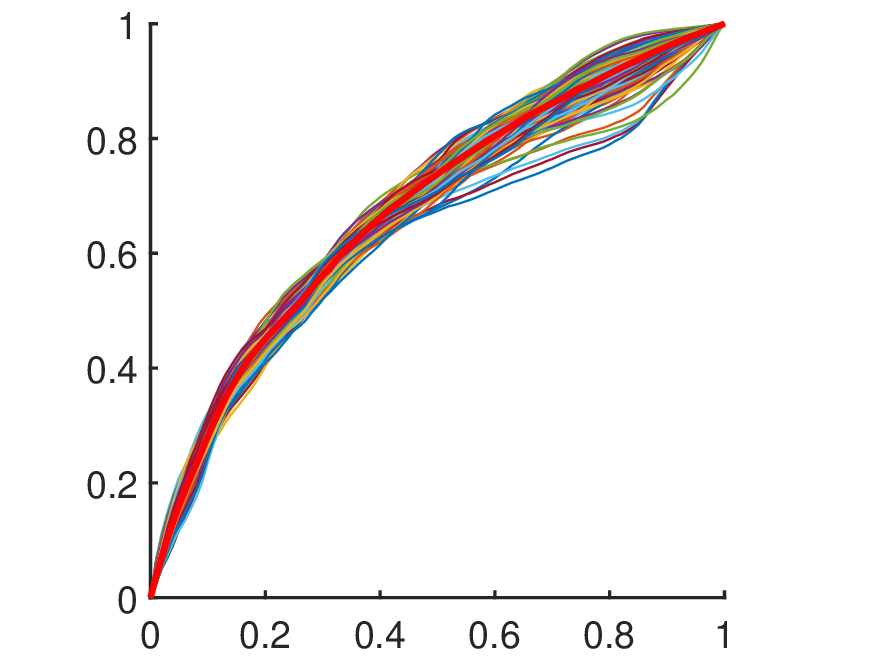}
	\caption{$\sigma = 10$}
\end{subfigure}\hspace{-0.5cm}
\begin{subfigure}[h]{0.25\textwidth}
	\includegraphics[width=\textwidth]{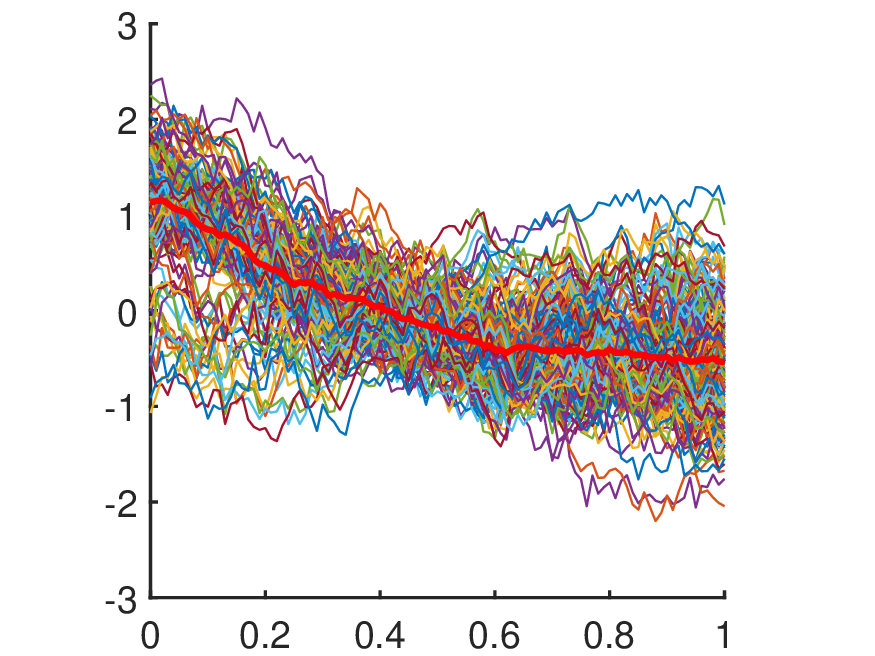}
	\includegraphics[width=\textwidth]{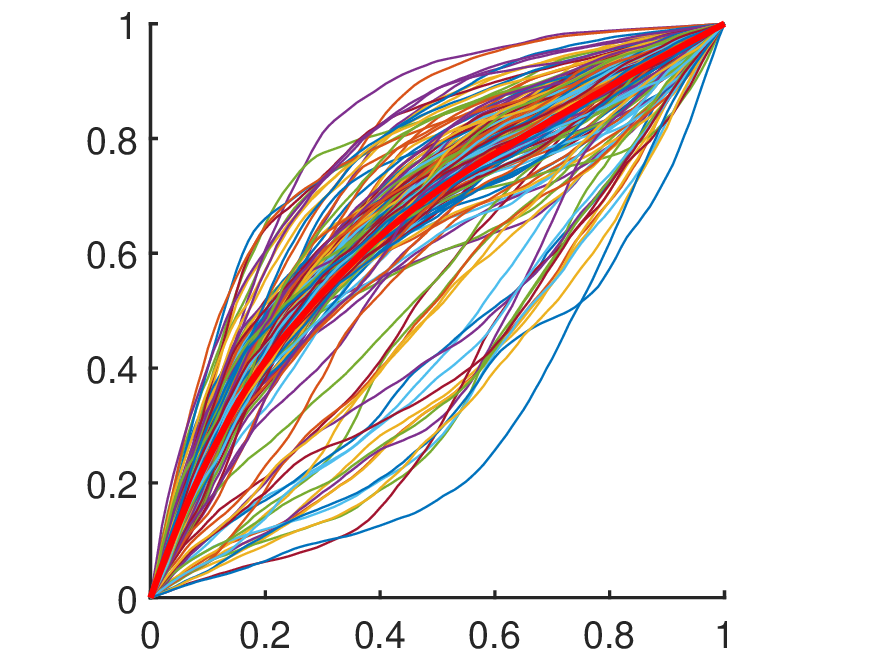}
	\caption{$\sigma = 20$}
\end{subfigure}\hspace{-0.5cm}
\begin{subfigure}[h]{0.25\textwidth}
	\includegraphics[width=\textwidth]{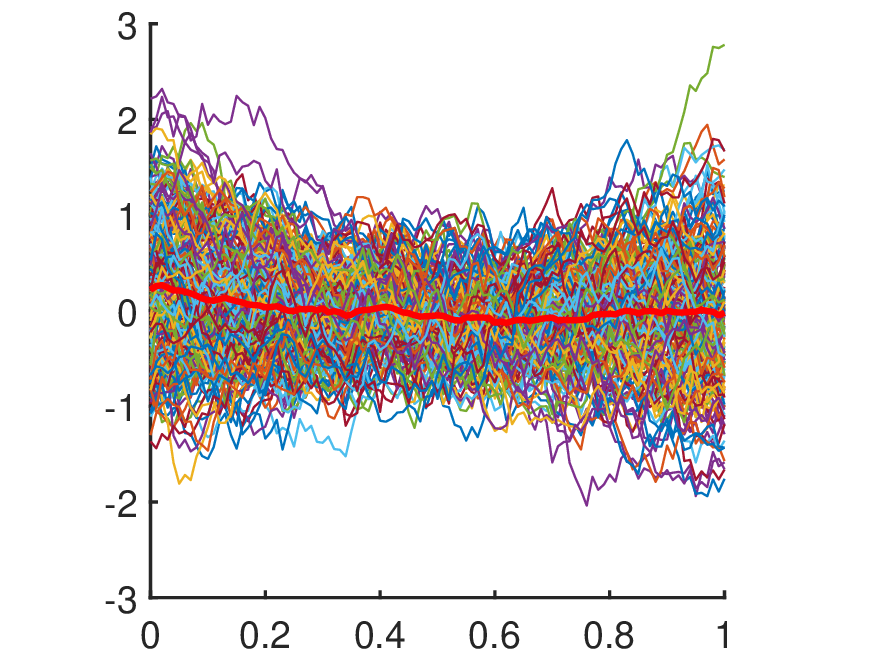}
	\includegraphics[width=\textwidth]{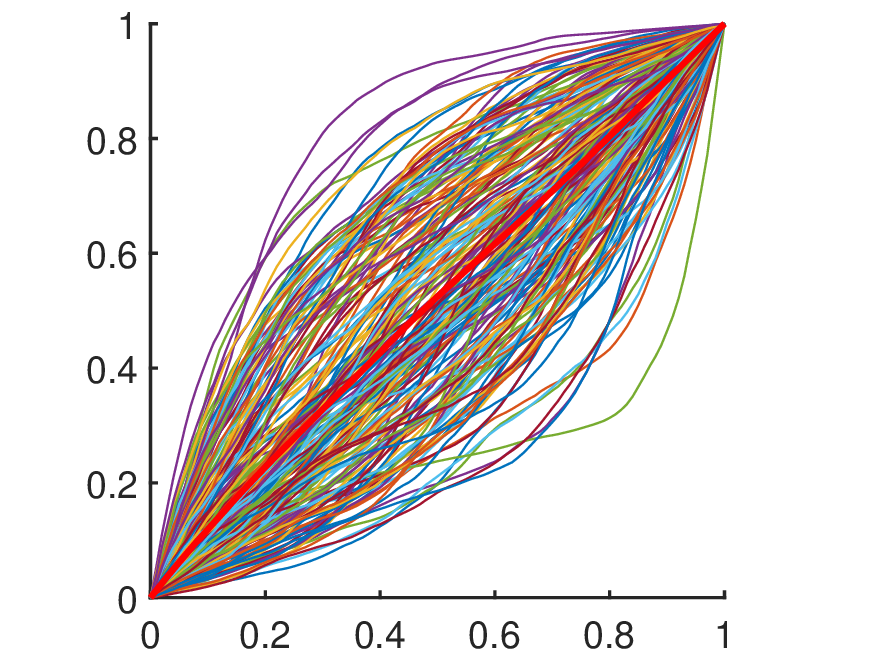}
	\caption{$\sigma = 50$}
\end{subfigure}
\caption{Result on Simulation 2 with different $\sigma$. In the first row, the curves represent the sampled CLR-transformed warping functions, with the red bold curve indicating their mean. The second row displays the corresponding warping functions. (a) $\sigma =10$, (b) $\sigma =20$, (c) $\sigma =50$.}
\label{fig:sim2_sigma}
\end{figure}

\textbf{Simulation 2*: }
To illustrate the impact of $\sigma$, we reuse Simulation 2 in Sec \ref{gp} and keep all parameters identical to the previous setup except for $\sigma$. Previously, in order to emphasize the likelihood term, we assigned a small value of 2 to $\sigma$. This resulted in the sampled warping closely resembling those obtained via the dynamic programming, which aims at maximizing the likelihood term. At the same time, the sampled warpings exhibited minimal variability, leading to aligned $f_2$ effectively matching $f_1$ in phase. Manipulating the value of $\sigma$ produces different registration results, as illustrated in Figure \ref{fig:sim2_sigma}. The first row presents the sampled CLR-transformed warping functions, while the second row shows the corresponding warping functions. Panels (a), (b), and (c) correspond to $\sigma$ values of 10, 20, and 50, respectively.  We can observe that as $\sigma$ increases, the sampled CLR-transformed warpings begin to more closely reflect the prior distribution. This results in the mean functions moving increasing toward to 0. Correspondingly, the average of these sampled warpings gradually approaches the identity warping. 
Furthermore, as $\sigma$ increases, the point-wise variance among the time points of the CLR-transformed functions grows, resulting in a broader band for the sampled CLR-transformed warping functions. This, in turn, causes the sampled warpings to expand in width as well.

\subsection{Effect of the covariance operator $C_h$}
The existence of an isometric isomorphism between the $\mathbb{L}^2$ subspace $H(0,1)$ and warping space $\Gamma_1$ allows for the covariance operator to be defined directly within the $\mathbb{L}^2$ space, which simplifies the process and enhances understanding of the chosen covariance. This is in contrast to the SRVF framework, which does not hold a one-to-one correspondence between the tangent space and warping space, and the covariance of functions on the tangent space may not accurately represent the covariability in the time domain of the original warping functions. 

In \cite{lu2017bayesian}, the covariance operator is specified through a pre-defined orthonormal basis such as a Fourier basis. In contrast, our approach allows for a more direct customization on the covariance. The choice of the covariance may not appear critical, especially when the focus of the function registration is solely on the likelihood term.  However, the scenario changes when there is a need to integrate the prior information.  For instance, a proper prior may provide desirable smoothness on the time warping, whereas a likelihood term in general cannot achieve that goal.  Moreover, to provide different variability across the time domain, one can simply tailor the covariance function accordingly by assigning higher variance values to intervals requiring less restriction and lower values to those needing more restrictions. Such degree of customization is challenging to achieve with a pre-given basis method. 

In constructing Gaussian process prior, it is desirable to make the covariance kernel continuous.  This is based on two main considerations: Firstly, the warping functions is used to align two functions that are absolutely continuous, so the warping also is expected to be continuous. Secondly, to apply the Karhunen-Lo\`eve (KL) expansion, it is essential to ensure the continuity of the covariance kernel. Moreover, if the covariance kernel is continuous, its basis functions will also be continuous and therefore belong to $H(0, 1)$, a $\mathbb{L}^2$ subspace.  Consequently, unlike the penalized registration in \citep{ma2024stochastic} where a diagonal covariance can be adopted directly, we need to ensure that the covariance kernel is continuous. A simple method to achieve this goal is by applying a smoothing procedure. This is illustrated with the following example. 

\begin{figure}[ht!]
\centering
\begin{subfigure}[h]{0.3\textwidth}
	\includegraphics[width=\textwidth]{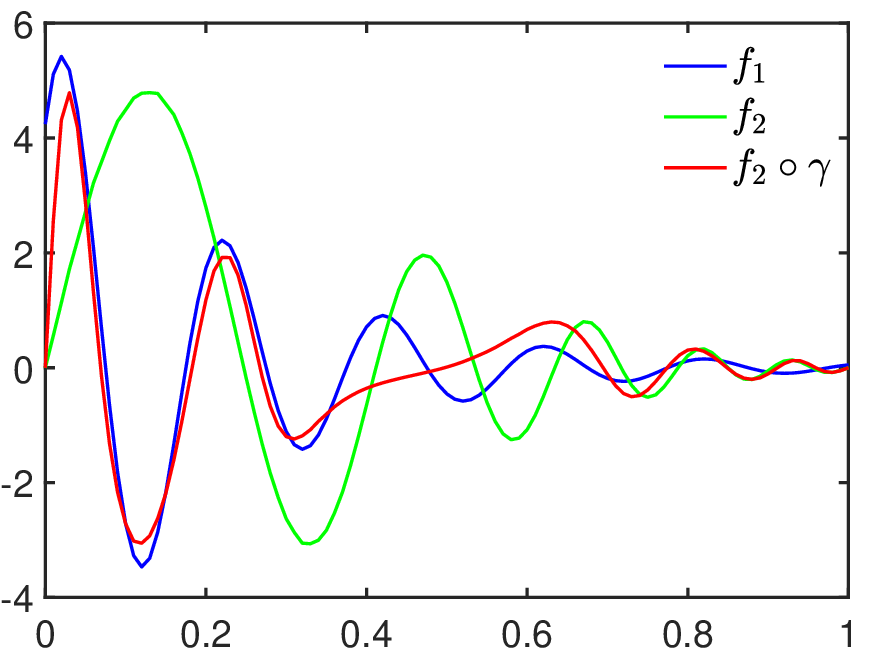}
	\caption{observations}
\end{subfigure}
\begin{subfigure}[h]{0.3\textwidth}
	\includegraphics[width=\textwidth]{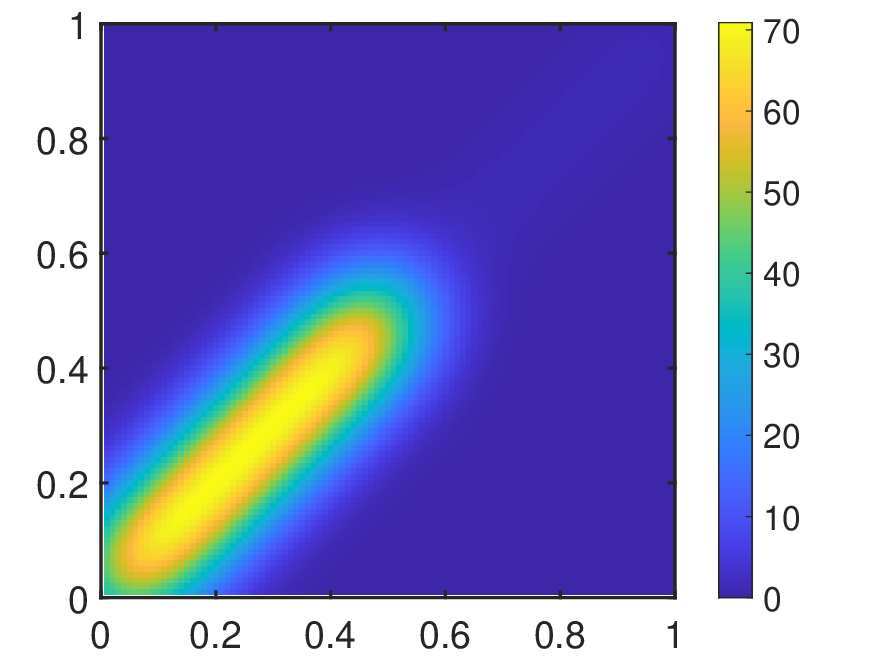}
	\caption{$C_h$}
\end{subfigure}
\vfill
\begin{subfigure}[h]{0.3\textwidth}
	\includegraphics[width=\textwidth]{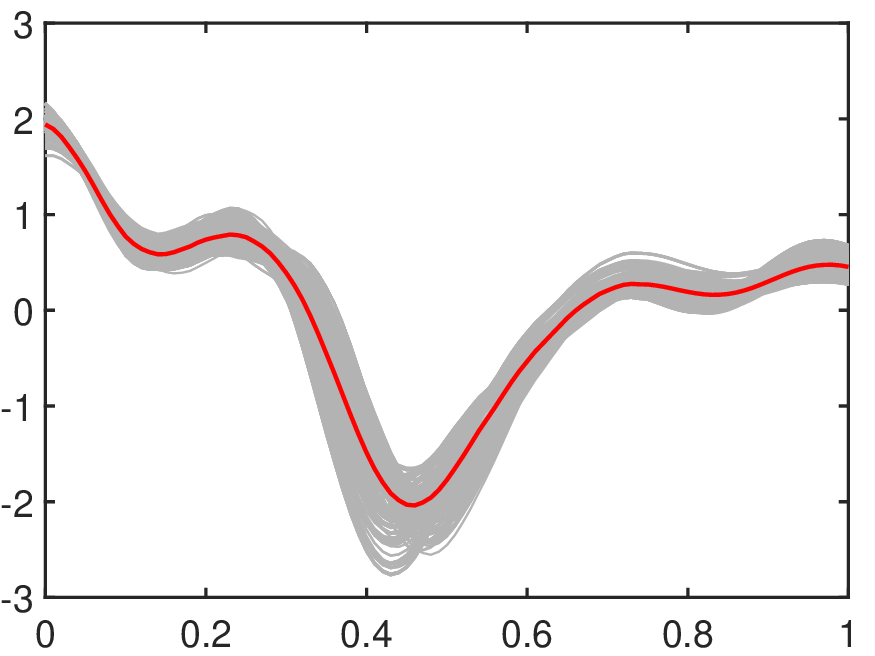}
	\caption{sample in $H(0,1)$}            
\end{subfigure}	
\begin{subfigure}[h]{0.3\textwidth}
	\includegraphics[width=\textwidth]{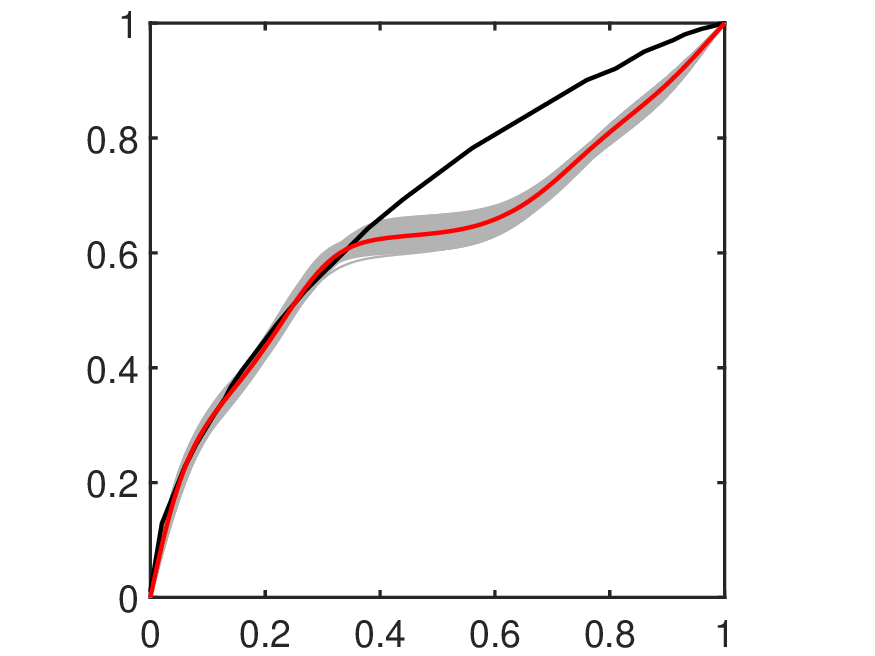}
	\caption{sampled warpings}
\end{subfigure}	

\caption{Results on Simulation 4. (a) Original and aligned functions. The blue and green curves are the two given functions $f_1$ and $f_2$.  The red curve is the aligned $f_2$ with the averaged warping. (b) The heatmap of $C_h$. (c) Sampled functions in $H(0,1)$, and their mean is shown in red. (d) Sampled warping functions in the Bayesian alignment are shown in grey shade, and their mean is shown as the red curve. The bold black curve is the warping estimated using the dynamic programming method.} 
\label{fig:sim3}
\end{figure}

\textbf{Simulation 4: }  To illustrate the impact of $C_h$, we reuse observations generated in Simulation 2 in Sec \ref{gp}.  The functions $f_1(t)$ and $f_2(t)$ are shown in Figure \ref{fig:sim3}(a). In Simulation 2, we ignore the magnitudes of the functions and align all peaks and valleys. However, if our objective is to align only the peaks with higher magnitudes, treating lower magnitude portions as noise, we can use a diagonal-like covariance. That is, we provide increased flexibility in aligning the first half of the function while restricting flexibility in the second half. Therefore, to build such a Gaussian process prior, we take the bivariate kernel function $C(s,t)=r(t)\delta(s-t)$, where $r(t)=\begin{cases}
5 & \text {if $0 \le t \le 0.5$} \\
0.1 & \text {if $0.5 < t \le 1$} 
\end{cases}$, and $\delta(\cdot)$ is the Dirac delta function.  Then, we apply a 2-D Gaussian smoother with bandwidth 0.08 to get the covariance function $C_h$.   This formulation allows a non-uniform variation of the CLR-transformed warping $h$ across the time domain. Specifically, the variance is substantial in $[0, 1/2]$, and then decreases rapidly to near 0 in $[1/2, 1]$.  This is illustrated with a heat-map graph in Figure \ref{fig:sim3}(b). It can be seen that the weights of the prior term are non-homogeneous across the time domain. Finally, we have intentionally set $\sigma$ to a moderate value of 5 to achieve a balanced impact between the likelihood term and the prior term in the posterior. 

By employing Algorithm \ref{alg:pwbayesian}, the sampled CLR-transformed warping functions in $H(0,1)$ are shown Panel (c) in Figure \ref{fig:sim3} with their corresponding warping functions in Panel (d).  It can be seen that a distinctive pattern emerges in the sampled warping functions: In the first half of time, these functions notably mirror the $\gamma$ obtained via the dynamic programming. However, an intriguing shift occurs in the second half, where the dispersion diminishes in comparison to the first half. Notably, the sampled warping in the second half time tends to resemble the identity warping. This phenomenon can be explained by the fact that, in the second half, the sampled CLR-transformed warping experiences a reduction in variance, resulting in a smaller overall variability. Therefore, the second half of the CLR-transformed function converges closely to 0, causing the corresponding warping part to exhibit characteristics of the identity warping.

From this simulation, it becomes evident that we have the flexibility to create various covariance to modulate the alignment result we wish to achieve. Essentially, this signifies that, with prior information or desirable constraints, our method empowers us to define a requisite covariance structure, tailoring it to specific requirements and enhancing the adaptability of our alignment framework.

\subsection{Impact of $\mu_h$: Optimize alignment by choosing various means}
\subsubsection{Clustering of the alignment}
An inherent advantage of Bayesian registration lies in its ability to sample various warping functions, distinguishing it from the dynamic programming method in the Fisher-Rao framework that searches for a singular optimal solution. This not only enhances the versatility of the registration method but also allows for a more comprehensive exploration of potential warpings, catering to a wider range of scenarios, and offering increased adaptability in addressing complex registration challenges. 

Consider a scenario with two functions, one exhibiting a bi-modal pattern and the other a uni-modal pattern, as previously demonstrated in \citep{kurtek2017geometric}. With the dynamic programming method, only one warping choice is available -- aligning the uni-modal function with either the first or second mode of the bi-modal function. However, by applying the MCMC algorithm within the Bayesian registration framework, we can obtain the posterior of the warping functions.  This distribution includes all possible warpings that can be utilized in the alignment process, which is notably forming two distinct clusters.  With this collection of diverse warping functions, clustering methods such as $K$-means can be used to categorize them into distinct groups. $K$-means can be directly used with the conventional $\mathbb{L}^2$ metric on the CLR-transformed warping functions.  This is in contrast to previous $K$-means methods for the SRVF of time warping, where the Karcher mean is estimated via an iterative procedure \citep{kurtek2017geometric}.  We will use one simulation to demonstrate the effectiveness of our clustering approach as follows:  

\begin{figure}[ht!]
\centering
\begin{subfigure}[h]{0.3\textwidth}
	\includegraphics[width=\textwidth]{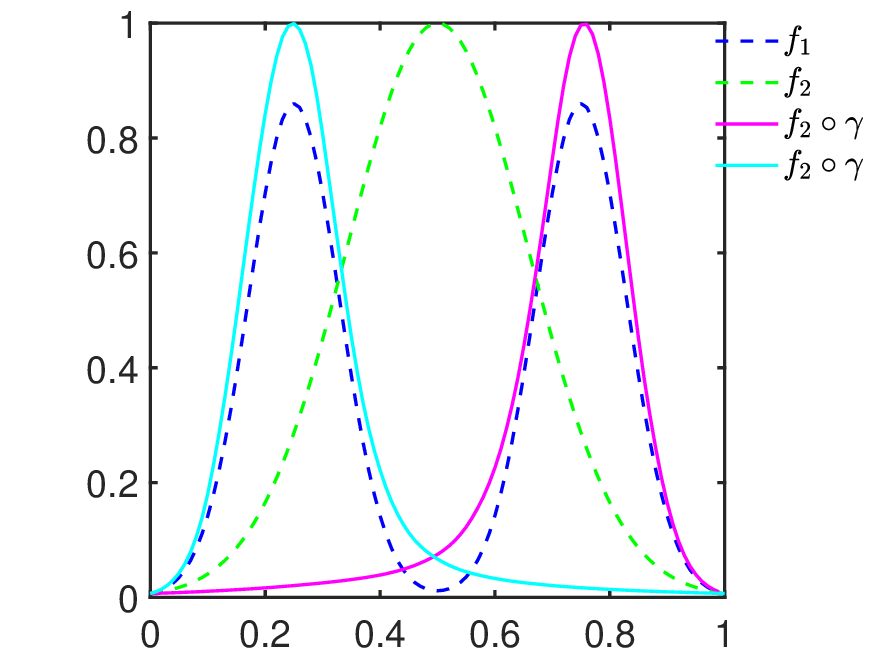}
	\caption{observations}
\end{subfigure}
\begin{subfigure}[h]{0.3\textwidth}
	\includegraphics[width=\textwidth]{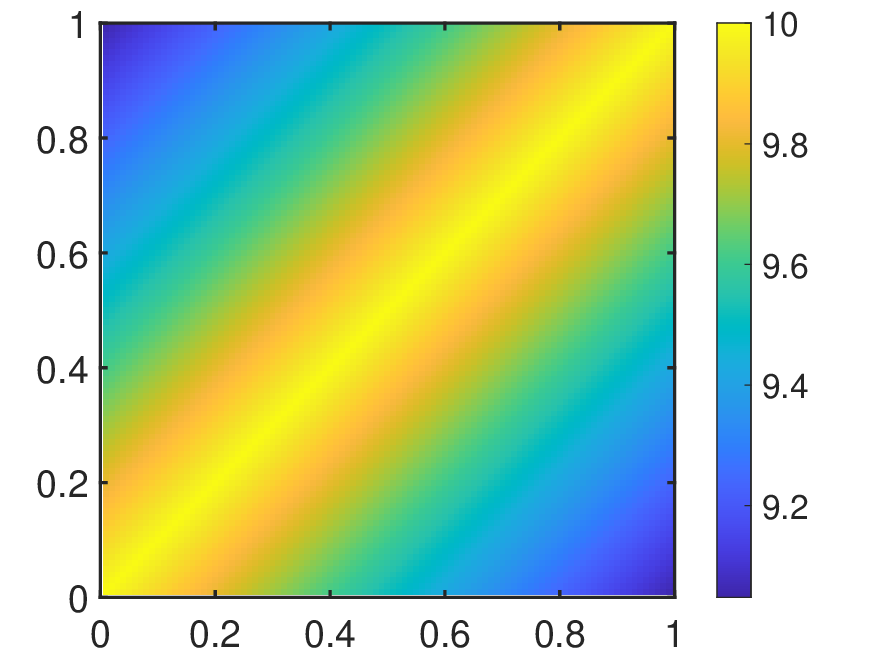}
	\caption{$C_h$}
\end{subfigure}
\vfill
\begin{subfigure}[h]{0.3\textwidth}
	\includegraphics[width=\textwidth]{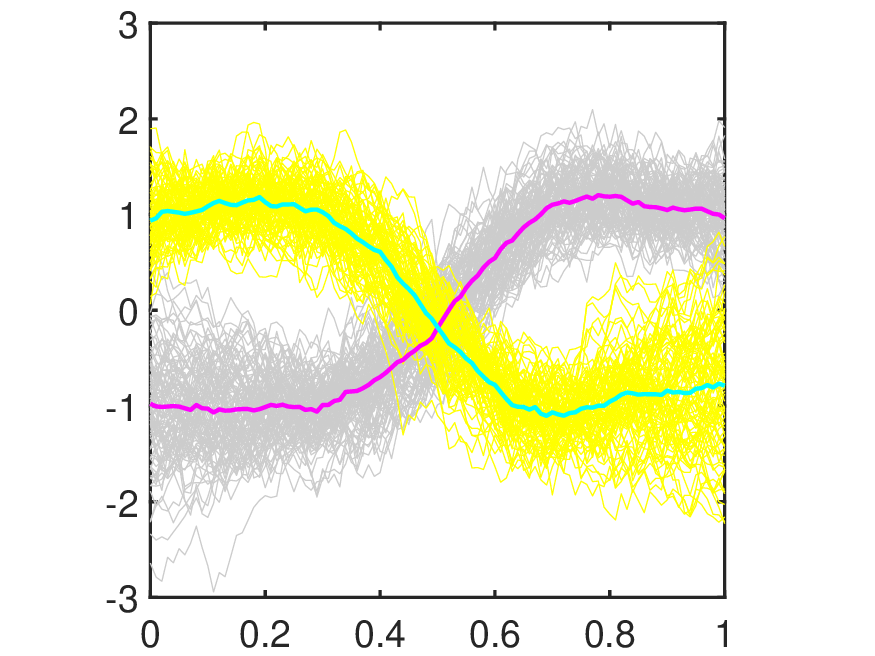}
	\caption{sample in $H(0,1)$}
\end{subfigure}
\begin{subfigure}[h]{0.3\textwidth}
	\includegraphics[width=\textwidth]{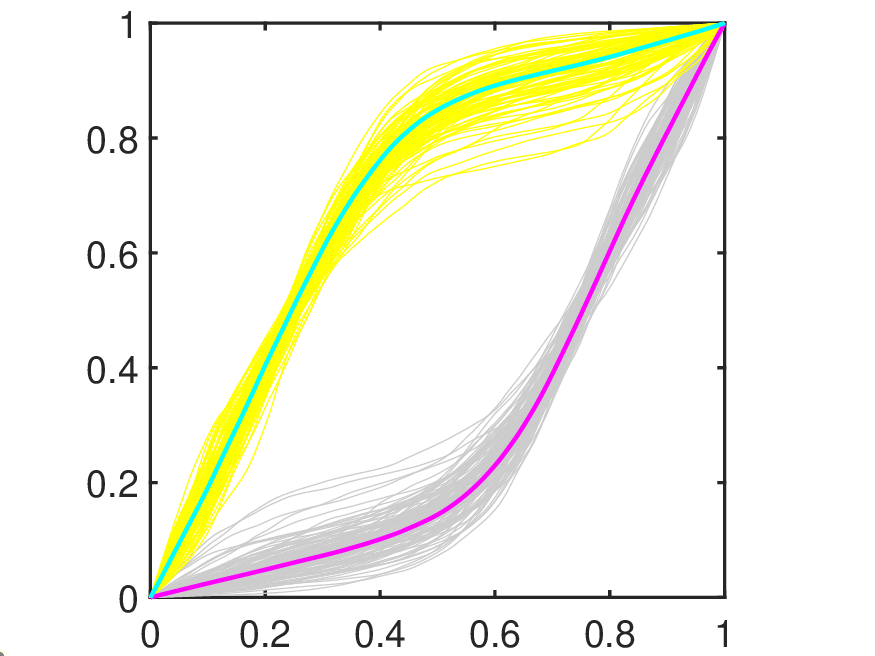}
	\caption{sampled warpings}
\end{subfigure}

\caption{Result on Simulation 5. (a) Original functions and alignment functions. The blue and green dotted curves represent the two given functions $f_1$ and $f_2$.  For comparative purposes, the superimposed magenta and cyan solid curves are the aligned $f_2$ with the mean of the sampled warpings for two clusters obtained through $K$-means. They correspond to the two warpings depicted in Panel (d). (b) The heatmap of $C_h$. (c) Sampled functions in $H(0, 1)$, categorized into two clusters. The cyan and magenta curves represent the mean warpings for the yellow and grey groups, respectively. (d) Obtained warping functions corresponding to functions in Panel (c).}
\label{fig:2v1}
\end{figure}

\textbf{Simulation 5: } We generate one bi-modal function $f_1(t)=0.86e^{-80(t-0.25)^2}+0.86e^{-80(t-0.75)^2}$ and one uni-modal function $f_2(t)=e^{-20(t-0.5)^2}$. The functions $f_1(t)$ and $f_2(t)$ are shown in Figure \ref{fig:2v1}(a), respectively. Moreover, we set $C_h(s,t)=10\cdot0.999^{100|s-t|}$ (see its heat-map graph in Figure \ref{fig:2v1}(b)). 
All the sampled warpings in $H(0,1)$ are plotted in Figure \ref{fig:2v1}(c), revealing a clear separation into two distinct groups. By applying the $K$-means method, the functions in the two groups are identified and shown in grey and yellow in the plot.  The mean warping for each cluster is separately depicted, with magenta representing the grey group and cyan indicating the mean warping for the yellow group.  The samples and means can also be displayed in $\Gamma_1$ as warping functions in Figure \ref{fig:2v1}(d).
Utilizing these two mean warping functions, we also show the aligned $f_2$ in magenta and cyan in Panel (a), respectively. Notably, the cyan warping aligns the peak of $f_2$ with the first peak of $f_1$, while the magenta warping aligns the peak of $f_2$ with the second peak of $f_1$.

\subsubsection{Control on the alignment}
In the Bayesian registration, the prior term encapsulates our initial knowledge about phase variability before observing any data. The choice of the prior term is a critical aspect of the Bayesian inference, and it involves a trade-off between incorporating prior knowledge and letting the data speak for itself. Different choices of prior can lead to varying results in the posterior. We have discussed the impact of Gaussian and non-Gaussian process priors, the role of the parameter $\sigma$, and the effect of covariance operators.  Furthermore, manipulating the prior allows us to dictate which peak we aim to align. This can be achieved by using different mean function $\mu_h$ and adjusting the scale of $C_h$. An example will be presented below to illustrate the effectiveness of this method.

\begin{figure}[ht!]
\centering
\begin{subfigure}[h]{0.21\textwidth}
	\includegraphics[width=\textwidth]{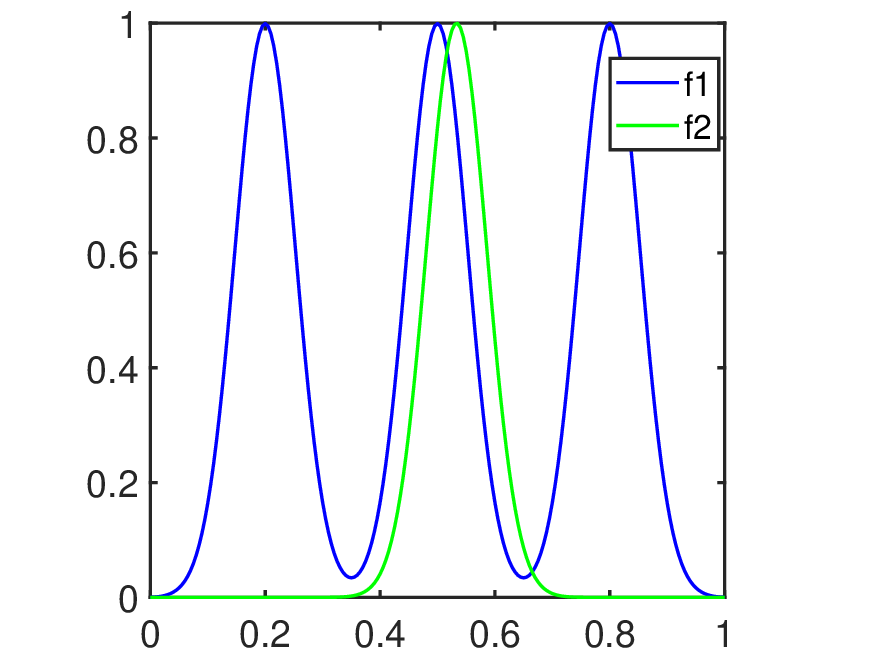}
	\caption{obs}
\end{subfigure}\hspace{-0.5cm}
\begin{subfigure}[h]{0.21\textwidth}
	\includegraphics[width=\textwidth]{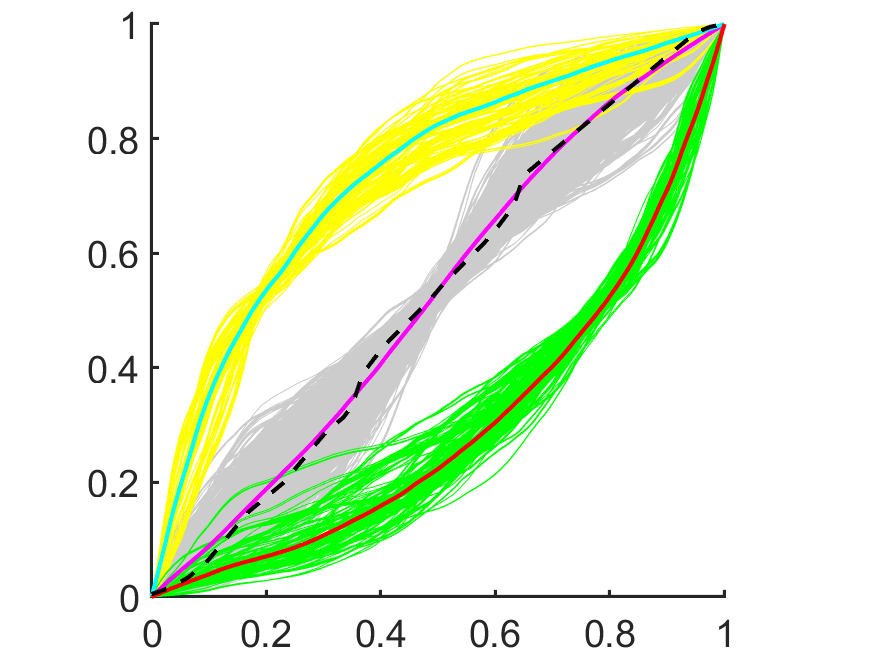}
	\includegraphics[width=\textwidth]{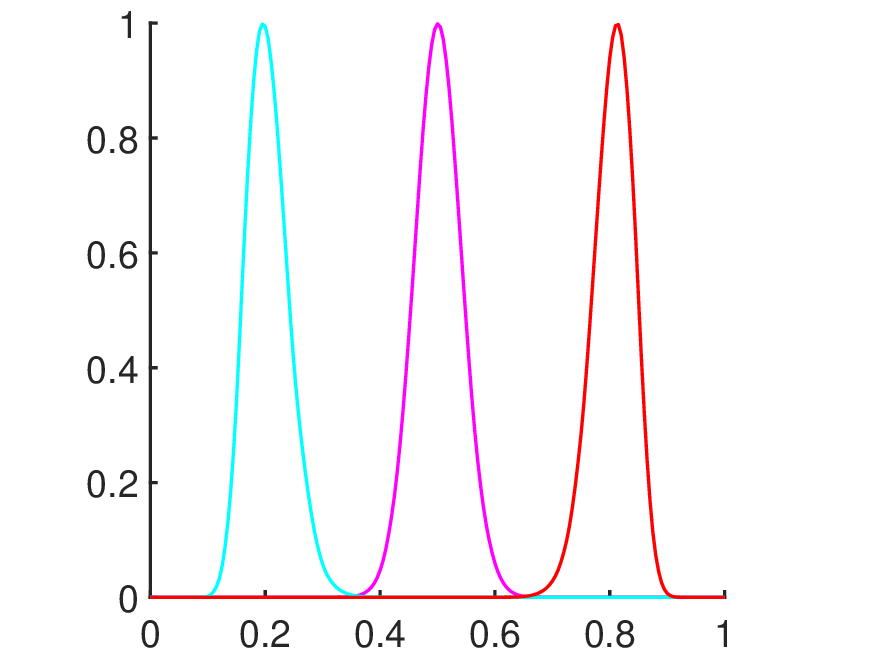}
	\captionsetup{justification=centering}
	\caption{$\mu_h = 0$\\$\sigma_{sc}=8$}
\end{subfigure}\hspace{-0.5cm}
\begin{subfigure}[h]{0.21\textwidth}
	\includegraphics[width=\textwidth]{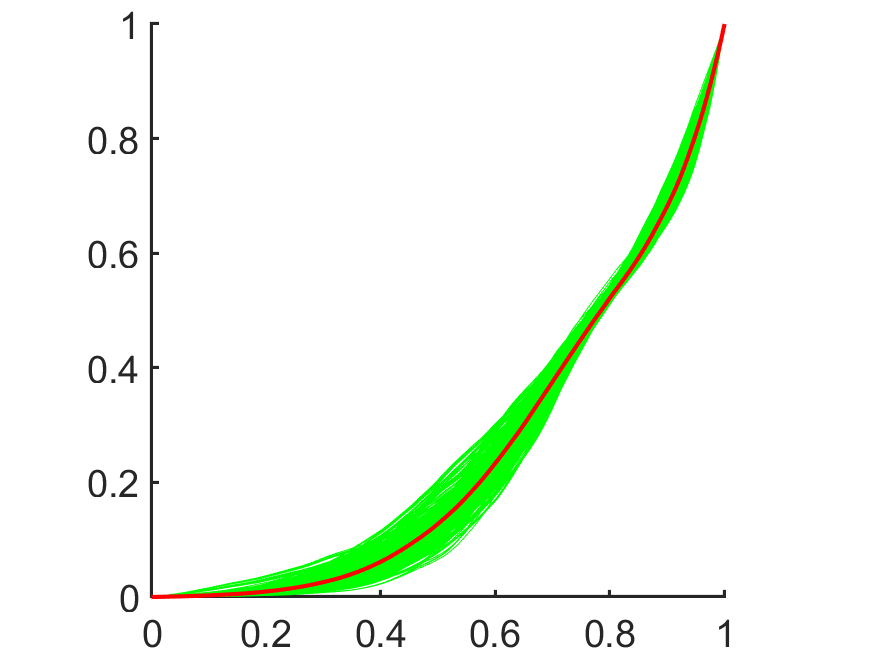}
	\includegraphics[width=\textwidth]{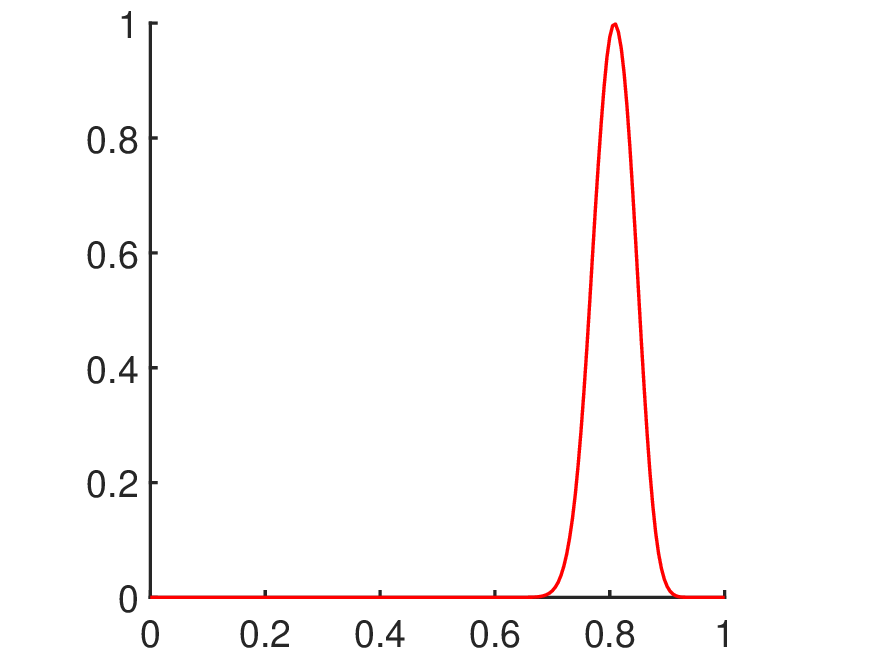}
	\captionsetup{justification=centering}
	\caption{$\mu_h = 2(t-0.5)$\\$\sigma_{sc}=2$}
\end{subfigure}\hspace{-0.5cm}
\begin{subfigure}[h]{0.21\textwidth}
	\includegraphics[width=\textwidth]{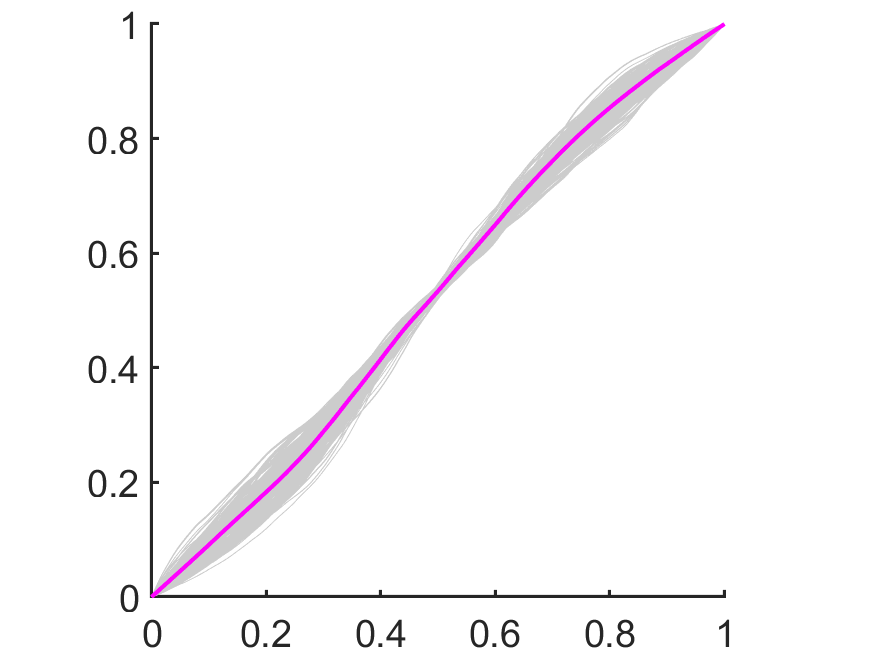}
	\includegraphics[width=\textwidth]{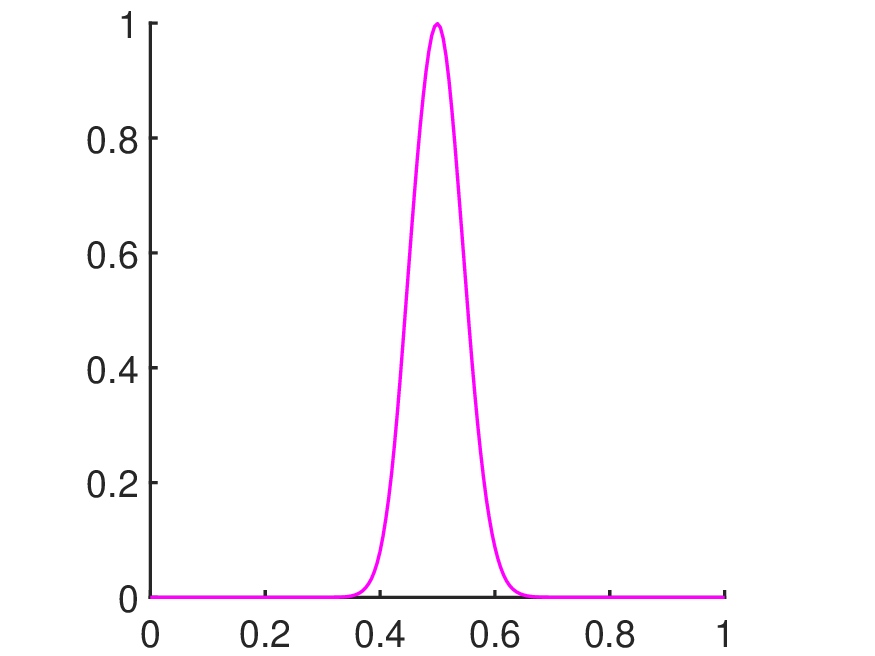}
	\captionsetup{justification=centering}
	\caption{$\mu_h = 0$\\$\sigma_{sc}=2$}
\end{subfigure}	\hspace{-0.5cm}
\begin{subfigure}[h]{0.21\textwidth}
	\includegraphics[width=\textwidth]{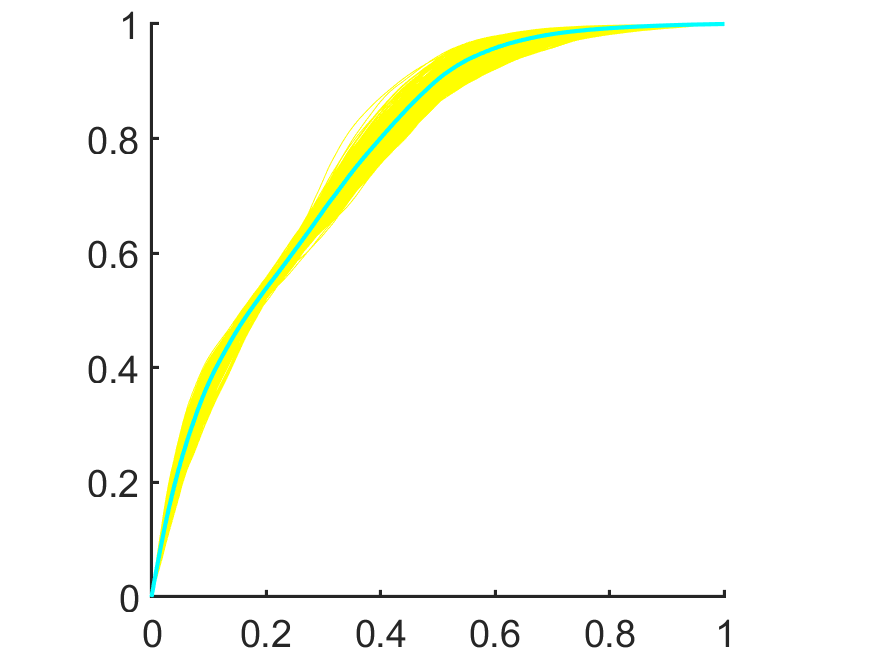}
	\includegraphics[width=\textwidth]{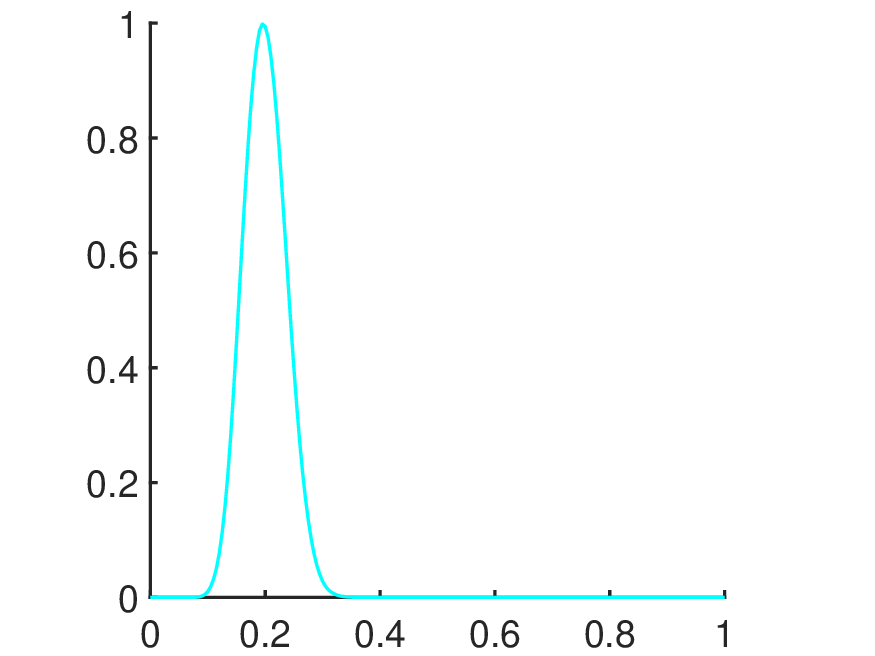}
	\captionsetup{justification=centering}
	\caption{$\mu_h = -2(t-0.5)$\\$\sigma_{sc}=2$}
\end{subfigure}
\caption{Result on peak controlling. (a) Original observations. The blue and green curves represent the two given functions $f_1$ and $f_2$, respectively. (b) Top row: sampled warping functions via a Gaussian process prior with $\mu_h=0$ and $\sigma_{sc} = 8$. The warping functions are categorized into three clusters, colored in yellow, grey, and green, and the means are colored in cyan, magenta, and red, respectively. Bottom row: the cyan, red and magenta curves are the aligned $f_2$ with the mean warpings in these three clusters, respectively. (c) Same as (b) except that $\mu_h =2(t-0.5)$ and $\sigma_{sc} = 2$, (d) Same as (b) except that $\mu_h =0$ and $\sigma_{sc} = 2$, (e) Same as (b) except that $\mu_h =-2(t-0.5)$ and $\sigma_{sc} = 2$.  }
\label{fig:3v1}
\end{figure}

\textbf{Simulation 6: } We here use one simulation to demonstrate the algorithm and showcase the effectiveness the alignment control by adjusting the prior mean. We at first simulate two functions, i.e., $f_1(t)= \exp(-180(t-0.2)^2) + \exp(-180(t-0.5)^2) + \exp(-180(t-0.8)^2), t\in [0, 1]$ and $f_2(t)= \exp(-180(t-0.53)^2), t\in [0, 1]$. In this simulation, our objective is to align the single-peaked $f_2$ with the triple-peaked $f_1$, shown in Figure \ref{fig:3v1}(a). Conventional dynamic programming offers a singular optimal alignment, lacking the capability to choose a specific peak to align. However, the introduction of a new Gaussian process prior enables us to precisely control this process. 

At first, we set the mean function $\mu_h = 0$ and  $C_h(s,t) = \sigma_{sc} \cdot 0.999^{200|s-t|}, s, t \in[0,1]$, where $\sigma_{sc} = 8$. This prior enables identification of warpings that align $f_2$ with each of these three peaks of $f_1$, same as Simulation 5. The outcomes of this approach are presented in Panel (b), where the plot in first row showcases sampled warpings that form three distinct groups. By employing the $K$-means clustering, these warpings can be categorized into three clusters. Utilizing the mean warping of these clusters for function alignment makes $f_2$ to align with the first, second, and the third peaks of $f_1$, respectively, as shown in the second row. 

To choose a specific peak to align, the first step is to reduce the scale of the covariance operator by setting $\sigma_{sc}=2$, a relatively smaller value. This adjustment aims to restrict the variability of the sampled CLR-transformed warpings and make them pinpointed to a specific peak for alignment. 
We have three options to set the mean function $\mu$: 1) By setting the mean $\mu_h = 2(t-0.5)$, we compel the single-peaked curve $f_2$ to align with the last peak of the triple-peaked curve $f_1$. This is accomplished through a designed warping function that starts flat and becomes steeper later, as illustrated by the green curves in Panel (c). 2) On the other hand, if the mean function is set to $0$, it aligns with the middle peak of $f_1$. Here, the warping function remains very close to the identity warping as depicted in Panel (d), leading to a minimal alignment adjustment. 3) Finally, setting the mean function to $-2(t-0.5)$ targets the first peak of $f_1$. In this scenario, the warping function initially has a steeper profile, and then flattens, as shown in Panel (e). This flexibility in mean and covariance function adjustment provides a powerful tool for precisely controlling the alignment outcome.

\subsection{Comparison with previous methods}

We have conducted a comprehensive study of the new Bayesian framework and emphasized its advantages over the state-of-the-art methods.  In this subsection, we will summarize these comparisons.  We will begin by comparing our framework with the dynamic programming, which is a superior alignment method by maximizing the likelihood term. Following this, we will compare our method with previous Bayesian registration methods.  \\

\noindent {\bf With the Dynamic Programming (DP):}  It is pointed out that if there exists a true warping between the two functions, i.e., $f_2 = f_1 \circ \gamma$, then DP is able to identify this true warping, albeit with discretization error. However, in practical applications, this condition may not always hold. 
\begin{enumerate}
\item DP is a deterministic optimization method that is optimal when considering only the likelihood term.  However, when prior information on time warping is desirable, a Bayesian framework is more appropriate which naturally takes into account such information using a prior model. 
\item The DP method may have computational issues: 1) DP requires dense discretization for accurate results. It may not perform well with sparse discretization, particularly for functions with sharp features. 2) DP has quadratic computational complexity, which can lead to efficiency issues when the number of discrete points is large.  3) The DP procedure lacks smoothness consideration, which may generate undesirable optimal warping. In contrast, the Bayesian method can naturally generate smooth warpings from the prior model, and the computation is linear with respect to the number of discretization points. 
\item DP does not provide variability in its estimates.  In contrast, our MCMC method can generate samples of the posterior, and the variability can be easily addressed.          
\item DP lacks flexibility in cases when multiple alignments are close to be optimal. This limitation can be effectively addressed by using the Bayesian samples.
\end{enumerate}

\noindent  {\bf With other (SRVF-based) Bayesian methods:}  The new Bayesian framework in this paper adopts the same likelihood model as in the previous SRVF-based methods.  The difference is at the prior model (see Figure \ref{fig:srvf} for a visual comparison between our CLR-based prior model and the SVRF-based prior model).   
\begin{enumerate}
\item Our prior is under a linear framework, where there exits an isometric isomorphism between the warping space and conventional $\mathbb{L}^2$ space. This allows us to use the functions in the $\mathbb{L}^2$ space to accurately represent the warping functions, without any approximation.  This makes all computation precise and accurate (e.g., estimation of the mean, $K$-means clustering).
\item 
Due to the one-to-one correspondence between the warping space and the $\mathbb{L}^2$ subspace, our sampling process in the MCMC algorithm is efficient and accurate. In contrast, SRVF-based methods need to take a verification procedure to remove sampled warping functions which do not meet the definition (e.g., not strictly-increasing). This issue becomes more pronounced with drastic warpings and many proposed warpings falling outside $\Gamma_1$ must be discarded.  
\item We have the flexibility to model the prior using various stochastic processes, not limited to just Gaussian processes.  This provides a more powerful representation on the prior and achieve more desirable alignments. 
\end{enumerate}

\noindent \textbf{Remark 3:} In our Bayesian registration, the likelihood term can directly use the observational functions instead of their SRVF functions. Specifically, we can assume that $f_1-f_2\circ \gamma$, i.e., the alignment difference on the original observations, follows a Gaussian process \citep{ramsay1998curve}.   This adjustment addresses the limitation encountered when functional observations are highly noisy and robust SRVF estimations are not available  \citep{tucker2021multimodal}. Some preliminary study is shown in \ref{f_regist}.  

\section{Multiple function registration}
\label{Sec: mul}

We have introduced a new prior in the Bayesian registration for a pair of functions.  In practical applications, we often have multiple functional observations, so that multiple function alignment is of more importance. In this section, we will extend the pairwise registration described in Section \ref{sec: pairwise} to dealing with multiple registrations. To simultaneously register multiple functions $\{f_i\}_{i=1}^n$, we at first compute their SRVFs $\{q_i\}_{i=1}^n$.  Our goal is to find a template $\bar q$, to which all $\{q_i\}_{i=1}^n$ are aligned.  To simplify the modeling procedure, we focus on adopting Gaussian processes as prior models in this paper.  More general non-Gaussian processes can also be explored as in the pairwise case.


For multiple function registration,  we can model each warping function $\gamma_i$ between $q_i$ and $\bar q$ independently using a Gaussian process prior with mean function 0 and the covariance operator $C_{h}$, similar to the procedure in the pairwise case. In addition, the template $\bar{q}$ can also be modeled by assigning a Gaussian process prior with a mean function of 0 and a covariance operator $C_q$. The prior measure is the product measure $\mu_0\equiv (GP(0, C_{h}))^{n}\times GP(0, C_{q})$. Therefore, the posterior measure denoted by $\mu$, is absolutely continuous with respect to the prior with density is given as: 

\begin{eqnarray}
\frac{d \mu}{d \mu_0}(\gamma_{h_{1:n}}, \bar{q}) &\propto& \pi(q_{1:n}|\gamma_{h_{1:n}}, \bar{q}) \nonumber \\
&\propto& \prod_{i=1}^{n}\exp\Big(-\frac{1}{2\sigma^2} \|\bar{q}([t])-(q_i,\gamma_{h_i})([t])\|^2\Big).
\end{eqnarray}  

Similar to Algorithm \ref{alg:pwbayesian}, we can employ a Metropolis-Hastings algorithm to sample from the posterior. In addition, akin to the Fisher-Rao registration, we also need an extra step to centralize of the estimated warping after each iteration.  In each iterative step, assume the current mean template is $\bar q$ and the warpings from $q_i$ to $\bar q$ are $\gamma_i, i = 1, \cdots, n$.  As the estimation is done in the conventional $\mathbb{L}^2$ space, the mean warping of $\{\gamma_i\}_{i=1}^n$ can be simply computed as $\bar{\gamma} = \psi_B^{-1}{(\frac{1}{n}\sum_{1}^{n}\psi_B{(\gamma_i)})}$.  Then, the centered mean template SRVF can be obtained as $(\bar q, {\bar{\gamma}}^{-1})$. Using this new template, we can easily find that the updated warpings from $q_i$ to $\bar q$ is updated to  $\gamma_i\circ {\bar{\gamma}}^{-1}, i = 1, \cdots, n$. By using Lemma \ref{lemma1}, it is straightforward to verify that the mean of $\{\gamma_i\circ {\bar{\gamma}}^{-1}\}_{i=1}^n$ is the identity warping $\gamma_{id}(t) = t$. 

Furthermore, the center of the estimated warping can be generalized to any warping function in $\Gamma_1$ other than the identity. For example, to make the mean of estimated warping be a given $\gamma^* \in \Gamma_1$, we can simply add an additional composition with $\gamma^*$, i.e., the warping $\gamma_i$ is updated to $\gamma_i\circ {\bar{\gamma}}^{-1}\circ\gamma^*, i = 1, \cdots, n$. This can also be verified by using Lemma \ref{lemma1}. In this case, the centered mean template SRVF can be obtained as $(q, {\bar{\gamma}}^{-1}\circ\gamma^*)$. The specific steps are outlined in Algorithm \ref{alg:mulbayesian_G} as follows.

\begin{algorithm}[ht!]
\caption{Multiple Bayesian registration with Gaussian process priors}
\begin{algorithmic} 
	\Require SRVFs $\{q_i\}_{i=1}^n$, pre-given distribution for $\beta_1, \beta_2 \in [0, 1]$, covariance $C_h$ for the warping prior, pre-given center $\gamma^*$, covariance $C_q$ for the template prior 
	\State Random pick $h^{(1)}_i \sim GP(0, C_h), i = 1, \cdots, n$
	\State Calculate $\bar{q}^{(1)} = \frac{1}{n}\sum_{1}^{n}q_i$
	\For{$k = 1: N$}
	\For{$i = 1: n$}
	\State Pick $\beta_1$ from the pre-given distribution 
	\State Propose $h'_{i}= \sqrt{1-\beta_1^2}h_i^{(k)} + \beta_1\xi$, where $\xi\sim GP(0, C_{h})$
	\State Set $h_i^{(k+1)} =\begin{cases}
		h'_i & \text {with probability$\rho = min\Big(1, \frac{\pi(q_i|\gamma_{h_{1:i-1}^{(k+1)}},{\gamma}_{h'_i},\gamma_{i+1:n}^{(k)}, \bar{q}^{(k)})}{\pi(q_i|\gamma_{h_{1:i-1}^{(k+1)}},{\gamma}_{h_i^{(k)}},\gamma_{i+1:n}^{(k)}, \bar{q}^{(k)})}\Big)$} \\
		h_i^{(k)} & \text {with probability $1 -\rho$} 
	\end{cases}$
	\EndFor
	\State $\bar{\gamma} = \psi_B^{-1}(\frac{1}{n}\sum_{1}^{n}h_i^{(k+1)})$
	\State $\gamma_{1:n}^{(k+1)} \leftarrow \gamma_{1:n}^{(k+1)}\circ\bar{\gamma}^{-1}\circ\gamma^*$ 
	\State $\bar{q}^{(k)} \leftarrow (\bar{q}^{(k)}, \bar{\gamma}^{-1}\circ\gamma^*)$
	\State Pick $\beta_2$ from the pre-given distribution 
	\State Propose $\bar{q}'= \sqrt{1-\beta_2^2}\bar{q}^{(k)} + \beta_2 \zeta, \zeta\sim GP(0, C_{q})$
	\State Set $\bar{q}^{(k+1)}=\begin{cases}
		\bar{q}' & \text {with probability$\rho = min\Big(1, \frac{\pi(q_{1:n}|\gamma_{h1:n}^{(k+1)}, \bar{q}')}{\pi(q_{1:n}|\gamma_{h1:n}^{(k+1)}, \bar{q}^{(k)})}\Big)$} \\
		\bar{q}^{(k)} & \text {with probability $1 -\rho$} 
	\end{cases}$
	\EndFor
	\State Output $\{\gamma_{1:n}^{(k)}\}_{k=2}^{N+1}$ and $\{\bar q^{(k)}\}_{k=2}^{N+1}$  
\end{algorithmic}
\label{alg:mulbayesian_G}
\end{algorithm}

\begin{figure}[ht!]
\centering
\begin{subfigure}[h]{0.24\textwidth}
	\includegraphics[width=\textwidth]{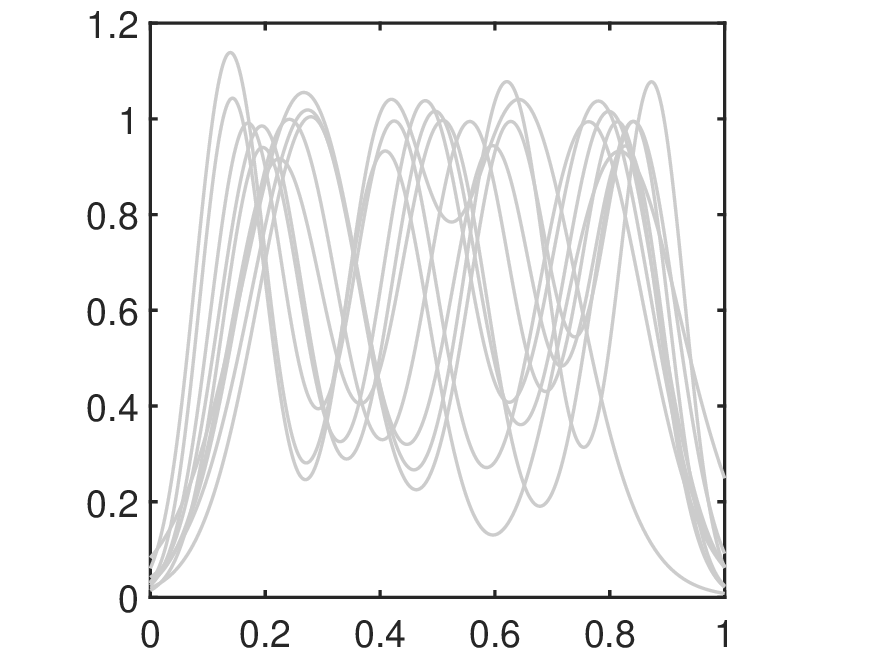}
	\caption{observations}
\end{subfigure}\hspace{-0.5cm}
\begin{subfigure}[h]{0.24\textwidth}
	\includegraphics[width=\textwidth]{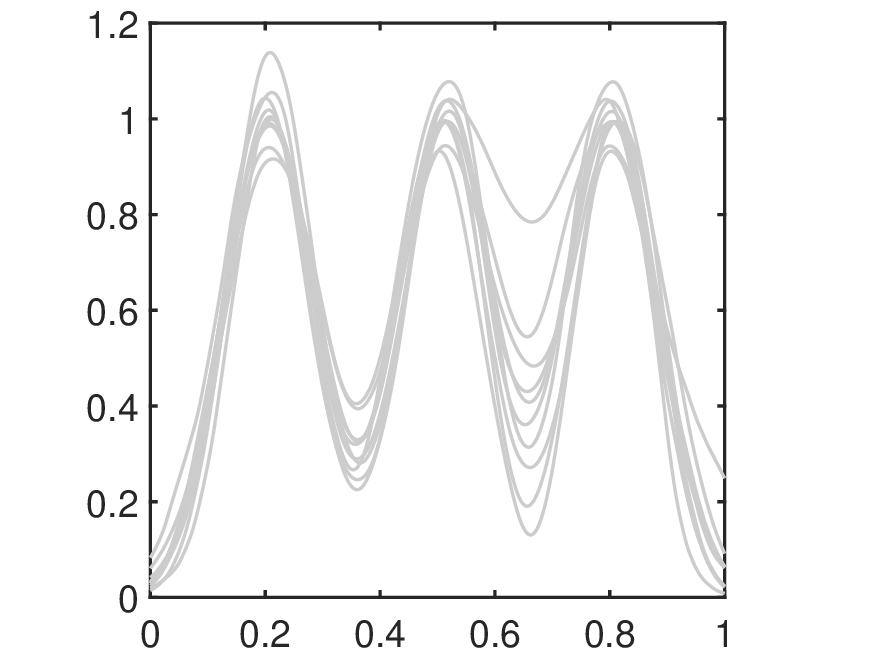}
	\caption{alignment}
\end{subfigure}\hspace{-0.5cm}
\begin{subfigure}[h]{0.24\textwidth}
	\includegraphics[width=\textwidth]{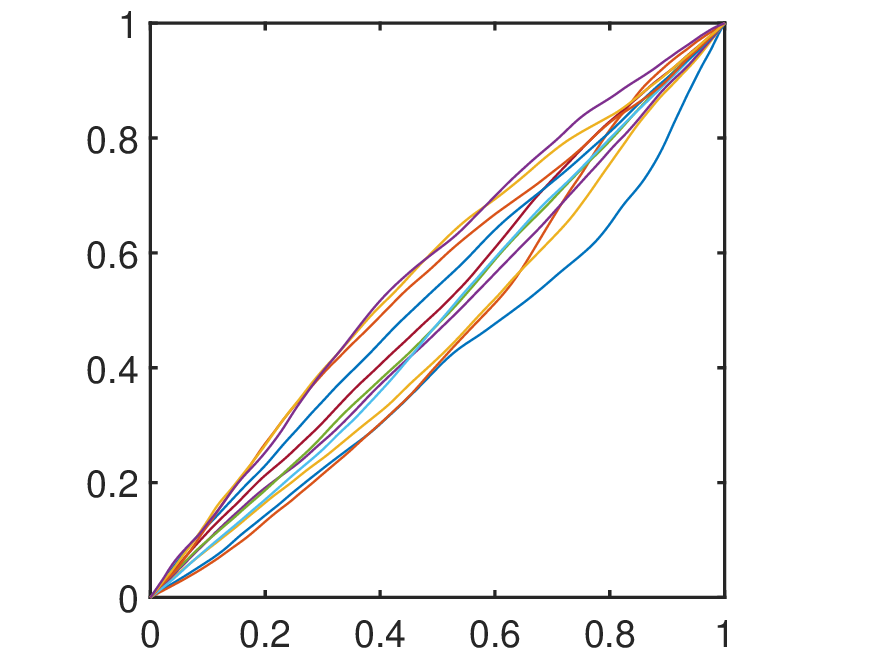}
	\caption{warping}
\end{subfigure}
\vfill
\begin{subfigure}[h]{0.24\textwidth}
	\includegraphics[width=\textwidth]{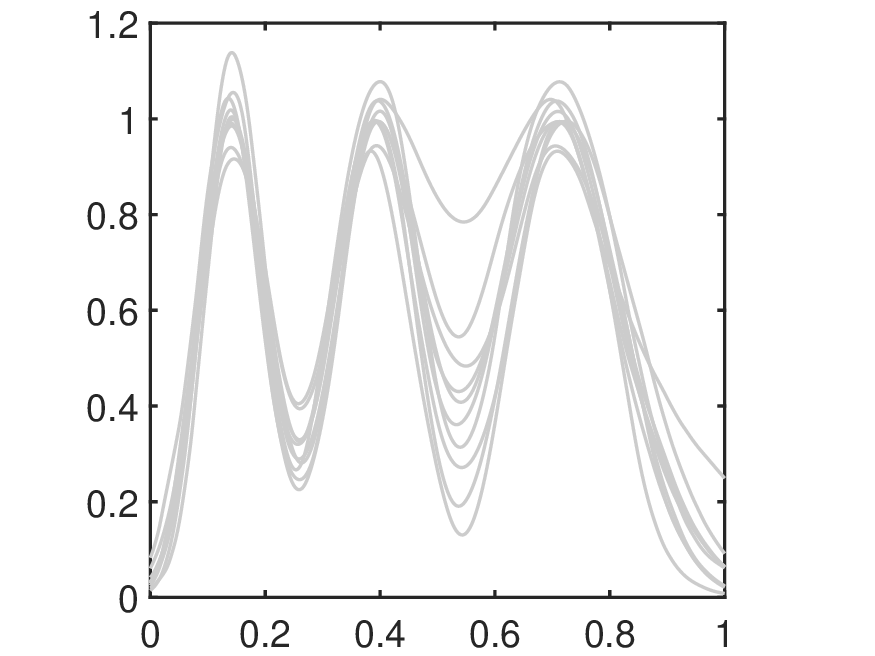}
	\captionsetup{justification=centering}
	\caption{alignment with \\$h_{\gamma^*}=0.5-t$}
\end{subfigure}\hspace{-0.5cm}
\begin{subfigure}[h]{0.24\textwidth}
	\includegraphics[width=\textwidth]{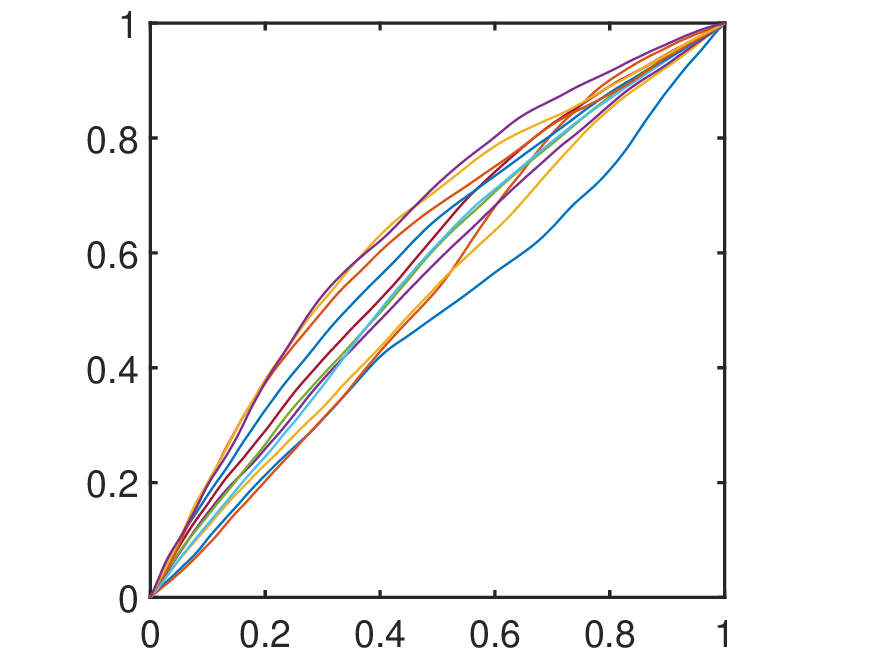}
	\captionsetup{justification=centering}
	\caption{warping with \\ $h_{\gamma^*}=0.5-t$}
\end{subfigure}\hspace{-0.5cm}
\begin{subfigure}[h]{0.24\textwidth}
	\includegraphics[width=\textwidth]{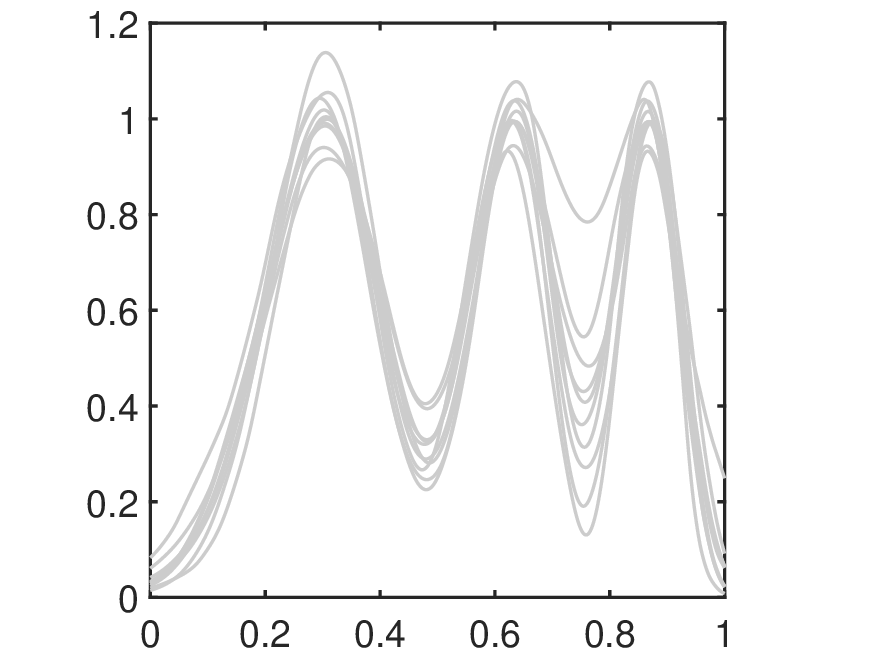}
	\captionsetup{justification=centering}
	\caption{alignment with \\$h_{\gamma^*}=t-0.5$}
\end{subfigure}\hspace{-0.5cm}
\begin{subfigure}[h]{0.24\textwidth}
	\includegraphics[width=\textwidth]{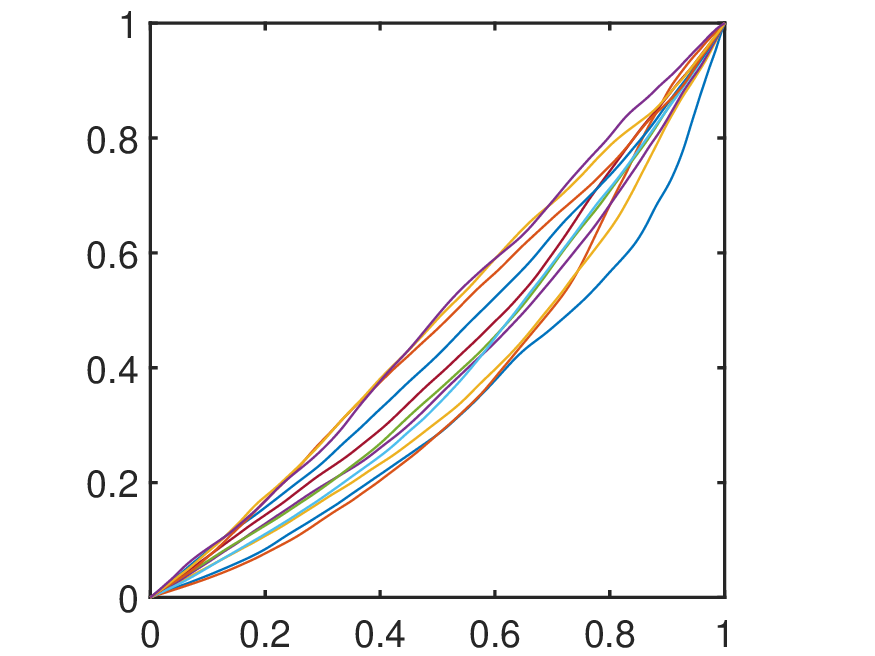}
	\captionsetup{justification=centering}
	\caption{warping with \\$h_{\gamma^*}=t-0.5$}
\end{subfigure}
\caption{Registration results on Multiple function alignment}
\label{fig:mul_align}
\end{figure}

\textbf{Simulation 7}:  We here use one simulation to illustrate Algorithm \ref{alg:mulbayesian_G}. The individual functions are given by $y_i(t) = z_{i,1}e^{-(t-c_{i,1})^2/2} + z_{i,2}e^{-(t-c_{i,2})^2/2} + z_{i,3}e^{-(t-c_{i,3})^2/2}, i=1, 2, \cdots, 11$, where $z_{i,1}$,$z_{i,2}$ and $z_{i,3}$ are $i.i.d$ normal with mean one and standard deviation 0.05, $c_{i,1}$,$c_{i,2}$ and $c_{i,3}$ are $i.i.d$ normal with standard deviation 0.02, and mean 0.2, 0.5, and 0.8, respectively. Each of these functions then warped according to: $\gamma_i(t) = \frac{e^{a_it}-1}{e^{a_i}-1}$ for $a_i\neq 0$, otherwise, $\gamma_i = \gamma_{id}$, where $a_i = \frac{i-11}{10}$, and the observed functions are computed using $f_i(t) = y_i(\gamma_i(t)), i=1, 2, \cdots, 11$. These 11 functions form the original data and are shown in Panel (a) of Figure \ref{fig:mul_align}.   At first, we set the center $\gamma^*$ as the identity warping and use Algorithm \ref{alg:mulbayesian_G} to obtain a set of sampled warpings for each function observation. We then take the mean of these sampled warpings to perform the function registration, resulting in the alignment shown in Panel (b) for the aligned functions and in Panel (c) for the corresponding warping functions. In contrast to the approach in \cite{lu2017bayesian}, our method simplifies the computation of the warping mean and centers the orbit effortlessly. 

Similar to the pairwise registration, we possess the flexibility to choose the center at any desired location. For example, setting $\gamma^* = \psi_B^{-1}(h_{\gamma^*})$ with $h_{\gamma^*}(t) = 0.5-t$ yields warping functions that exhibit an overall convex pattern, resulting in aligned functions with each peak occurring earlier than its corresponding peak in Panel (b), as shown in Panels (d) and (e). Conversely, setting $\gamma^* = \psi_B^{-1}(h_{\gamma^*})$ with $h_{\gamma^*}(t) = t-0.5$ produces concave warping functions, leading to aligned functions with each peak occurring later than its corresponding peak in Panel (b), as displayed in Panels (f) and (g). 

\begin{figure}[ht!]
\centering
\begin{subfigure}[h]{0.22\textwidth}
	\includegraphics[width=\textwidth]{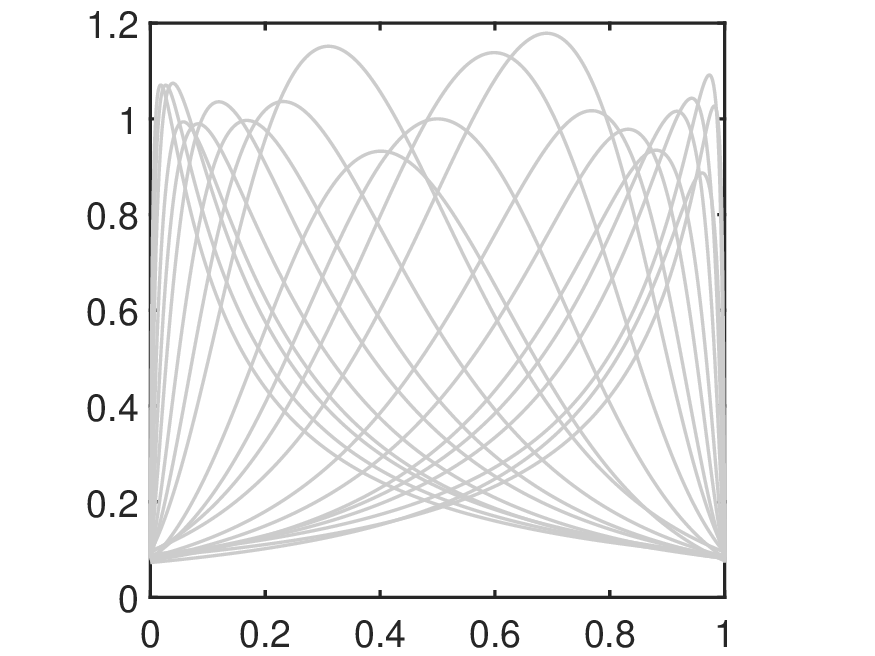}
	\caption{observations}
\end{subfigure}\hspace{-0.5cm}
\begin{subfigure}[h]{0.22\textwidth}
	\includegraphics[width=\textwidth]{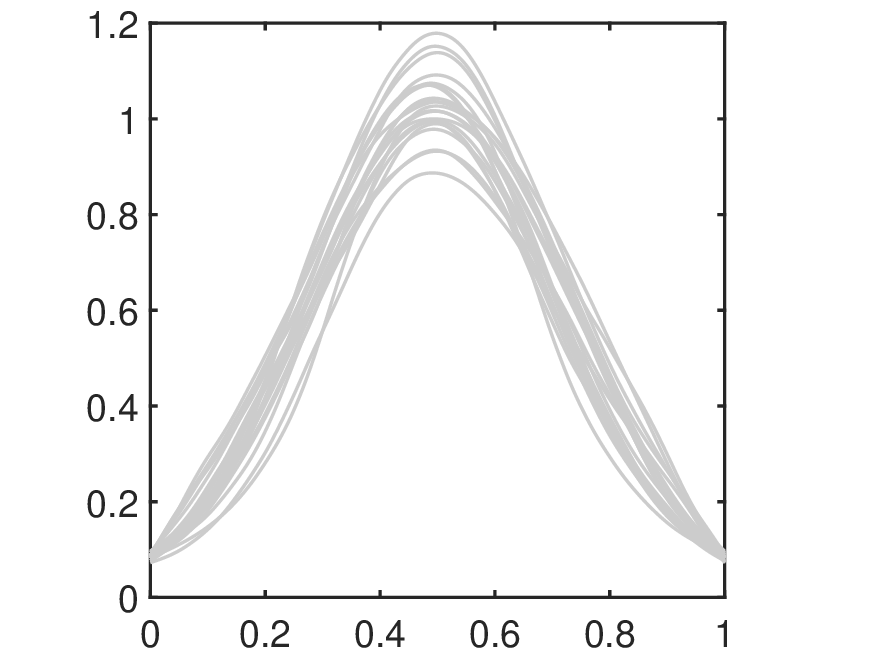}
	\caption{clr alignment}
\end{subfigure}\hspace{-0.5cm}
\begin{subfigure}[h]{0.22\textwidth}
	\includegraphics[width=\textwidth]{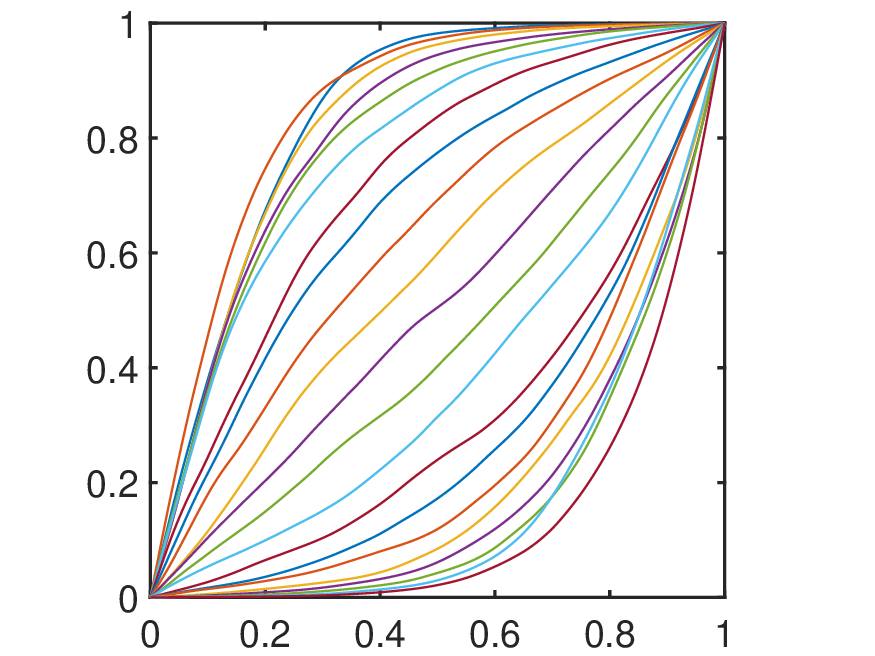}
	\caption{clr warping}
\end{subfigure}\hspace{-0.5cm}
\begin{subfigure}[h]{0.22\textwidth}
	\includegraphics[width=\textwidth]{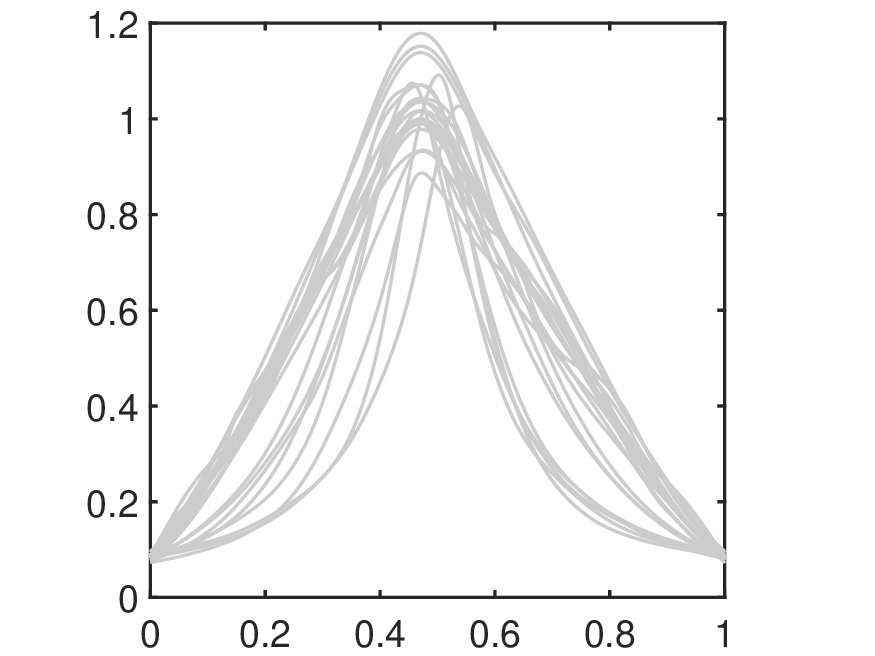}
	\caption{srvf alignment}
\end{subfigure}\hspace{-0.5cm}
\begin{subfigure}[h]{0.22\textwidth}
	\includegraphics[width=\textwidth]{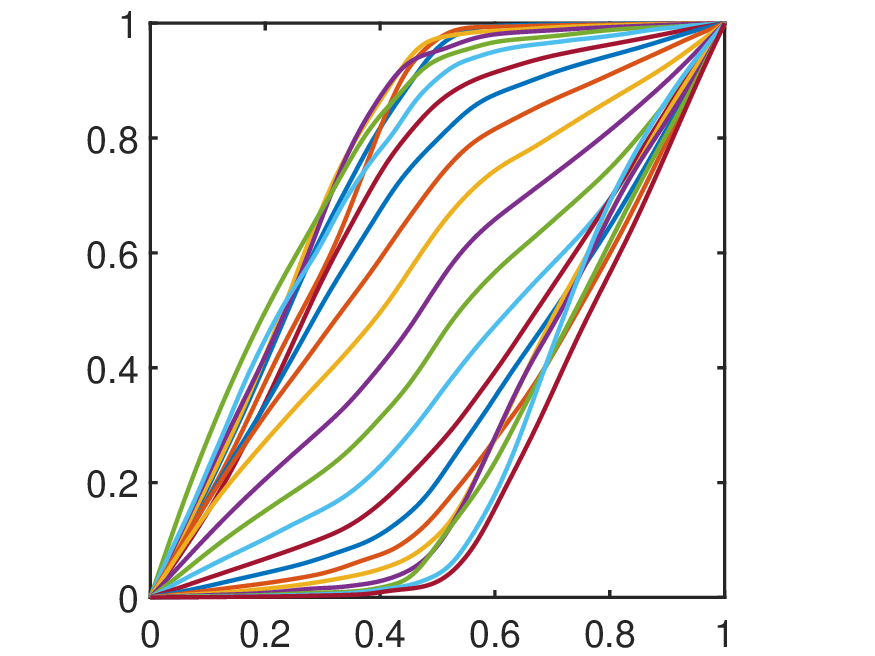}
	\caption{srvf warping}
\end{subfigure}
\caption{Comparison of registration results on Multiple function alignment}
\label{fig:mul_align2}
\end{figure}

\begin {table}[ht!]
\centering
\begin{tabular}{c c c c c c c c c c c c c c c c c||} 
\hline
& $f_1$ & $f_3$ &  $f_5$ & $f_7$ & $f_9$ & $f_{11}$ & $f_{13}$ & $f_{15}$ & $f_{17}$ & $f_{19}$ & $f_{21}$ \\ [0.5ex] 
\hline\hline
No. discards& 1301 & 1010 & 872 & 777 & 694 & 557 & 614 & 655  & 764 & 1007 & 1193 \\
\hline
\end{tabular}
\caption{Number of discarded proposals in Multiple function alignment}
\label{tab:mul_align}
\end {table}

\textbf{Simulation 8}:  In this simulation, we compare Bayesian registration results with warping proposed using CLR-framework and SRVF-framework. We at first simulation a function $g(t) = e^{-10(t-0.5)^2}, t\in[0, 1]$. This function is then warped according to: $\gamma_i(t) = \frac{e^{a_it}-1}{e^{a_i}-1}$ for $a_i\neq 0$, otherwise, $\gamma_i = \gamma_{id}$, where $a_i = \frac{4(i-11)}{5}$, and the observed functions are computed using $f_i(t) = g(\gamma_i(t)), i=1, 2, \cdots, 21$. These 21 functions form the original data and are shown in Panel (a) of Figure \ref{fig:mul_align2}.   First, we use Algorithm \ref{alg:mulbayesian_G} to obtain a set of sampled warpings for each function observation. We then take the mean of these sampled warpings to perform the function registration, resulting in the alignment shown in Panel (b) for the aligned functions and in Panel (c) for the corresponding warping functions. Next, we use the SVRF-based method introduced in \cite{lu2017bayesian}, the aligned functions and the corresponding warpings are displayed in Panel (d) and (e), respectively. To ensure fairness, we set the same number of sampling rounds, and all other parameters are kept the same. 

It can be observed that both methods yield good alignment results when the warping is moderate. However, when the warping becomes drastic, the SRVF-based method cannot achieve desired alignment.  This issue arises because the SRVF-based method necessitates the removal of sampled warpings that fall outside. With a limited number of iterations, it becomes difficult to sample extreme warpings, ultimately hindering the acquisition of the desired posterior. This limitation is inherent to the SRVF framework and not easily resolved.  Additionally,  another drawback of the SRVF-based method is the discard of out-of-bound samples, which decreases the efficiency of the MCMC sampling. We set the number of iteration to be 2000, ensuring that we get 2000 valid warping proposals for each function observation. Under the CLR framework, only 2000 iterations are required for each observation. However, that is not the case with the SRVF-based method. To illustrate this inefficiency, we document the number of discards in the SRVF-method in Table \ref{tab:mul_align}. For example, to obtain 2000 valid warping proposals for $f_1$, we had to propose 3301 times, whereas for $f_{11}$, we had to propose 2557 times to achieve  2000 valid warping proposals. Notably, $f_1$ involves a more drastic warping compared to the $f_{11}$. It clearly indicates that the number of required sampling attempts increases as the warping becomes more drastic.

\section{Real data application}
\label{Sec:realdata}
\begin{figure}[ht!]
\centering
\begin{subfigure}[h]{0.3\textwidth}
	\includegraphics[width=\textwidth]{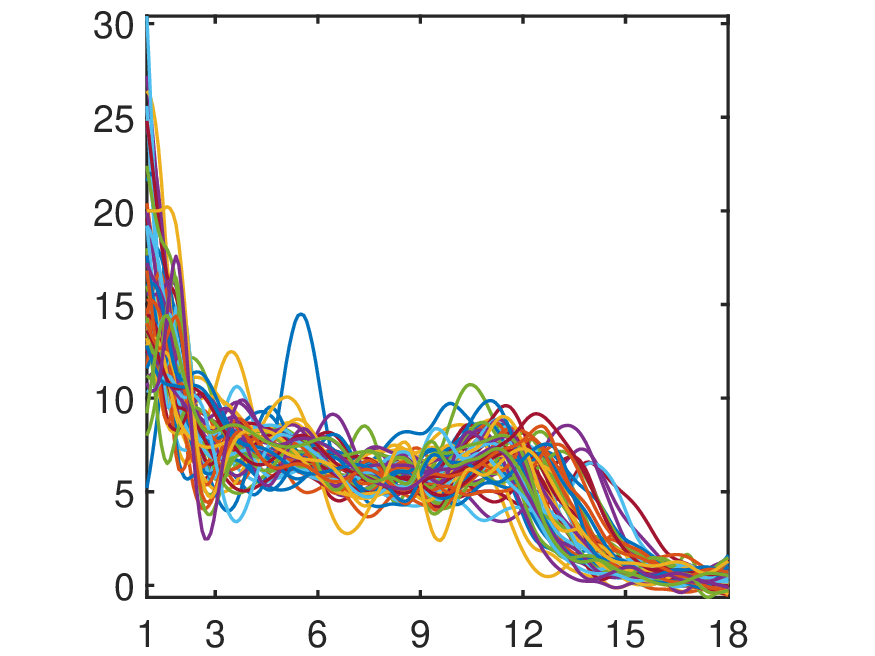}
	\caption{obs}
\end{subfigure}\hspace{-0.4cm}
\begin{subfigure}[h]{0.3\textwidth}
	\includegraphics[width=\textwidth]{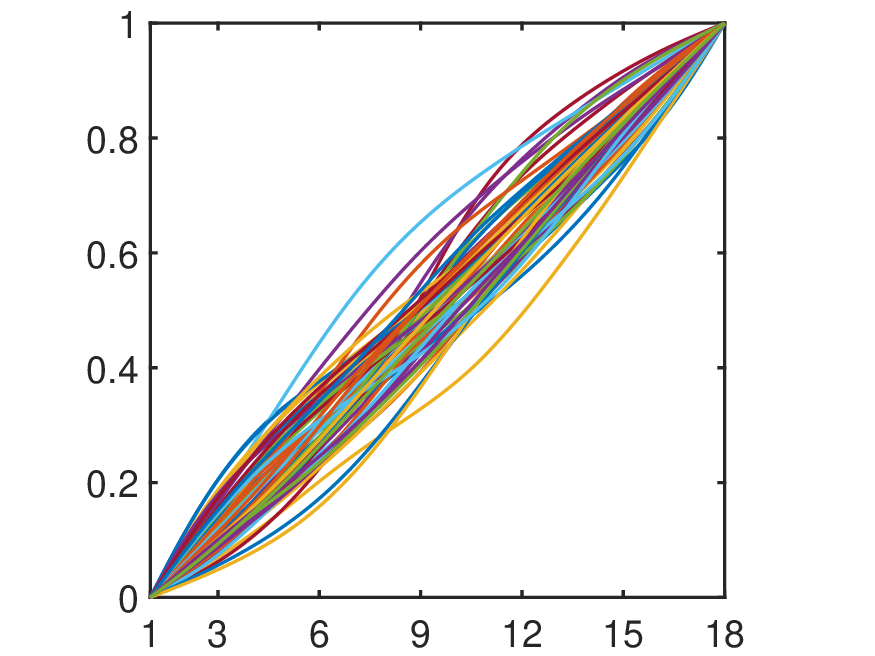}
	\caption{iso warping}
\end{subfigure}\hspace{-0.4cm}
\begin{subfigure}[h]{0.3\textwidth}
	\includegraphics[width=\textwidth]{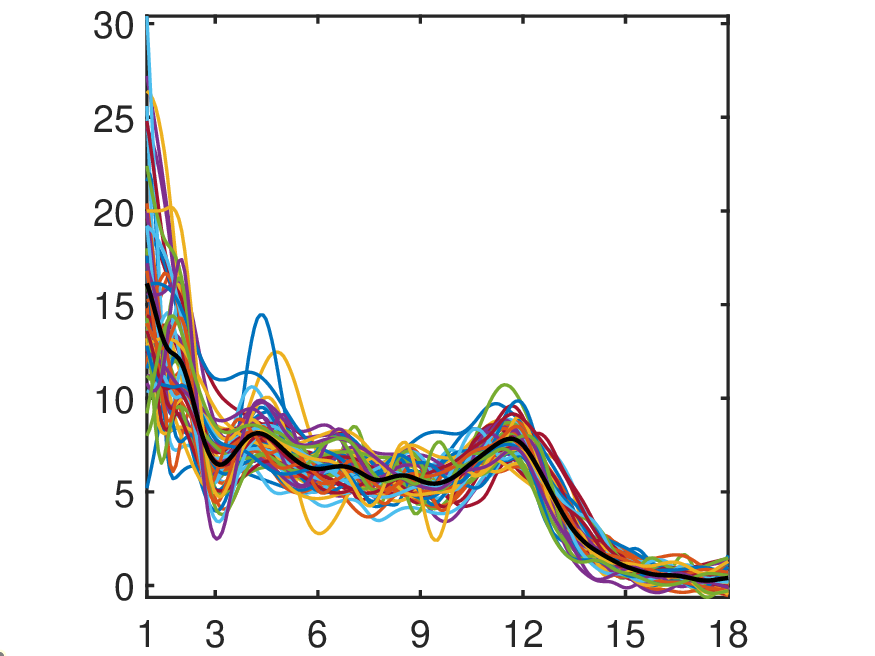}
	\caption{iso alignment}
\end{subfigure}
\begin{subfigure}[h]{0.3\textwidth}
	\includegraphics[width=\textwidth]{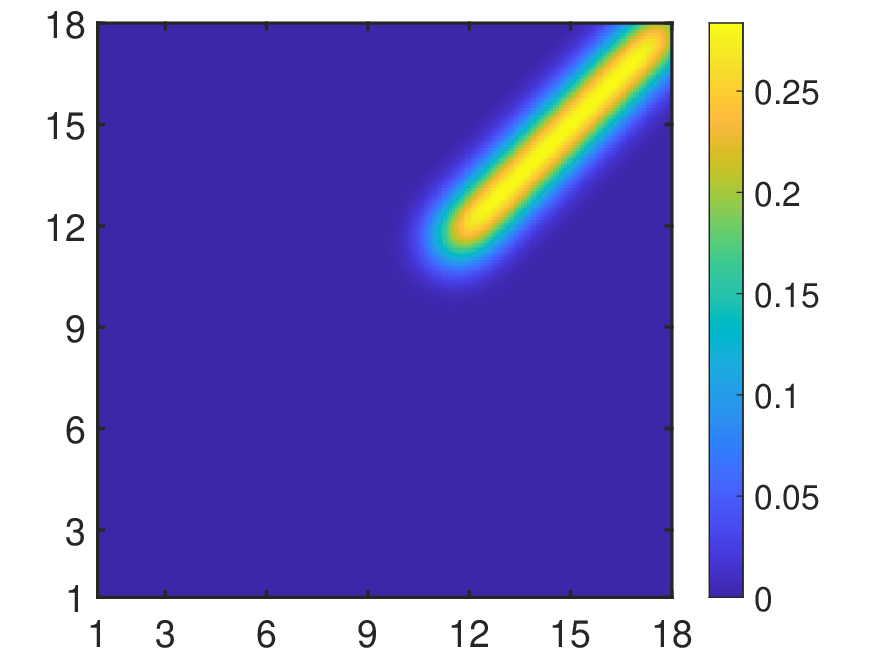}
	\caption{$C_h$}
\end{subfigure}\hspace{-0.4cm}
\begin{subfigure}[h]{0.3\textwidth}
	\includegraphics[width=\textwidth]{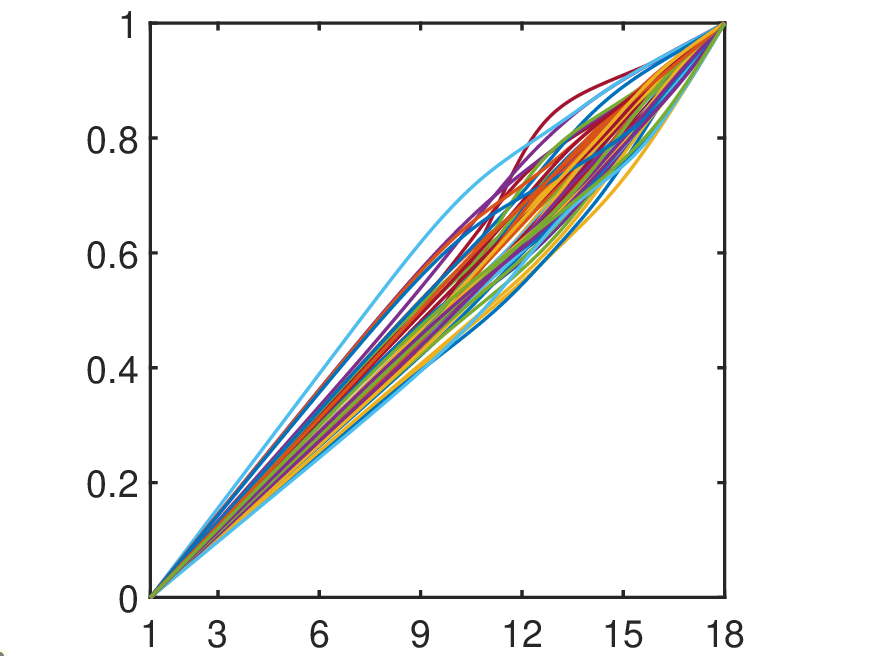}
	\caption{diag warping}
\end{subfigure}\hspace{-0.4cm}
\begin{subfigure}[h]{0.3\textwidth}
	\includegraphics[width=\textwidth]{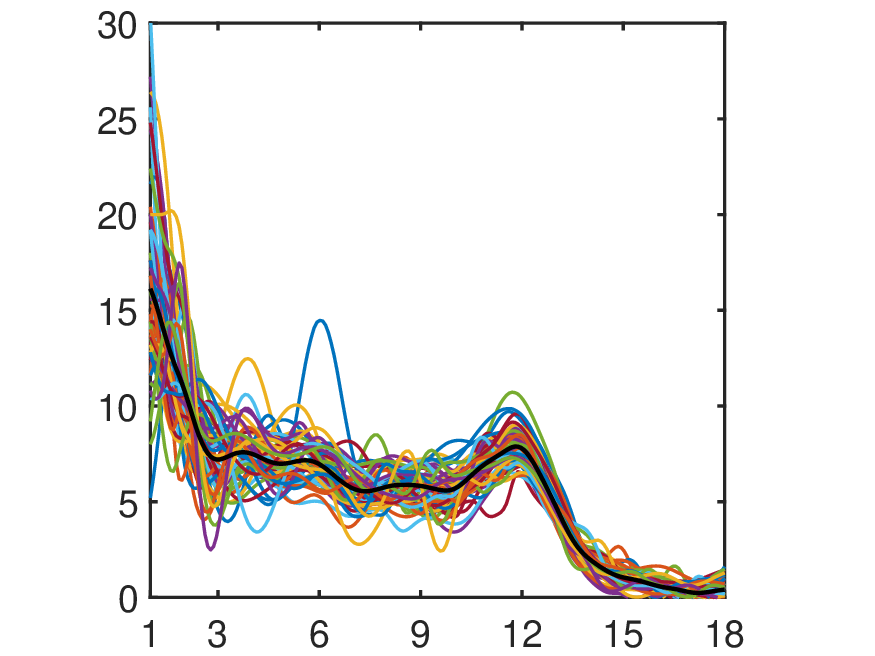}
	\caption{diag alignment}
\end{subfigure}
\caption{Registration results on the Berkeley growth data for the female group. (a) original 54 observed growth rate functions, (b), the mean of sampled warping functions for each observations (c) the registered growth rate functions, with the corresponding mean shown in bold black. (d) the heatmap of $C_h$, (e)-(f) same as (b)-(c) except using covariance $C_h$ in (d) for prior warping. }
\label{fig:female}
\end{figure}

In this section, we apply our method to the well-known Berkeley Growth curve data (available at https://rdrr.io/cran/fda/man/growth.html), which records the heights of 54 girls at 31 time points from ages 1 to 18. We are interested in comparing the growth rates, which is the first derivatives of the growth curves, as they offer valuable insights into the number and timing of growth spurts during childhood and adolescence. To achieve this, we utilize the proposed Bayesian registration with a Gaussian process prior, initially setting the covariance of the Gaussian process prior to resemble an isotropic-like structure. Figure \ref{fig:female} presents the outcomes of our registration analysis. In particular, Panel (a) displays the initial, unaligned growth curves. Panels (b) and (c) depict the means of sampled warping functions and corresponding aligned functions, respectively. Through our analysis, we obtain a clearer perspective on the growth patterns compared to the original, unaligned observations. Initially, after birth, the growth rate begins to decline, stabilizing around 3 years old. This is followed by a minor growth spurt, which, however, is short-lived, and then the rates stabilize around 5. Then, another  growth spurt occurs around 12 years old, which is apparently significant and actually often how the term ``growth spurt'' is referred to. 

Because the first, minor growth spurt does not appear to be particularly important, we can adjust the covariates to focus solely on aligning the second, significant one.  We modify the covariates of the Gaussian prior as illustrated in Panel (d). The resulting means of sampled warping functions and the corresponding aligned functions are displayed in Panels (e) and (f), respectively. 
The resulting warpings now closely resemble the identity warping in the first half of the time domain, causing the corresponding aligned functions to remain very similar to the original, unaligned observations. Despite this, the second half of the curves is still well aligned, as evidenced by the clear discernibility of the second growth spurts.  A key aspect of our approach is the restriction of the warping flexibility in the initial segment of the growth curve. By setting a lower variance in this early phase, we achieve an optimal alignment that closely resembles identity warping. This strategy allows us to focus on the second growth rate sprout, which is typically more significant, while effectively minimizing the influence of noise present in the initial part of the curve. Overall, our method provides a nuanced understanding of the growth patterns and their variations.

\section{Discussion}
\label{Sec: disc}
In this paper, we have introduced a new prior based on the CLR-transformed representations of warping functions for Bayesian registration.  We conduct a comprehensive exploration to examine the effects of different priors, including Gaussian prior with various means and covariances, as well as non-Gaussian priors. We demonstrate how the choice of prior can effectively control alignment.  Additionally, we extend our approach to multiple function registration and compare it with the SRVF-based method, highlighting the advantages of our CLR-based method. The main benefits of our new prior method over the previous nonlinear approximation methods include a clearer understanding of the prior model and more efficient sampling proposal.  This is largely because the warping space and conventional $\mathbb{L}^2$ subspace is isometrically isomorphic. Moreover, centering of the orbit is also much more efficient. Finally, we apply our framework to a real-world dataset, the Berkeley growth data, to demonstrate its effectiveness.

There are few directions for future work. Firstly, the primary strength of Bayesian registration lies in its capacity to integrate various warping priors to explore potential registrations.  We have provided several simulated and real data examples to emphasize the impacts of different warpings.  However, our understanding of the effects of the Gaussian process prior, particularly how the covariance between two different time points of the warping prior influences the alignment result, is still incomplete. Consequently, identifying the optimal prior in real application remains a challenging and critically important problem that requires further research. Secondly, our selection of covariance in this manuscript is primarily based on visual calibration, applied on a case-by-case basis, and does not offer a universal strategy. Future work could focus on automating this process and potentially incorporating it into the Bayesian framework to develop a more generalized approach to selecting covariance. Finally, this paper continues to employ the SRVF-framework within the context of likelihood terms. However, we suggest the potential of using the original functions to derive the likelihood terms, which could address the issue of the noisy observations. This possibility is briefly mentioned in the Appendix, but it presents promising avenue for future exploration.

\bibliographystyle{plainnat}  
\bibliography{references}  

\begin{thebibliography}{19}
\providecommand{\natexlab}[1]{#1}
\providecommand{\url}[1]{\texttt{#1}}
\expandafter\ifx\csname urlstyle\endcsname\relax
  \providecommand{\doi}[1]{doi: #1}\else
  \providecommand{\doi}{doi: \begingroup \urlstyle{rm}\Url}\fi

\bibitem[Beskos et~al.(2011)Beskos, Pinski, Sanz-Serna, and
  Stuart]{beskos2011hybrid}
Alexandros Beskos, Frank~J Pinski, Jes{\'u}s~Mar{\i}a Sanz-Serna, and Andrew~M
  Stuart.
\newblock Hybrid monte carlo on hilbert spaces.
\newblock \emph{Stochastic Processes and their Applications}, 121\penalty0
  (10):\penalty0 2201--2230, 2011.

\bibitem[Cheng et~al.(2014)Cheng, Dryden, Hitchcock, and Le]{cheng2014analysis}
Wen Cheng, Ian~L Dryden, David~B Hitchcock, and Huiling Le.
\newblock Analysis of proteomics data: Bayesian alignment of functions.
\newblock \emph{Electronic Journal of Statistics}, 8\penalty0 (2):\penalty0
  1734--1741, 2014.

\bibitem[Cheng et~al.(2016)Cheng, Dryden, and Huang]{cheng2016bayesian}
Wen Cheng, Ian~L Dryden, and Xianzheng Huang.
\newblock Bayesian registration of functions and curves.
\newblock \emph{Bayesian Analysis}, 11\penalty0 (2):\penalty0 447--475, 2016.

\bibitem[Cotter et~al.(2013)Cotter, Roberts, Stuart, and White]{cotter2013mcmc}
Simon~L Cotter, Gareth~O Roberts, Andrew~M Stuart, and David White.
\newblock Mcmc methods for functions: modifying old algorithms to make them
  faster.
\newblock 2013.

\bibitem[Egozcue et~al.(2006)Egozcue, D{\'\i}az-Barrero, and
  Pawlowsky-Glahn]{egozcue2006hilbert}
Juan~Jos{\'e} Egozcue, Jos{\'e}~Luis D{\'\i}az-Barrero, and Vera
  Pawlowsky-Glahn.
\newblock Hilbert space of probability density functions based on aitchison
  geometry.
\newblock \emph{Acta Mathematica Sinica}, 22\penalty0 (4):\penalty0 1175--1182,
  2006.

\bibitem[Eilers(2004)]{eilers2004parametric}
Paul~HC Eilers.
\newblock Parametric time warping.
\newblock \emph{Analytical chemistry}, 76\penalty0 (2):\penalty0 404--411,
  2004.

\bibitem[Gervini and Gasser(2004)]{gervini2004self}
Daniel Gervini and Theo Gasser.
\newblock Self-modelling warping functions.
\newblock \emph{Journal of the Royal Statistical Society: Series B (Statistical
  Methodology)}, 66\penalty0 (4):\penalty0 959--971, 2004.

\bibitem[Happ et~al.(2019)Happ, Scheipl, Gabriel, and Greven]{happ2019general}
Clara Happ, Fabian Scheipl, Alice-Agnes Gabriel, and Sonja Greven.
\newblock A general framework for multivariate functional principal component
  analysis of amplitude and phase variation.
\newblock \emph{Stat}, 8\penalty0 (1):\penalty0 e220, 2019.

\bibitem[James(2007)]{james2007curve}
Gareth~M James.
\newblock Curve alignment by moments.
\newblock \emph{The Annals of Applied Statistics}, 1\penalty0 (2):\penalty0
  480--501, 2007.

\bibitem[Kurtek(2017)]{kurtek2017geometric}
Sebastian Kurtek.
\newblock A geometric approach to pairwise bayesian alignment of functional
  data using importance sampling.
\newblock \emph{Electronic Journal of Statistics}, 11\penalty0 (1):\penalty0
  502--531, 2017.

\bibitem[Lu et~al.(2017)Lu, Herbei, and Kurtek]{lu2017bayesian}
Yi~Lu, Radu Herbei, and Sebastian Kurtek.
\newblock Bayesian registration of functions with a gaussian process prior.
\newblock \emph{Journal of Computational and Graphical Statistics}, 26\penalty0
  (4):\penalty0 894--904, 2017.

\bibitem[Ma et~al.(2024)Ma, Zhou, and Wu]{ma2024stochastic}
Yijia Ma, Xinyu Zhou, and Wei Wu.
\newblock A stochastic process representation for time warping functions.
\newblock \emph{Computational Statistics \& Data Analysis}, page 107941, 2024.

\bibitem[Matuk et~al.(2021)Matuk, Bharath, Chkrebtii, and
  Kurtek]{matuk2021bayesian}
James Matuk, Karthik Bharath, Oksana Chkrebtii, and Sebastian Kurtek.
\newblock Bayesian framework for simultaneous registration and estimation of
  noisy, sparse, and fragmented functional data.
\newblock \emph{Journal of the American Statistical Association}, pages 1--17,
  2021.

\bibitem[Neal et~al.(2011)]{neal2011mcmc}
Radford~M Neal et~al.
\newblock Mcmc using hamiltonian dynamics.
\newblock \emph{Handbook of markov chain monte carlo}, 2\penalty0
  (11):\penalty0 2, 2011.

\bibitem[Ramsay and Li(1998)]{ramsay1998curve}
James~O Ramsay and Xiaochun Li.
\newblock Curve registration.
\newblock \emph{Journal of the Royal Statistical Society: Series B (Statistical
  Methodology)}, 60\penalty0 (2):\penalty0 351--363, 1998.

\bibitem[Srivastava and Klassen(2016)]{srivastava2016functional}
Anuj Srivastava and Eric~P Klassen.
\newblock \emph{Functional and shape data analysis}, volume~1.
\newblock Springer, 2016.

\bibitem[Srivastava et~al.(2011)Srivastava, Wu, Kurtek, Klassen, and
  Marron]{srivastava2011registration}
Anuj Srivastava, Wei Wu, Sebastian Kurtek, Eric Klassen, and James~Stephen
  Marron.
\newblock Registration of functional data using fisher-rao metric.
\newblock \emph{arXiv preprint arXiv:1103.3817}, 2011.

\bibitem[Tucker et~al.(2021)Tucker, Shand, and Chowdhary]{tucker2021multimodal}
J~Derek Tucker, Lyndsay Shand, and Kenny Chowdhary.
\newblock Multimodal bayesian registration of noisy functions using hamiltonian
  monte carlo.
\newblock \emph{Computational Statistics \& Data Analysis}, 163:\penalty0
  107298, 2021.

\bibitem[Wu and Srivastava(2014)]{wu2014analysis}
Wei Wu and Anuj Srivastava.
\newblock Analysis of spike train data: Alignment and comparisons using the
  extended fisher-rao metric.
\newblock \emph{Electronic Journal of Statistics}, 8\penalty0 (2):\penalty0
  1776--1785, 2014.

\end{thebibliography}

\newpage
\newpage 
\appendix

\section{Proof of Lemma \ref{lemma1}}
\label{app:lemma}
 By the CLR-transformation, 
\begin{eqnarray*}
	& &\psi_B(\gamma_i \circ \gamma_0)(t) \\
	&=& \log(\dot{\gamma_i \circ \gamma_0})(t) - \int_0^1 \log(\dot{\gamma_i \circ \gamma_0})(s) ds \\
	&=& \log(\dot{\gamma_i} (\gamma_0(t)) +  \log(\dot{\gamma_0}(t)) 
	- \int_0^1 \log(\dot{\gamma_i} (\gamma_0(s)) ds 
	- \int_0^1 \log(\dot{\gamma_0}(s)) ds \\
	&=& 	 \psi_B(\gamma_i)(\gamma_0(t)) + \psi_B(\gamma_0)(t) - \int_0^1 \log(\dot{\gamma_i} (\gamma_0(s)) ds  + \int_0^1 \log(\dot{\gamma_i}(s)) ds 
\end{eqnarray*}

Taking expectation on both sides, we have 
\begin{eqnarray*}
	& &\mathbb E(\psi_B(\gamma_i \circ \gamma_0)(t)) \\
	&=& 	\mathbb E( \psi_B(\gamma_i)(\gamma_0(t))) + \psi_B(\gamma_0)(t) - \mathbb E(\int_0^1 \log(\dot{\gamma_i} (\gamma_0(s)) ds  - \int_0^1 \log(\dot{\gamma_i}(s)) ds)  \\
	&=& \psi_B(\gamma^*)(\gamma_0(t)) + \psi_B(\gamma_0)(t) - \int_0^1 \mathbb E(\log(\dot{\gamma_i} (\gamma_0(s)))  - \log(\dot{\gamma_i}(s))) ds  \\
	&=& \psi_B(\gamma^* \circ \gamma_0)(t) + C
\end{eqnarray*}
where $C = \int_0^1 [\log(\dot{\gamma^*} (\gamma_0(s)))  - \log(\dot{\gamma^*}(s))  - \mathbb E(\log(\dot{\gamma_i} (\gamma_0(s)))  - \log(\dot{\gamma_i}(s)))] ds$. \\

We then take integration on both sides and have 
\begin{eqnarray*}
	0 &=& \int_0^1 \mathbb E(\psi_B(\gamma_i \circ \gamma_0)(t)) dt = \int_0^1 \psi_B(\gamma^* \circ \gamma_0)(t)dt  + C = 0 + C
\end{eqnarray*}

Therefore, $C = 0$ and 
$$ \mathbb E (\psi_B(\gamma_i \circ \gamma_0)) =  \psi_B(\gamma^* \circ \gamma_0).   $$  

\section{Proof of Proposition \ref{prop}}
\label{app:prop}
{\bf Proof:}. Let $q_i$ denote the SRVF of $f_i, i = 1, \cdots, n$ and $q_g$ denote the SRVF of $g$. Then $q_i = \sqrt{c_i} (q_g, \gamma_i)$. 
Assume the initial estimate is given in the form $q_0 =\sqrt{c_0} (q_g, \gamma_0)$ (e. g. we can take $q_0 = q_1$).  Because
$$\| q_0 - (q_i, \gamma)\| = \| \sqrt{c_0}(q_g,\gamma_0) - \sqrt{c_i}(q_g, \gamma_i \circ \gamma)\|,$$ 
the optimal warping from $q_i$ to $q_0$ is $\gamma_i^* = \gamma_i^{-1} \circ \gamma_0$.  By Lemma 1, as $\mathbb E(\psi_B(\gamma_i^{-1})) = 0$, we have $\mathbb E(\psi_B(\gamma_i^*)) = \psi_B(\gamma_0)$. Therefore, the sample mean $\psi_B(\hat \gamma_0) = \frac{1}{n}\sum_{i=1}^n \psi_B(\gamma_i^*)$ is a consistent estimator of 
$\psi_B(\gamma_0)$.   That is, 
$$\hat \gamma_0 \rightarrow \gamma_0 \ \ (a.s.) \ \ \ (n \rightarrow \infty) $$

Then, we can apply the estimated warpings on the observations $f_i$.  That is, 
\begin{eqnarray*}
	\tilde f_i =  f_i \circ \gamma_i^{**}  
	&=& c_i g \circ \gamma_i  \circ \gamma_i^* \circ \hat \gamma_0^{-1} + e_i \\
	&=& c_i g \circ \gamma_i  \circ \gamma_i^{-1} \circ \gamma_0 \circ \hat \gamma_0^{-1}  +e_i \\
	&=& c_i g  \circ  \gamma_0 \circ \hat \gamma_0^{-1}  +e_i 
\end{eqnarray*}

Taking average over the sample, we have
\begin{eqnarray*}
	\hat g_n = \frac{1}{n} \sum_{i=1}^n \tilde f_i 
	&=& \left( \frac{1}{n} \sum_{i=1}^n  c_i \right ) g  \circ  (\gamma_0 \circ \hat \gamma_0^{-1})
	+  \left( \frac{1}{n} \sum_{i=1}^n  e_i \right )   
\end{eqnarray*}

By Assumption \ref{assp:cont} on $g$ and the result that $\hat \gamma_0 \rightarrow \gamma_0 \ \ (a.s.)$, we have 
\begin{eqnarray*}
	\hat g_n &\rightarrow&  1 \cdot g + 0 = g   \ \ (a.s.) \ \ \ (n \rightarrow \infty) 
\end{eqnarray*}

\hfill $\square$

\section{Direct Functional Registration on Original Functions}
\label{f_regist}
\begin{figure}[ht!]
	\centering
	\begin{subfigure}[h]{0.25\textwidth}
		\includegraphics[width=\textwidth]{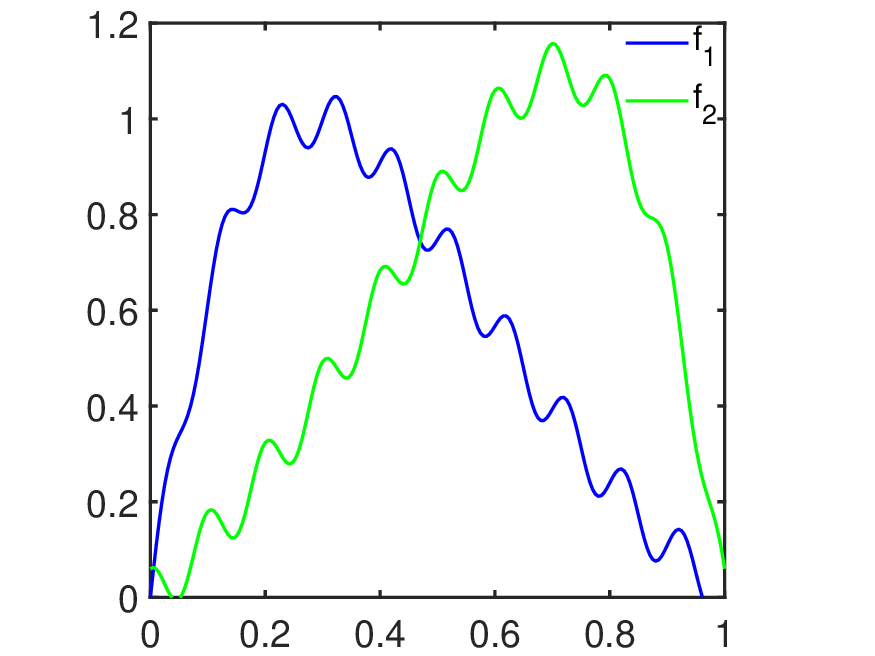}
		\includegraphics[width=\textwidth]{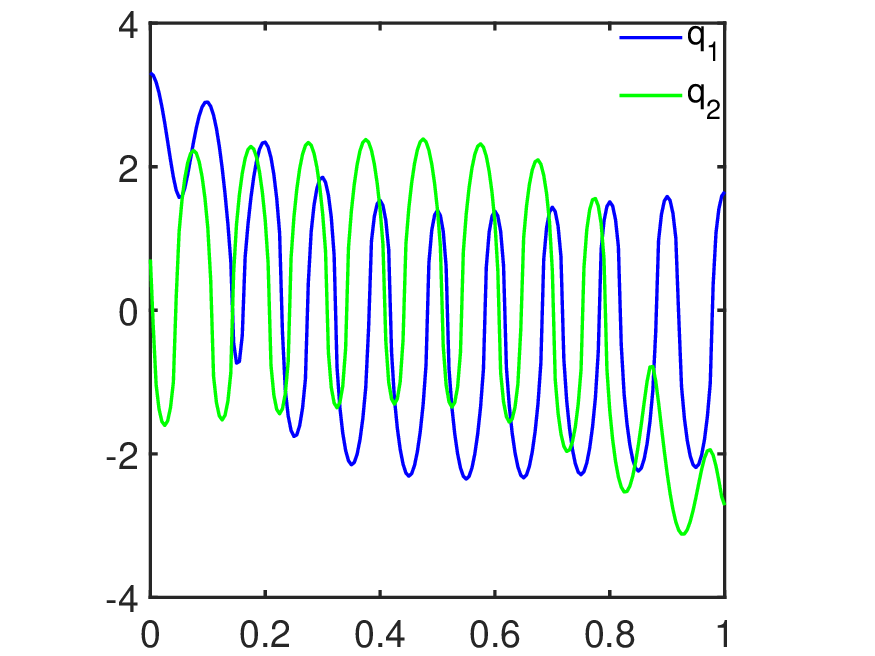}
		\caption{observations\\ SRVFs}
	\end{subfigure}\hspace{-0.5cm}
	\begin{subfigure}[h]{0.25\textwidth}
		\includegraphics[width=\textwidth]{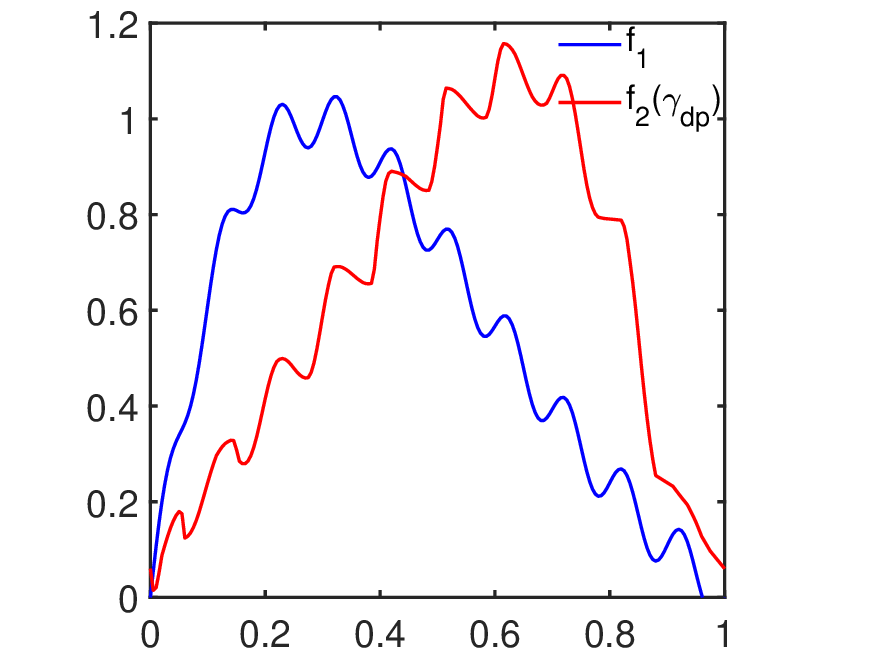}
		\includegraphics[width=\textwidth]{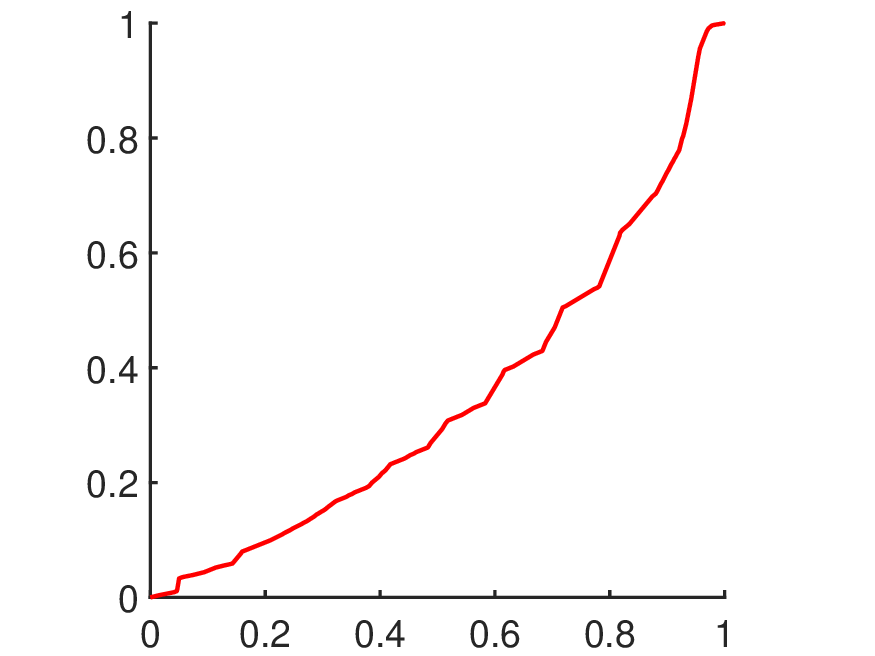}
		\captionsetup{justification=centering}
		\caption{DP aligned \\ warpings}
	\end{subfigure}\hspace{-0.5cm}
	\begin{subfigure}[h]{0.25\textwidth}
		\includegraphics[width=\textwidth]{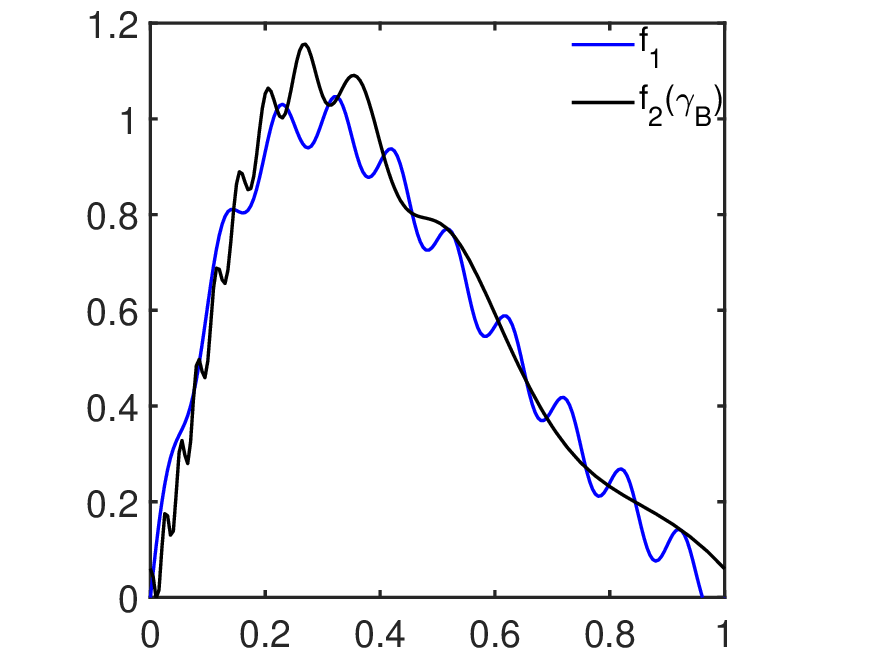}
		\includegraphics[width=\textwidth]{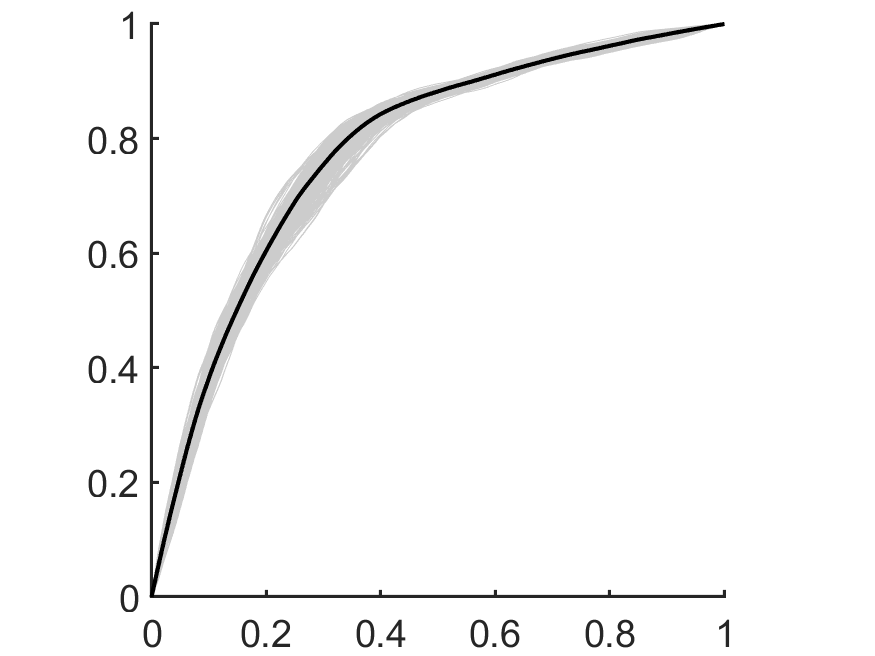}
		\captionsetup{justification=centering}
		\caption{Bayesian aligned \\ warpings}
	\end{subfigure}
	\caption{Comparison of SRVF framework and Original f framework when noise presents.}
	\label{fig:usef}
\end{figure}

The utilization of the SRVF framework, highlighted by \cite{srivastava2016functional}, provides a valuable advantage in alleviating the pinching effect. However, performing function registration in the SRVF space introduces another challenge. Specifically, in cases where the function lacks smoothness, leveraging the SRVF framework may excessively emphasize noise rather than the primary peak of interest.  See in Figure \ref{fig:usef}.  Panel (a) displays the original noisy observations, $f_1$ and $f_2$, in the first row, and their corresponding SRVF functions in the second row. When aligning these functions via the SRVF framework, there's a potential risk of missing the primary peak pattern. This is illustrated in first row of Panel (b), where the blue and red curves represent $f_1$ and the aligned $f_2$, respectively. The associated warping function is depicted in the second row's subfigure of Panel (b). The alignment outcomes using the original f function are visible in Panel (c). This approach effectively aligns the major peak patterns while disregarding minor noise.

By leveraging the smoothing effect provided by the warping prior in a Bayesian framework, it becomes possible to address pinching challenges without the need for transformation into the SRVF space. Moreover, directly utilizing the original function allows for resolving the noise-related issue mentioned earlier.

Now, For two function $f_1,\, f_2$, we can directly assume a zero mean Gaussian process for the difference of these two functions, i.e., ${(f_1-f_2\circ \gamma)|\gamma}\sim GP$. If we use $f_1([t])$ and $(f_2,\gamma)([t])$ to denote vectors evaluated at the same finite points on the domain of $f_1(t)$ and $(f_2\circ \gamma)(t)$, respectively, then the joint distribution of these finite differences $f_1([t])-(f_2,\gamma)([t])|\gamma$ is a multivariate normal distribution based on the Gaussian process assumption, i.e., $\Big\{f_1([t])-(f_2\circ\gamma)([t])|\gamma\Big\}\sim N_k(0_k,\sigma I_{k\times k})$, where $k$ is the number of points. The likelihood is given as: 
\begin{eqnarray}
	\pi(f_1,f_2|\gamma) &\propto& \exp\Big(-\frac{1}{2\sigma^2} \|f_1([t])-(f_2\circ\gamma)([t])\|^2\Big). \nonumber 
\end{eqnarray}  

Same as before, we propose to use a Gaussian process prior, denoted by $\mu_0$, to model the transformed warping functions in the $\mathbb L^2$ space, i.e.,$h(t) = \log(\dot{\gamma}(t))-\int_{0}^{1}\log(\dot{\gamma}(s))ds\sim GP(\mu_h, C_h)$. Note that the prior distribution of $\gamma$ are probability measures on the space $H(0, 1)$. The posterior measure, denoted by $\mu$, is absolutely continuous with respect to the prior $\mu_0$ with density can be written as following:
\begin{eqnarray}
	\frac{d\mu}{d \mu_0}(\gamma_h) &\propto& \pi(f_1,f_2|\gamma_h)  \nonumber\\
	&\propto& \exp\Big(-\frac{1}{2\sigma^2} \|f_1([t])-(f_2\circ\gamma_h)([t])\|^2\Big) \nonumber
\end{eqnarray}  
By directly applying Bayesian registration to the original functions of the previously noisy data, the alignment results are shown in the first row of Figure \ref{fig:usef}(c), and the corresponding warping is displayed in the second row.
\end{document}